\documentclass[a4paper,12pt]{article}
\usepackage{amssymb}
\usepackage{amsmath}
\usepackage[utf8]{inputenc}
\usepackage{fullpage}
\usepackage{boxedminipage}
\usepackage{listings}
\usepackage{minitoc}
\usepackage[pdftex]{graphicx}
\usepackage{graphicx}
\makeatletter \@addtoreset{equation}{section} \makeatother
\renewcommand{\theequation}{\thesection.\arabic{equation}}
\newif\ifpdf \ifx\pdfoutput\undefined \pdffalse
\else \pdfoutput=1 \pdftrue \fi \ifpdf \else \fi
\input epsf

\begin{document}

\ifpdf\DeclareGraphicsExtensions{.pdf, .jpg, .tif} \else%
\DeclareGraphicsExtensions{.eps, .jpg} \fi
\begin{titlepage}

    \thispagestyle{empty}
    \begin{flushright}
        \hfill{CERN-PH-TH/2006-108} \\
         \hfill{LNF-06-16-P}\\
        \hfill{UCLA/06/TEP/18}\\
    \end{flushright}

    %\vspace{5pt}
    \begin{center}
        { \LARGE{\bf Charge Orbits\\\vspace{5pt} of Symmetric Special Geometries\\ \vspace{8pt} and Attractors}}

        \vspace{28pt}

        {\bf Stefano Bellucci$^\clubsuit$, Sergio Ferrara$^{\diamondsuit\clubsuit\flat}$, Murat G\"{u}naydin$^{\spadesuit}$\\\vspace{5pt} and \ Alessio Marrani$^{\heartsuit\clubsuit}$}

        \vspace{5pt}

        {$\clubsuit$ \it INFN - Laboratori Nazionali di Frascati, \\
        Via Enrico Fermi 40,00044 Frascati, Italy\\
        \texttt{bellucci,marrani@lnf.infn.it}}

        \vspace{1pt}

        {$\diamondsuit$ \it Physics Department,Theory Unit, CERN, \\
        CH 1211, Geneva 23, Switzerland\\
        \texttt{sergio.ferrara@cern.ch}}

        \vspace{1pt}

        {$\flat$ \it Department of Physics and Astronomy,\\
        University of California, Los Angeles, CA USA\\
        \texttt{ferrara@physics.ucla.edu}}

        \vspace{1pt}

         {$\spadesuit$ \it Department of Physics, Penn State University\\
        University Park, PA16802, USA\\
        \texttt{murat@phys.psu.edu}}

        \vspace{1pt}

        {$\heartsuit$ \it Museo Storico della Fisica e\\
        Centro Studi e Ricerche ``Enrico Fermi"\\
        Via Panisperna 89A, 00184 Roma, Italy}

        %\vspace{15pt}

        \vspace{7pt}

        {ABSTRACT}
    \end{center}

    \vspace{5pt}

We study the critical points of the black hole scalar potential $V_{BH}$ in $%
N=2$, $d=4$ supergravity coupled to $n_{V}$ vector multiplets, in an
asymptotically flat extremal black hole background described by a
$2\left( n_{V}+1\right) $-dimensional dyonic charge vector and
(complex) scalar fields which are coordinates of a special
K\"{a}hler manifold.

For the case of homogeneous symmetric spaces, we find three general
classes of regular attractor solutions with non-vanishing
Bekenstein-Hawking entropy. They correspond to three
(inequivalent) classes of orbits of the charge vector, which is in a $%
2\left( n_{V}+1\right) $-dimensional representation $R_{V}$ of the $U$%
-duality group. Such orbits are non-degenerate, namely they have
non-vanishing quartic invariant (for rank-3 spaces). Other than the $\frac{1%
}{2}$-BPS one, there are two other distinct non-BPS classes of
charge orbits, one of which has vanishing central charge.

The three species of solutions to the $N=2$ extremal black hole
attractor equations give rise to different mass spectra of the
scalar fluctuations, whose pattern can be inferred by using
invariance properties of the critical points of $V_{BH}$ and some
group theoretical considerations on homogeneous symmetric special
K\"{a}hler geometry.

    \vspace{150pt}

\end{titlepage}
\newpage\baselineskip6 mm
\tableofcontents
\section{\label{Intro}Introduction}

The charge configurations of stationary, spherically symmetric,
asymptotically flat ``classical'' extremal black holes in $d=4$ dimensions
are expressed by the (electric-magnetic field strengths) representation $%
R_{V}$ of the duality group $G_{4}\equiv G$ of the underlying $d=4$
supergravity (SUGRA) theory. It has been known for some time \cite{FG} that
in the case of scalar manifolds which are symmetric spaces such ``charge
vectors'' belong to distinct classes of orbits of the representation $R_{V}$%
, i.e. that the $R_{V}$-representation space of $G$ can be actually divided
in disjoint classes of orbits. Such orbits are classified by suitable
constraints on the (lowest order) $G$-invariant quantity $\mathcal{I}$ built
out of the representation $R_{V}$.

Moreover, for $N\geqslant 3$, $d=4$ SUGRAs the scalar manifold of the theory
is an homogeneous symmetric space $G/H$, and $R_{V}$ is a real symplectic
representation of $G$. Thus, for such SUGRAs some relations between the
coset expressions of the aforementioned orbits and different real
(non-compact) forms of the stabilizer $H$ of the scalar manifold $G/H$ can
be established (even though in the cases having $\mathcal{I}=0$ such
relations do not rely on a critical Attractor Mechanism).

The work of \cite{FG} on the classification of the orbits of U-duality%
\footnote{%
Here $U$-duality is referred to as the ``continuous'' version, valid for
large values of the charges, of the $U$-duality groups introduced by Hull
and Townsend \cite{HT}.} groups in maximal supergravity and $N=2$ MESGTs in
five and four dimensions also suggested that four dimensional U-duality
groups may act as spectrum generating conformal symmetry groups in the
charge spaces of the corresponding five dimensional theories . In \cite{GKN}
this idea was developed further and extended to the proposal that the 3
dimensional U duality group $E_{8(8)}$ of maximal supergravity must
similarly act as a spectrum generating quasiconformal symmetry group in the
charge space of black hole solutions in four dimensions extended by an extra
coordinate interpreted as black hole entropy. This proposal extends
naturally to 3-dimensional U-duality groups of $N=2$ MESGTs acting as
spectrum generating quasiconformal symmetry groups in four dimensions \cite
{MG2005}. Remarkably, the quantization of the geometric quasiconformal
action yields directly the minimal unitary representation of the
corresponding U-duality group \cite{GKN2,GP1,GP2}. More recently it was
conjectured that the indexed degeneracies of certain $N=8$ and $N=4$ BPS
black holes are given by some automorphic forms related to the minimal
unitary representations of the corresponding 3 dimensional U-duality groups
\cite{pioline2005}. Motivated by these results and conjectures stationary
and spherically symmetric solutions of $N\geq 2$ supergravities with
symmetric scalar manifolds were studied in \cite{GNPW}. By using the
equivalence of four dimensional attractor flow with the geodesic motion on
the scalar manifold of the corresponding three dimensional theory the
authors of \cite{GNPW} argued further that the three-dimensional U-duality
groups must act as spectrum generating symmetry groups for BPS black hole
degeneracies in 4 dimensions.

\ In this paper we will study in full generality the critical points
(generically referred to as attractors) of the black hole scalar potential $%
V_{BH}$ for all $N=2$ symmetric special geometries in $d=4$. These extrema
describe the regular configurations (BPS as well as non-BPS) of $N=2,6,8$
SUGRAs, corresponding to a finite, non-vanishing quartic invariant $I=I_{4}$
and thus to extremal black holes with classical non-vanishing entropy $%
S_{BH}\neq 0$ . The related orbits in the $R_{V}$ of the $d=4$ duality group
$G$ will correspondingly be referred to as non-degenerate orbits. The
attractor equations for BPS configurations were first studied in \cite
{FKS,FK1,FK2,Strom}. Flow Equations for the general case were given in \cite
{FGK}, and recently non-BPS attractors have been found for $N=2$ and $N=8$
theories \cite{GIJT}-\cite{FKlast}.

Attractor solutions and their non-degenerate charge orbits in $d=5$ have
been recently classified for the case of all rank-2 symmetric spaces in \cite
{FG2}.\bigskip\

Let us start by considering $N=8$, $d=4$ SUGRA. For such a theory the
duality group is the Cremmer and Julia's \cite{CJ} one $G=E_{7(7)}$ and the
stabilizer is $H=SU(8)$, which is at the same time the $\mathcal{R}$%
-symmetry group of the $N=8$ supersymmetry algebra and the maximal compact
subgroup (m.c.s.) of $E_{7(7)}$. The dimension of the resulting real scalar
manifold $\frac{E_{7(7)}}{SU(8)}$ is $70$. The complex antisymmetric central
charge matrix $Z_{AB}$ ($A=1,....,N=8$) sits in the real symplectic,
fundamental representation $R_{V}=\mathbf{56}$ of $E_{7(7)}$. The two
classes of non-degenerate charge orbits of the $\mathbf{56}$ of $E_{7(7)}$
are classified by the unique Cartan-Cremmer-Julia quartic $E_{7(7)}$%
-invariant \cite{Cartan,CJ} $\mathcal{I}=I_{4}$ constructed from the $%
\mathbf{56}$ of $E_{7(7)}$ \cite{FG}. Depending on $sgn\left( I_{4}\right) $%
, one gets:
\begin{eqnarray}
I_{4} &>&0:\mathcal{O}_{\frac{1}{8}-BPS}=\frac{E_{7(7)}}{E_{6(2)}}=\frac{G}{%
\mathcal{H}_{0}}\text{ \ \ \ }\frac{1}{8}\text{-BPS;}  \label{1} \\
&&  \notag \\
I_{4} &<&0:\mathcal{O}_{non-BPS}=\frac{E_{7(7)}}{E_{6(6)}}=\frac{G}{\widehat{%
\mathcal{H}}_{0}}\text{ \ \ \ non-BPS.}  \label{2}
\end{eqnarray}

The real dimension of both orbits is $dim\left( E_{7}\right) -dim\left(
E_{6}\right) =55$. No other classes of non-degenerate orbits exist in this
case; this is essentially related to the fact that no other real
(non-compact) forms of $E_{6}$ exist in $E_{7(7)}$ beside $E_{6(2)}$ and $%
E_{6(6)}$. The $\frac{1}{8}$-BPS and non-BPS non-degenerate orbits
correspond to the maximal (non-compact) subgroup of $E_{7(7)}$ to be $%
E_{6(2)}\otimes U(1)$ and $E_{6(6)}\otimes SO(1,1)$, respectively.

Actually, $N=8$ non-degenerate orbits turn out to be classified by $5$
moduli-dependent parameters \cite{Cvetic,ADFFT,FK2}, $4$ positive
eigenvalues $\rho _{1,}...,\rho _{4}$ and an overall phase $\varphi $.
Indeed, by using the fact that under $SU(8)$ the symplectic, fundamental
real representation $R_{V}$ of $E_{7(7)}$ decomposes as $\mathbf{56}=\mathbf{%
28}+\overline{\mathbf{28}}$, one can see $Z_{AB}$ as a complex basis in the $%
\mathbf{56}$. Consequently, $Z_{AB}$ can be skew-diagonalized \cite{FSZ} by
performing an $SU(8)$ rotation\footnote{%
Actually, such a skew-diagonalization procedure is nothing but an
application of the Bloch-Messiah-Zumino Theorem \cite{BM,Zumino}\textbf{\ }%
to the case of $N=8$, $d=4$ SUGRA.}. Such a procedure corresponds to nothing
but a change of reference frame in the $\mathbf{56}$-representation space of
$E_{7(7)}$:

\begin{gather}
\text{Generic frame : }Z_{AB}  \notag \\
\notag \\
\downarrow SU(8)\text{ rotation}  \notag \\
\notag \\
\text{``Normal'' frame : }Z_{AB,normal}=e^{i\varphi /4}\left(
\begin{array}{cccc}
\rho _{1} &  &  &  \\
& \rho _{2} &  &  \\
&  & \rho _{3} &  \\
&  &  & \rho _{4}
\end{array}
\right) \otimes {\large \epsilon ,}  \notag \\
\notag \\
\rho _{1},\rho _{2},\rho _{3},\rho _{4},\in \mathbb{R}^{+},\text{ }\varphi
\in \left[ 0,8\pi \right) ,
\end{gather}
where $\epsilon $ is the 2-dim. symplectic metric
\begin{equation}
{\large \epsilon \equiv }\left(
\begin{array}{cc}
0 & -1 \\
1 & 0
\end{array}
\right) .
\end{equation}
By looking at $Z_{AB,normal}$, it is immediate to conclude that the overall
symmetry of such a moduli-dependent skew-diagonal $8\times 8$ complex matrix
in a generic point of the real 70-dim. scalar manifold $\frac{E_{7(7)}}{SU(8)%
}$ is $(SU(2))^{4}$. Thence, quartic $E_{7(7)}$-invariant $I_{4}$ can be
written in the ``normal'' frame as follows \cite{FM}:
\begin{eqnarray}
&&
\begin{array}{l}
I_{4,normal}\left( \rho _{1},\rho _{2},\rho _{3},\rho _{4},\varphi \right) =
\\
\\
=\left[ \left( \rho _{1}+\rho _{2}\right) ^{2}-\left( \rho _{3}+\rho
_{4}\right) ^{2}\right] \left[ \left( \rho _{1}-\rho _{2}\right) ^{2}-\left(
\rho _{3}-\rho _{4}\right) ^{2}\right] +8\rho _{1}\rho _{2}\rho _{3}\rho
_{4}(cos\varphi -1).
\end{array}
\notag \\
&&  \label{I4}
\end{eqnarray}

$N=8$ extremal black hole attractor equations \cite{FKlast} have only 2
distinct classes of regular solutions as expected from the analysis of \cite
{FG}:\smallskip

\textbf{1. }$\frac{1}{8}$\textbf{-BPS solution:}
\begin{equation}
\rho _{1}=\rho _{\frac{1}{8}-BPS}\in \mathbb{R}_{0}^{+},\varphi \in \left[
0,8\pi \right) ,\rho _{2}=\rho _{3}=\rho _{4}=0.  \label{primera-1}
\end{equation}
As given by Eq. (\ref{1}), the corresponding orbit in the $\mathbf{56}$ of $%
E_{7(7)}$ is $\mathcal{O}_{\frac{1}{8}-BPS}=\frac{E_{7(7)}}{E_{6(2)}}$, with
$I_{4,normal,\frac{1}{8}-BPS}=\rho _{\frac{1}{8}-BPS}^{4}>0$ and classical
entropy given by the Bekenstein-Hawking entropy-area formula \cite{BH1}
\begin{equation}
S_{BH,\frac{1}{8}-BPS}=\pi \sqrt{I_{4,normal,\frac{1}{8}-BPS}}=\pi \rho _{%
\frac{1}{8}-BPS}^{2}.  \label{S-BPS}
\end{equation}
\smallskip

\textbf{2. non-BPS solution: }
\begin{equation}
\rho _{1}=\rho _{2}=\rho _{3}=\rho _{4}=\rho _{non-BPS}\in \mathbb{R}%
_{0}^{+},\text{ \ }\varphi =\pi .  \label{segunda-1}
\end{equation}
As given by Eq. (\ref{2}), the corresponding orbit in the $\mathbf{56}$ of $%
E_{7(7)}$ is $\mathcal{O}_{non-BPS}=\frac{E_{7(7)}}{E_{6(6)}}$, with $%
I_{4,normal,non-BPS}=-16\rho _{non-BPS}^{4}<0$ and classical entropy
\begin{equation}
S_{BH,non-BPS}=\pi \sqrt{-I_{4,normal,non-BPS}}=4\pi \rho _{non-BPS}^{2}.
\label{S-non-BPS}
\end{equation}

The deep meaning of the extra factor $4$ in Eq. (\ref{S-non-BPS}) as
compared with Eq. (\ref{S-BPS}) can be clearly explained when considering
the so-called ``$stu$ interpretation'' of $N=8$ regular critical points \cite
{FKlast}. \smallskip

It is interesting to note that the symmetry gets enhanced at the particular
points of $\frac{E_{7(7)}}{SU(8)}$ given by the aforementioned regular
solutions. In general, \textit{the invariance properties of the regular
solutions to attractor eqs. are given by the m.c.s. of the stabilizer of the
corresponding charge orbit}. In the considered case $N=8$, at $\frac{1}{8}$%
-BPS and non-BPS critical point(s) the following symmetry enhancements
respectively hold:
\begin{eqnarray}
&&
\begin{array}{l}
\frac{1}{8}\text{-BPS}:(SU(2))^{4}\longrightarrow SU(2)\otimes
SU(6)=h_{0}=m.c.s.\left( \mathcal{H}_{0}=E_{6(2)}\right) ;
\end{array}
\\
&&  \notag \\
&&
\begin{array}{l}
\text{non-BPS}:(SU(2))^{4}\longrightarrow USp(8)=\widehat{h}%
_{0}=m.c.s.\left( \widehat{\mathcal{H}}_{0}=E_{6(6)}\right) .
\end{array}
\end{eqnarray}
\smallskip

The aim of the present work is to extend the results holding for $N=8$, $d=4$
SUGRA to the particular class of $N=2$, $d=4$ symmetric Maxwell-Einstein
SUGRA theories (MESGTs) \cite{GST1,GST2,GST3}. Such a class consists of $N=2$%
, $d=4$ SUGRAs sharing the following properties:

$i)$ beside the SUGRA multiplet, the matter content is given only by a
certain number $n_{V}$ of Abelian vector multiplets;

$ii)$ the space of the vector multiplets' scalars is an homogeneous
symmetric special K\"{a}hler manifold, i.e. a special K\"{a}hler manifold
with coset structure $\frac{G}{H_{0}\otimes U(1)}$, where $G$\ is a
semisimple non-compact Lie group and $H_{0}\otimes U(1)$\ is its m.c.s.;

$iii)$ the charge vector in a generic (dyonic) configuration with $n_{V}+1$
electric and $n_{V}+1$ magnetic charges sits in a real (symplectic)
representation $R_{V}$ of $G$ of $dim\left( R_{V}\right) =2\left(
n_{V}+1\right) $.

By exploiting such special features and relying on group theoretical
considerations, we will be able to relate the coset expressions of the
various distinct classes of non-degenerate orbits (of dimension $2n_{V}+1$)
in the $R_{V}$-representation space of $G$ to different real (non-compact)
forms of the compact group $H_{0}$. Correspondingly, we will solve the $N=2$
extremal black hole attractor eqs. for all such classes, also studying the
scalar mass spectrum of the theory corresponding to the obtained
solutions.\medskip

The plan of the paper is as follows.

In Sect. \ref{N=2-Orbits} we review some basic facts about $N=2$, $d=4$
symmetric MESGTs, their non-degenerate charge orbits and the relations with
the regular solutions of the $N=2$ extremal black hole attractor equations.

In Sect. \ref{Freud} we determine the orbits of the U-duality groups of $N=2$
MESGTs whose scalar manifolds are symmetric spaces, acting on the
representation $R_{V}$ of charges. With the exception of the $\frac{SU(1,n+1)%
}{SU(1+n)\otimes U(1)}$ family, all such MESGTs have their origin in
five-dimensional $N=2$ MESGTs defined by Euclidean Jordan algebras of degree
three. The five-dimensional correspondence between the vector fields (and
hence charges) and the elements of the Jordan algebra extends to a
four-dimensional correspondence between the field strengths (and their
magnetic duals) and the elements of Freudenthal triple system defined over
the corresponding Jordan algebra \cite{GST1,MG2005,GP2}. The automorphism
groups of the FTSs are isomorphic to the $U$-duality groups of the
corresponding four-dimensional $N=2$ MESGTs. Using the action of the
automorphism group on the considered FTSs, we determine the orbits with
non-vanishing quartic norms, which correspond to the quartic invariants of
the MESGTs. We find three classes of such non-degenerate charge orbits, two
with positive norm and one with a negative norm\footnote{%
This is to be contrasted with the symmetric situation in $d=5$, where one
finds two orbits with positive cubic norm and two with negative norm, which
are pairwise isomorphic \cite{FG,FG2}.}.

Thence, the first two Subsects. of Sect. \ref{N=2-Attractors} are devoted to
the general analysis of the three classes of regular extremal black hole
attractors of $N=2$, $d=4$ symmetric ''magical'' MESGTs, and of the
corresponding classes of non-degenerate charge orbits in the symplectic
representation space of the relevant $d=4$ duality group. In particular, the
$\frac{1}{2}$-BPS solutions are treated in Subsect. \ref{N=2-Attractors-BPS}%
, while the two general species of non-BPS $Z\neq 0$ and non-BPS $Z=0$
attractors are respectively considered in Subsubsects. \ref
{N=2-Attractors-non-BPS-1} and \ref{N=2-Attractors-non-BPS-2}.

Subsect. \ref{Orb-Attr-O-H} deal with the two noteworthy cases of $N=2$, $%
d=4 $ symmetric ''magical'' MESGTs based on the manifolds $\frac{E_{7(-25)}}{%
E_{6}\otimes U(1)}$ (Subsubsect. \ref{Orb-Attr-O}) and $\frac{SO^{\ast }(12)%
}{U(6)}$ (Subsubsect. \ref{Orb-Attr-H}). This latter homogeneous symmetric
special K\"{a}hler manifold is considered also in Subsubsect. \ref{dual-2-6}%
, where its dual role in the interplay between $N=2 $ and $N=6$, $d=4$
SUGRAs is pointed out.

The splittings of the mass spectra of $N=2$, $d=4$ symmetric ''magical''
MESGTs along their three classes of non-degenerate charge orbits are studied
in Sect. \ref{N=2-Spectra}.

Finally, the concluding Sect. \ref{Conclusion} contains some general remarks
and observations, as well as an outlook of possible future further
developments along the considered research directions.

Two Appendices conclude the paper; they treat the regular attractors,
non-degenerate charge orbits and critical mass spectra of the $N=2$, $d=4$
symmetric MESGTs based on the sequences $\frac{SU(1,1+n)}{U(1)\otimes SU(1+n)%
}$ (Appendix I) and $\frac{SU(1,1)}{U(1)}\otimes \frac{SO(2,2+n)}{%
SO(2)\otimes SO(2+n)}$ (Appendix II).

\section{\label{N=2-Orbits}BPS and non-BPS Attractors :\newline
~Charge Orbits of $N=2$, $d=4$ MESGTs}

The symmetric special K\"{a}hler manifolds of $N=2$, $d=4$ MESGTs have been
classified in the literature \cite{CVP,dWVVP}. \textbf{\ }With the exception
of the family whose prepotential is quadratic , all such theories can be
obtained by dimensional reduction of the $N=2$, $d=5$ MESGTs that were
constructed in \cite{GST1,GST2,GST3}. The MESGTs with symmetric manifolds
that originate from 5 dimensions all have cubic prepotentials determined by
the norm form of the Jordan algebra of degree three that defines them \cite
{GST1,GST2,GST3}. They include the two infinite sequences
\begin{eqnarray}
&&
\begin{array}{l}
I:\text{\ }\frac{SU(1,1+n)}{U(1)\otimes SU(1+n)},\text{ }r=1;
\end{array}
\\
&&  \notag \\
&&
\begin{array}{l}
II:\frac{SU(1,1)}{U(1)}\otimes \frac{SO(2,2+n)}{SO(2)\otimes SO(2+n)},r=3,
\end{array}
\end{eqnarray}
where $r$ stands for the rank of the coset. Of these two infinite families,
the first family is the one whose prepotentials are quadratic. The second
family has a five dimensional origin and its associated Jordan algebras are
not simple. It is referred to as the generic Jordan family since it exists
for any $n$. The first elements of such sequences (obtained for $n=0$)
respectively correspond to the following manifolds and holomorphic
prepotential functions in special coordinates:
\begin{eqnarray}
&&
\begin{array}{l}
I_{0}:\frac{SU(1,1)}{U(1)},\text{ }F(t)=\frac{i}{4}\left( t^{2}-1\right) ;
\end{array}
\\
&&  \notag \\
&&
\begin{array}{l}
II_{0}:\frac{SU(1,1)\otimes SO(2,2)}{U(1)\otimes SO(2)\otimes SO(2)}=\left(
\frac{SU(1,1)}{U(1)}\right) ^{3},F(s,t,u)=stu.
\end{array}
\label{stu}
\end{eqnarray}
In general, all manifolds of type $I$ correspond to quadratic prepotentials (%
$C_{ijk}=0$), as well as all manifolds of type $II$ correspond to cubic
prepotentials (in special coordinates $F=\frac{1}{3!}d_{ijk}t^{i}t^{j}t^{k}$
and therefore $C_{ijk}=e^{K}d_{ijk}$, where $K$ denotes the K\"{a}hler
potential and $d_{ijk}$ is a completely symmetric rank-3 constant tensor).
The 3-moduli case $II_{0}$ is the well-known $stu$ model \cite{BKRSW}, whose
noteworthy \textit{triality symmetry} has been recently related to quantum
information theory \cite{Duff,KL,Levay}.

\textbf{\ }Beside the infinite sequence $II$, there exist four other MESGTs
defined by simple Jordan algebras of degree three with the following rank-3
coset manifolds:
\begin{eqnarray}
&&
\begin{array}{l}
III:\frac{E_{7(-25)}}{E_{6}\otimes U(1)};
\end{array}
\\
&&  \notag \\
&&
\begin{array}{l}
IV:\frac{SO^{\ast }(12)}{U(6)};
\end{array}
\\
&&  \notag \\
&&
\begin{array}{l}
V:\frac{SU(3,3)}{S\left( U(3)\otimes U(3)\right) }=\frac{SU(3,3)}{%
SU(3)\otimes SU(3)\otimes U(1)};
\end{array}
\\
&&  \notag \\
&&
\begin{array}{l}
VI:\frac{Sp(6,\mathbb{R})}{U(3)}.
\end{array}
\end{eqnarray}
The $N=2$, $d=4$ MESGTs whose geometry of scalar fields is given by the
manifolds $III$-$VI$ are called ``magical'', since their symmetry groups are
the groups of the famous Magic Square of Freudenthal, Rozenfeld and Tits
associated with some remarkable geometries \cite{Freudenthal2,magic}. The
four $N=2$, $d=4$ ``magical'' MESGTs $III$-$VI$, as their 5-d. versions, are
defined by four simple Jordan algebras $J_{3}^{\mathbb{O}}$, $J_{3}^{\mathbb{%
H}}$, $J_{3}^{\mathbb{C}}$ and $J_{3}^{\mathbb{R}}$ of degree 3 with
irreducible norm forms, namely by the Jordan algebras of Hermitian $3\times
3 $ matrices over the four division algebras, i.e. respectively over the
octonions $\mathbb{O}$, quaternions $\mathbb{H}$, complex numbers $\mathbb{C}
$ and real numbers $\mathbb{R}$ \cite
{GST1,GST2,GST3,Jordan,Jacobson,Guna1,GPR}.

By denoting with $n_{V}$ the number of vector multiplets coupled to the
SUGRA one, the total number of Abelian vector fields in the considered $N=2$%
, $d=4$ MESGT is $n_{V}+1$; correspondingly, the real dimension of the
corresponding scalar manifold is $2n_{V}=dim\left( G\right) -dim\left(
H_{0}\right) -1$. Since the $2\left( n_{V}+1\right) $-dim. vector of
extremal black hole charge configuration is given by the fluxes of the
electric and magnetic field-strength two-forms, it is clear that $dim_{%
\mathbb{R}}\left( R_{V}\right) =2\left( n_{V}+1\right) $.

Since $H_{0}$ is a proper compact subgroup of the duality semisimple group $%
G $, we can decompose the $2\left( n_{V}+1\right) $-dim. real symplectic
representation $R_{V}$ of $G$ in terms of complex representations of $H_{0}$%
, obtaining in general the following decomposition scheme:
\begin{equation}
R_{V}\longrightarrow R_{H_{0}}+\overline{R_{H_{0}}}+\mathbf{1}_{\mathbb{C}}+%
\overline{\mathbf{1}}_{\mathbb{C}}=R_{H_{0}}+\mathbf{1}_{\mathbb{C}}+c.c.,
\label{decomp1}
\end{equation}
where ``$c.c.$'' stands for the complex conjugation of representations, and $%
R_{H_{0}}$ is a certain complex representation of $H_{0}$.\footnote{%
As will be discussed in Sect. 3, this decomposition reflects the
decomposition of the corresponding Freudenthal triple system with respect to
the underlying Jordan algebra.}\textbf{\ }

The basic data of the cases $I$-$VI$ listed above are summarized in Tables 1
and 2.
\begin{table}[t]
\begin{center}
\begin{tabular}{|c||c|c|}
\hline
& $I$ & $II$ \\ \hline\hline
$G$ & $SU(1,1+n)$ & $SU(1,1)\otimes SO(2,2+n)$ \\ \hline
$H_{0}$ & $SU(1+n)$ & $SO(2)\otimes SO(2+n)$ \\ \hline
$r$ & $1$ & $3$ \\ \hline
$dim_{\mathbb{R}}\left( \frac{G}{H_{0}\otimes U(1)}\right) $ & $2\left(
n+1\right) $ & $2\left( n+3\right) $ \\ \hline
$n_{V}$ & $n+r=n+1$ & $n+r=n+3$ \\ \hline
$R_{V}$ & $\left( \mathbf{2}\left( \mathbf{n+2}\right) \right) _{\mathbb{R}}$
& $\left( \mathbf{2}\left( \mathbf{n+4}\right) \right) _{\mathbb{R}}$ \\
\hline
$R_{H_{0}}$ & $\left( \mathbf{n+1}\right) _{\mathbb{C}}$ & $\left( \mathbf{%
n+2+1}\right) _{\mathbb{C}}$ \\ \hline
$dim_{\mathbb{R}}\left( R_{V}\right) $ & $2\left( n+2\right) $ & $2\left(
n+4\right) $ \\ \hline
$dim_{\mathbb{R}}\left( R_{H_{0}}\right) $ & $2\left( n+1\right) $ & $%
2\left( n+3\right) $ \\ \hline
$
\begin{array}{c}
R_{V} \\
\downarrow \\
R_{H_{0}}+\mathbf{1}_{\mathbb{C}}+ \\
+c.c.
\end{array}
$ & $
\begin{array}{c}
\left( \mathbf{2}\left( \mathbf{n+2}\right) \right) _{\mathbb{R}} \\
\downarrow \\
\left( \mathbf{n+1}\right) _{\mathbb{C}}+\mathbf{1}_{\mathbb{C}}+ \\
+c.c.
\end{array}
$ & $
\begin{array}{c}
\left( \mathbf{2}\left( \mathbf{n+4}\right) \right) _{\mathbb{R}} \\
\downarrow \\
\left( \mathbf{n+2+1}\right) _{\mathbb{C}}+\mathbf{1}_{\mathbb{C}}+ \\
+c.c.
\end{array}
$ \\ \hline
\end{tabular}
\end{center}
\caption{\textbf{Basic data of the two sequences of symmetric $N=2$, $d=4$
MESGTs}}
\end{table}
\begin{table}[h]
\begin{center}
\begin{tabular}{|c||c|c|c|c|}
\hline
& $J_{3}^{\mathbb{O}}\leftrightarrow III$ & $J_{3}^{\mathbb{H}%
}\leftrightarrow IV$ & $J_{3}^{\mathbb{C}}\leftrightarrow V$ & $J_{3}^{%
\mathbb{R}}\leftrightarrow VI$ \\ \hline\hline
$G$ & $E_{7(-25)}$ & $SO^{\ast }(12)$ & $SU(3,3)$ & $Sp(6,\mathbb{R})$ \\
\hline
$H_{0}$ & $E_{6}$ & $SU(6)$ & $SU(3)\otimes SU(3)$ & $SU(3)$ \\ \hline
$r$ & $3$ & $3$ & $3$ & $3$ \\ \hline
$dim_{\mathbb{R}}\left( \frac{G}{H_{0}\otimes U(1)}\right) $ & $54$ & $30$ &
$18$ & $12$ \\ \hline
$n_{V}$ & $27$ & $15$ & $9$ & $6$ \\ \hline
$R_{V}$ & $\mathbf{56}_{\mathbb{R}}$ & $\mathbf{32}_{\mathbb{R}}$ & $\mathbf{%
10}_{\mathbb{R}}$ & $\mathbf{14}_{\mathbb{R}}^{\prime }$ \\ \hline
$R_{H_{0}}$ & $\mathbf{27}_{\mathbb{C}}$ & $\mathbf{15}_{\mathbb{C}}$ & $%
\left( \mathbf{3,3}^{\prime }\right) _{\mathbb{C}}$ & $\mathbf{6}_{\mathbb{C}%
}$ \\ \hline
$dim_{\mathbb{R}}\left( R_{V}\right) $ & $56$ & $32$ & $20$ & $14$ \\ \hline
$dim_{\mathbb{R}}\left( R_{H_{0}}\right) $ & $54$ & $30$ & $18$ & $12$ \\
\hline
$
\begin{array}{c}
R_{V} \\
\downarrow \\
R_{H_{0}}+\mathbf{1}_{\mathbb{C}}+ \\
+c.c.
\end{array}
$ & $
\begin{array}{c}
\mathbf{56}_{\mathbb{R}} \\
\downarrow \\
\mathbf{27}_{\mathbb{C}}+\mathbf{1}_{\mathbb{C}}+ \\
+c.c.
\end{array}
$ & $
\begin{array}{c}
\mathbf{32}_{\mathbb{R}} \\
\downarrow \\
\mathbf{15}_{\mathbb{C}}+\mathbf{1}_{\mathbb{C}}+ \\
+c.c.
\end{array}
$ & $
\begin{array}{c}
\mathbf{10}_{\mathbb{R}} \\
\downarrow \\
\left( \mathbf{3,3}^{\prime }\right) _{\mathbb{C}}+\mathbf{1}_{\mathbb{C}}+
\\
+c.c.
\end{array}
$ & $
\begin{array}{c}
\mathbf{14}_{\mathbb{R}}^{\prime } \\
\downarrow \\
\mathbf{6}_{\mathbb{C}}+\mathbf{1}_{\mathbb{C}}+ \\
+c.c.
\end{array}
$ \\ \hline
\end{tabular}
\end{center}
\caption{\textbf{Basic data of the four ``magical'' symmetric $N=2$, $d=4$
MESGTs}. {$\mathbf{14}_{\mathbb{R}}^{\prime }$ is the\textbf{\ }rank-3
antisymmetric tensor representation of $Sp(6,\mathbb{R})$.} In $\left(
\mathbf{3,3}^{\prime }\right) _{\mathbb{C}}$ the prime distinguishes the
representations of the two distinct $SU(3)$ groups.}
\end{table}

It was shown in \cite{FG} that $\frac{1}{2}$-BPS orbits of $N=2$, $d=4$
symmetric MESGTs are coset spaces of the form
\begin{equation}
\begin{array}{l}
\mathcal{O}_{\frac{1}{2}-BPS}=\frac{G}{H_{0}}, \\
\\
dim_{\mathbb{R}}\left( \mathcal{O}_{\frac{1}{2}-BPS}\right) =dim\left(
G\right) -dim\left( H_{0}\right) =2n_{V}+1=dim_{\mathbb{R}}\left(
R_{V}\right) -1.
\end{array}
\end{equation}
\bigskip

Now, in order to proceed further, we need to consider the $N=2$ extremal
black attractor eqs.; these are nothing but the criticality conditions for
the $N=2$ black hole scalar potential \cite{CDF,FK1}
\begin{equation}
V_{BH}\equiv \left| Z\right| ^{2}+G^{i\overline{i}}D_{i}Z\overline{D}_{%
\overline{i}}\overline{Z}  \label{VBH-def}
\end{equation}
in the corresponding special K\"{a}hler geometry \cite{FGK}:
\begin{equation}
\partial _{i}V_{BH}=0\Longleftrightarrow 2\overline{Z}D_{i}Z+iC_{ijk}G^{j%
\overline{j}}G^{k\overline{k}}\overline{D}_{\overline{j}}\overline{Z}%
\overline{D}_{\overline{k}}\overline{Z}=0,\forall i=1,...,n_{V}.
\label{AEs1}
\end{equation}
$C_{ijk}$ is the rank-3, completely symmetric, covariantly holomorphic
tensor of special K\"{a}hler geometry, satisfying (see e.g. \cite{CDFVP})
\begin{equation}
\overline{D}_{\overline{l}}C_{ijk}=0,~~~D_{[l}C_{i]jk}=0,  \label{propr-C}
\end{equation}
where the square brackets denote antisymmetrization with respect to the
enclosed indices.

For symmetric special K\"{a}hler manifolds the tensor $C_{ijk}$ is
covariantly constant:\textbf{\ }
\begin{equation}
D_{i}C_{jkl}=0,  \label{CERN1}
\end{equation}
which further implies \cite{GST2,CVP}\textbf{\ }
\begin{equation}
G^{k\overline{k}}G^{r\overline{j}}C_{r(pq}C_{ij)k}\overline{C}_{\overline{k}%
\overline{i}\overline{j}}=\frac{4}{3}G_{\left( q\right| \overline{i}%
}C_{\left| ijp\right) }.  \label{CERN2}
\end{equation}
This equation is simply the four dimensional version of the adjoint identity
satisfied by all Jordan algebras of degree three that define the
corresponding MESGTs in five dimensions.

$Z$ is the $N=2$ ``central charge'' function, whereas $\left\{
D_{i}Z\right\} _{i=1,...,n_{V}}$ is the set of its K\"{a}hler-covariant
holomorphic derivatives, which are nothing but the ``matter charges''
functions of the system. Indeed, the sets\footnote{%
We always consider ``classical'' frameworks, disregarding the actual
quantization of the ranges of the electric and magnetic charges $q_{0}$, $%
q_{i}$, $p^{0}$ and $p^{i}$. That is why we consider $\mathbb{R}^{2n_{V}+2}$
rather than the $\left( 2n_{V}+2\right) $-dim. charge lattice $\widehat{%
\Gamma }_{(p,q)}$.} $\left\{ q_{0},q_{i},p^{0},p^{i}\right\} \in \mathbb{R}%
^{2n_{V}+2}$ and $\left\{ Z,D_{i}Z\right\} \in \mathbb{C}^{n_{V}+1}$ (when
evaluated at purely $\left( q,p\right) $-dependent critical values of the
moduli) are two equivalent basis for the charges of the system, and they are
related by a particular set of identities of special K\"{a}hler geometry
\cite{CDF,BFM,AoB}. The decomposition (\ref{decomp1}) corresponds to nothing
but the splitting of the sets $\left\{ q_{0},q_{i},p^{0},p^{i}\right\} $ ($%
\left\{ Z,D_{i}Z\right\} $) of $2n_{V}+2$ ($n_{V}+1$) real\ (complex)
charges (``charge'' functions) in $q_{0},p^{0}$ ($Z$) (related to the
graviphoton, and corresponding to $\mathbf{1}_{\mathbb{C}}+c.c.$) and in $%
\left\{ q_{i},p^{i}\right\} $ ($\left\{ D_{i}Z\right\} $) (related to the $%
n_{V}$\ vector multiplets, and corresponding to $R_{H_{0}}+c.c.$).\medskip

In order to perform the subsequent analysis of orbits, it is convenient to
use ``flat'' $I$-indices by using the (inverse) $n_{V}$-bein $e_{I}^{i}$ of $%
\frac{G}{H_{0}\otimes U(1)}$:
\begin{equation}
D_{I}Z=e_{I}^{i}D_{i}Z.
\end{equation}
By switching to ``flat'' local $I$-indices, the special K\"{a}hler metric $%
G_{i\overline{j}}$ (assumed to be \textit{regular}, i.e. strictly positive
definite everywhere) will become nothing but the Euclidean $n_{V}$-dim.
metric $\delta _{I\overline{J}}$. Thus, the attractor eqs. (\ref{AEs1}) can
be ``flatted'' as follows:
\begin{equation}
\partial _{I}V_{BH}=0\Longleftrightarrow 2\overline{Z}D_{I}Z+iC_{IJK}\delta
^{J\overline{J}}\delta ^{K\overline{K}}\overline{D}_{\overline{J}}\overline{Z%
}\overline{D}_{\overline{K}}\overline{Z}=0,\forall I=1,...,n_{V}.
\label{AEs2}
\end{equation}
Note that $C_{IJK}$ becomes an $H_{0}$-invariant tensor \cite{ADFFF}. This
is possible because $C_{ijk}$ in special coordinates is proportional to the
invariant tensor $d_{IJK}$ of the $d=5$ $U$-duality group $G_{5}$. $G_{5}$
and $H_{0}$ correspond to two different real forms of the same Lie algebra
\cite{GST2}.

As it is well known, $\frac{1}{2}$-BPS attractors are given by the following
solution \cite{FGK} of attractor eqs. (\ref{AEs1}) and (\ref{AEs2}):
\begin{equation}
Z\neq 0,\text{ }D_{i}Z=0\Leftrightarrow D_{I}Z=0,\forall i,I=1,...,n_{V}.
\label{BPS}
\end{equation}

Since the ``flatted matter charges'' $D_{I}Z$\ are a vector of $R_{H_{0}}$,
Eq. (\ref{BPS}) directly yields that $\frac{1}{2}$-BPS solutions are
manifestly $H_{0}$-invariant. In other words, since the $N=2$, $\frac{1}{2}$%
-BPS orbits are of the form $\frac{G}{H_{0}}$, the condition for the $\left(
n_{V}+1\right) $-dim. complex vector $\left( Z,D_{i}Z\right) $ to be $H_{0}$%
-invariant is precisely given by Eq. (\ref{BPS}), defining $N=2$, $\frac{1}{2%
}$-BPS attractor solutions.\smallskip

Thus, as for the $N=8$ regular solutions (\ref{primera-1}) and (\ref
{segunda-1}), also for the $N=2$ $\frac{1}{2}$-BPS case \textit{the
invariance properties of the solutions at the critical point(s) are given by
the m.c.s. of the stabilizer of the corresponding charge orbit, }which in
the present case is the compact stabilizer itself. Thus, at $N=2$ $\frac{1}{2%
}$-BPS critical points the following enhancement of symmetry holds:
\begin{equation}
\mathcal{S}\longrightarrow H_{0},  \label{symm-en-BPS}
\end{equation}
where here and below $\mathcal{S}$ denotes the compact symmetry of a generic
orbit of the real symplectic representation $R_{V}$ of the $d=4$ duality
group $G$.\bigskip

However, all the scalar manifolds of $N=2$, $d=4$ symmetric MESGTs have
other species of regular critical points $V_{BH}$ (and correspondingly other
classes of non-degenerate charge orbits).

In recent months ($N=2$\ non-BPS) non-supersymmetric extremal black hole
attractors have been studied in the literature \cite{GIJT}\nocite{K1}\nocite
{TT}\nocite{G}\nocite{GJMT}\nocite{Ebra1}\nocite{K2}\nocite{Ira1}\nocite
{Spin}\nocite{BFM}\nocite{Sen1}-\cite{Ebra2}, also in the case of
non-homogeneous (and non-symmetric) special K\"{a}hler manifolds, e.g. in
the large volume limit of Calabi-Yau compactifications of Type IIA
superstrings \cite{TT}. Away from such a limit, exact non-BPS solutions have
been given for $n_{V}=1$ in the case of quintic in $\mathbb{C}P^{4}$
(working in the IIB mirror description) and of sixtic in $WP_{1,1,1,1,2}^{4}$
respectively in \cite{TT} and \cite{G}. While all treated cases are $N=2$
non-BPS, $Z\neq 0$ attractors, in \cite{G} the first (and, as far as we
know, the only) known example of $N=2$ non-BPS, $Z=0$ attractor is
presented. \textbf{\ }

Concerning the $N=2$, $d=4$ symmetric MESGTs, the rank-1 sequence $I$ has
one more, non-BPS class of orbits (with vanishing central charge), while all
rank-3 aforementioned cases $II$-$VI$ have two more distinct non-BPS classes
of orbits, one of which with vanishing central charge.

The results about the classes of non-degenerate charge orbits of $N=2$, $d=4$%
\ symmetric MESGTs are summarized in Table 3\footnote{%
It is here worth remarking that the column on the right of Table 2 of \cite
{FG} is not fully correct.
\par
Indeed, such a column coincides with the central column of Table 3 of the
present paper (by disregarding case $I$ and shifting $n\rightarrow n-2$ in
case $II$), listing the non-BPS, $Z\neq 0$ orbits of $N=2$, $d=4$ symmetric
MESGTs, which are all characterized by a strictly negative quartic $E_{7}$%
-invariant $I_{4}$. This does not match what is claimed in \cite{FG}, where
such a column is stated to list the particular class of orbits with $I_{4}>0$
and eigenvalues of opposite sign in pair.
\par
Actually, the statement of \cite{FG}\textbf{\ }holds true only for the case $%
I$ (which, by shifting $n\rightarrow n-1$, coincides with the last entry of
the column on the right of Table 2 of \cite{FG}). On the other hand, such a
case is the only one which cannot be obtained from $d=5$ by dimensional
reduction. Moreover, it is the only one not having non-BPS, $Z\neq 0$
orbits, rather it is characterized only by a class of non-BPS orbits with $%
Z=0$ and $I_{4}>0.$}.

\begin{table}[t]
\begin{center}
\begin{tabular}{|c||c|c|c|}
\hline
& $
\begin{array}{c}
\\
\frac{1}{2}\text{-BPS orbits } \\
~~\mathcal{O}_{\frac{1}{2}-BPS}=\frac{G}{H_{0}} \\
~
\end{array}
$ & $
\begin{array}{c}
\\
\text{non-BPS, }Z\neq 0\text{ orbits} \\
\mathcal{O}_{non-BPS,Z\neq 0}=\frac{G}{\widehat{H}}~ \\
~
\end{array}
$ & $
\begin{array}{c}
\\
\text{non-BPS, }Z=0\text{ orbits} \\
\mathcal{O}_{non-BPS,Z=0}=\frac{G}{\widetilde{H}}~ \\
~
\end{array}
$ \\ \hline\hline
$
\begin{array}{c}
\\
I \\
~
\end{array}
$ & $\frac{SU(1,n+1)}{SU(n+1)}~$ & $-$ & $\frac{SU(1,n+1)}{SU(1,n)}~$ \\
\hline
$
\begin{array}{c}
\\
II \\
~
\end{array}
$ & $\frac{SU(1,1)\otimes SO(2,2+n)}{SO(2)\otimes SO(2+n)}~$ & $\frac{%
SU(1,1)\otimes SO(2,2+n)}{SO(1,1)\otimes SO(1,1+n)}~$ & $\frac{%
SU(1,1)\otimes SO(2,2+n)}{SO(2)\otimes SO(2,n)}$ \\ \hline
$
\begin{array}{c}
\\
III \\
~
\end{array}
$ & $\frac{E_{7(-25)}}{E_{6}}$ & $\frac{E_{7(-25)}}{E_{6(-26)}}$ & $\frac{%
E_{7(-25)}}{E_{6(-14)}}~$ \\ \hline
$
\begin{array}{c}
\\
IV \\
~
\end{array}
$ & $\frac{SO^{\ast }(12)}{SU(6)}~$ & $\frac{SO^{\ast }(12)}{SU^{\ast }(6)}~$
& $\frac{SO^{\ast }(12)}{SU(4,2)}~$ \\ \hline
$
\begin{array}{c}
\\
V \\
~
\end{array}
$ & $\frac{SU(3,3)}{SU(3)\otimes SU(3)}$ & $\frac{SU(3,3)}{SL(3,\mathbb{C})}$
& $\frac{SU(3,3)}{SU(2,1)\otimes SU(1,2)}~$ \\ \hline
$
\begin{array}{c}
\\
VI \\
~
\end{array}
$ & $\frac{Sp(6,\mathbb{R})}{SU(3)}$ & $\frac{Sp(6,\mathbb{R})}{SL(3,\mathbb{%
R})}$ & $\frac{Sp(6,\mathbb{R})}{SU(2,1)}$ \\ \hline
\end{tabular}
\end{center}
\caption{\textbf{Non-degenerate orbits of $N=2$, $d=4$ symmetric MESGTs }}
\end{table}

\section{\label{Freud}Freudenthal Triple Systems\newline
~and\newline
~Orbits of Symmetric $N=2$, $d=4$ MESGTs}

In those symmetric $N=2$, $d=5$ MESGTs whose cubic norm form is taken to be
the norm form of an Euclidean Jordan algebra $\mathcal{J}$ of degree three,
there is an one-to-one correspondence between the vector fields (and hence
their charges) and the elements of $\mathcal{J}$ \cite{GST1,GST2,GST3}.

Under dimensional reduction to $d=4$, such a correspondence gets extended to
a correspondence between the field strengths of the vector fields (and their
magnetic duals) and the \textit{Freudenthal triple system} (FTS) \cite
{Freudenthal1,Freudenthal2} $\mathcal{F}\left( \mathcal{J}\right) $ defined
over $\mathcal{J}$ \cite{GST1,GKN,FG,MG2005,GP2}, and it can be realized as $%
2\times 2$ ``matrices'' :
\begin{equation}
\left(
\begin{array}{ccc}
F_{\mu \nu }^{0} &  & F_{\mu \nu }^{i} \\
&  &  \\
\widetilde{F}_{i}^{\mu \nu } &  & \widetilde{F}_{0}^{\mu \nu }
\end{array}
\right) \Longleftrightarrow \left(
\begin{array}{ccc}
\alpha &  & x \\
&  &  \\
y &  & \beta
\end{array}
\right) \in \mathcal{F}\left( \mathcal{J}\right) ,
\end{equation}
where $\alpha ,\beta \in \mathbb{R}$ and $x,y\in \mathcal{J}$. $F_{\mu \nu
}^{0}$ is the $d=4$ graviphoton field strength, i.e. the field strength of
the vector field coming from the $d=5$ graviton; $F_{\mu \nu }^{i}$ denote
the field strengths of the vector fields that already exist in $d=5$ (notice
that $i=1,...,\widehat{n}+1=n_{V}$, where $n_{V}$, $\widehat{n}$
respectively stand for the number of Abelian vector multiplets in $d=4$, $5$
dimensions) \cite{GST1,GST2,GST3}. %\textbf{(please, note the change of the
%notation and range of index along such multiplets...)}.

Consequently, one can associate an element of the FTS $\mathcal{F}\left(
\mathcal{J}\right) $ with the electric and magnetic charges $\left\{
q_{0},q_{i},p^{0},p^{i}\right\} \in \mathbb{R}^{2n_{V}+2}$ (fluxes of the
corresponding field strengths) of an $N=2$, $d=4$ (extremal) black hole:
\begin{equation}
\left(
\begin{array}{ccc}
p^{0} &  & p^{i} \\
&  &  \\
q_{i} &  & q_{0}
\end{array}
\right) \Longleftrightarrow \left(
\begin{array}{ccc}
\alpha &  & x \\
&  &  \\
y &  & \beta
\end{array}
\right) \in \mathcal{F}\left( \mathcal{J}\right) ,
\end{equation}
where $\alpha =p^{0}$, $\beta =q_{0}$, $x=p^{i}j_{i}$ and $y=q_{i}j^{i}$,
with $\left\{ j_{i}\right\} _{i=1,...,n_{V}}$ denoting the set of basis
vectors of $\mathcal{J}$.\medskip

In general, an FTS is defined as a vector space $\mathcal{M}$ with a
trilinear product $\left( P,Q,R\right) $ and a skew-symmetric bilinear form $%
\left\langle P,Q\right\rangle $, $P,Q,R\in \mathcal{M}$. One can always
modify the triple product $\left( P,Q,R\right) $ by adding terms like $%
\left\langle P,Q\right\rangle R$ in order to make it completely symmetric.
In our treatment we shall assume the triple product to be completely
symmetric, and we will follow the conventions of Brown \cite{Brown} , who
axiomatized the triple systems introduced by Freudenthal in his study of the
exceptional groups of the Magic Square \cite{Freudenthal1,Freudenthal2}%
.\smallskip

As reformulated by Ferrar \cite{Ferrar}, an FTS is a vector space $\mathcal{M%
}$ endowed with a trilinear product $\left( P,Q,R\right) $ and a
skew-symmetric bilinear form $\left\langle P,Q\right\rangle =-\left\langle
Q,P\right\rangle $, such that ($\forall P,Q,R,S\in \mathcal{M}$):\smallskip
\smallskip

\textbf{A1}. $\left( P,Q,R\right) $ is completely symmetric;\smallskip
\smallskip

\textbf{A2}.
\begin{equation}
\mathcal{Q}\left( P,Q,R,S\right) \equiv \left\langle P,\left( Q,R,S\right)
\right\rangle
\end{equation}
is a non-zero completely symmetric 4-linear form;\smallskip \smallskip

\textbf{A3}.
\begin{equation}
\left( \left( P,P,P\right) ,P,Q\right) =\left\langle Q,P\right\rangle \left(
P,P,P\right) +\left\langle Q,\left( P,P,P\right) \right\rangle P.
\end{equation}
\smallskip

For FTSs defined over an (Euclidean) Jordan algebra $\mathcal{J}$ of degree
three, it is customary to denote the elements of $\mathcal{M}$ as follows:
\begin{equation}
\begin{array}{l}
\mathcal{M}\ni P=\left(
\begin{array}{ccc}
\alpha &  & x \\
&  &  \\
y &  & \beta
\end{array}
\right) \equiv \left( \alpha ,\beta ,x,y\right) ; \\
\\
\mathcal{M}\ni Q=\left(
\begin{array}{ccc}
\gamma &  & w \\
&  &  \\
z &  & \delta
\end{array}
\right) \equiv \left( \gamma ,\delta ,w,z\right) ,
\end{array}
\end{equation}
where $\alpha ,\beta ,\gamma ,\delta \in \mathbb{R}$ and $x,y,w,z\in
\mathcal{J}$.

Thence
%\textbf{(why not adding also the definition of }$\left( P,Q,R\right) $%
%\textbf{\ in such a specific notation?)}
\begin{equation}
\left\langle P,Q\right\rangle \equiv \alpha \delta -\beta \gamma +T\left(
x,z\right) -T\left( y,w\right) ,
\end{equation}
where $T\left( x,z\right) \equiv Tr\left( x\circ z\right) $, with ``$Tr$''
denoting the trace of the ($3\times 3$ matrix representation of the)
elements of $\mathcal{J}$, and ``$\circ $'' standing for the symmetric
Jordan product\footnote{%
It should be mentioned that the symmetric Jordan product $\circ $ satisfies
the famous \textit{Jordan identity} \cite
{Jordan,Jacobson,Guna1,GPR,GST1,GST2,GST3} ($\forall x$, $y$ elements of the
Jordan algebra)
\begin{equation*}
x\circ \left( y\circ x^{2}\right) =\left( x\circ y\right) \circ x^{2}.
\end{equation*}
} in $\mathcal{J}$.

The quartic norm of an element $P\in \mathcal{M}$ is defined as
\begin{equation}
I_{4}\left( P\right) \equiv \mathcal{Q}\left( P,P,P,P\right) ,
\end{equation}
and it can be normalized such that\footnote{%
Eq. (\ref{II4}) corrects a misprint in Eq. (2-15) of \cite{FG}, by
substituting the factor $6$ with $4$ and introducing an overall sign.} \cite
{Faulkner}
\begin{equation}
I_{4}\left( P\right) =-\left\{ \alpha \beta -T\left( x,z\right) \right\}
^{2}-4\left\{ \alpha I_{3}\left( y\right) +\beta I_{3}\left( x\right)
-T\left( x^{\#},y^{\#}\right) \right\} ,  \label{II4}
\end{equation}
where $I_{3}$ denotes the cubic norm of $\mathcal{J}$ and ``$\#$'' stands
for the \textit{quadratic adjoint map} of $\mathcal{J}$. %\textbf{(Please
%check...I added such considerations in order to make contact with what is
%written after Eq. (2-15) of \cite{FG}).}
By introducing the \textit{Freudenthal cross (symmetric) product}
\begin{equation}
x\wedge y\equiv x\circ y-\frac{1}{2}\left[ Tr\left( x\right) y+Tr\left(
y\right) x\right] +\frac{1}{2}\left[ Tr\left( x\right) Tr\left( y\right)
-Tr\left( x\circ y\right) \right] ,
\end{equation}
the quadratic adjoint map of $x\in \mathcal{J}$ can be defined as
\begin{equation}
x^{\#}\equiv x\wedge x,
\end{equation}
and it can be shown that it has the iteration property \cite{McKrimmon}
\begin{equation}
x^{\#\#}=I_{3}\left( x\right) x.
\end{equation}
\medskip

The automorphism group $Aut\left( \mathcal{F}\left( \mathcal{J}\right)
\right) $ of the FTS defined over an Euclidan Jordan algebra $\mathcal{J}$
of degree three contains as proper subgroup the structure group $Str\left(
\mathcal{J}\right) $ of $\mathcal{J}$, i.e. the group which leaves the cubic
norm $I_{3}$ of $\mathcal{J}$ invariant up to a real overall scale factor $%
\lambda $. Furthermore, those transformations belonging to $Str\left(
\mathcal{J}\right) $ that leave the cubic norm $I_{3}$ invariant (i.e. such
that $\lambda =1$) form the reduced structure group \cite{Jacobson} $%
Str_{0}\left( \mathcal{J}\right) $ of $\mathcal{J}$, which is nothing but
the $U$-duality group $G_{5}$ of the corresponding $N=2$, $d=5$ symmetric
MESGT. Indeed, the symmetric $N=2$, $d=5$ MESGTs whose cubic norm form is
taken to be the norm form of an Euclidean Jordan algebra $\mathcal{J}$ of
degree three are endowed with homogeneous symmetric real\footnote{%
Sometimes in the literature the geometry of such manifolds is referred to as
\textit{real special }geometry, whereas the geometry of the manifolds of the
corresponding $d=4$ MESGTs obtained by dimensional reduction is sometimes
named \textit{very special K\"{a}hler} geometry.} manifolds of the form
\begin{equation}
\frac{G_{5}}{H_{5}}=\frac{Str_{0}\left( \mathcal{J}\right) }{Aut\left(
\mathcal{J}\right) ,}
\end{equation}
where $Aut\left( \mathcal{J}\right) =m.c.s.\left( Str_{0}\left( \mathcal{J}%
\right) \right) $ denotes the automorphism group of $\mathcal{J}$. For
further elucidation on the $\mathcal{J}$-related symmetric $N=2$, $d=5$
MESGTs, see \cite{GST1,GST2,GST3,GST4} and \cite{FG,FG2}.

With the exception of the family with the scalar manifold $\frac{SU(1,1+n)}{%
U(1)\otimes SU(1+n)}$, all symmetric $N=2$, $d=4$ MESGTs can be obtained
from the corresponding $d=5$ theories by dimensional reduction \cite
{GST1,GST2,GST3,GST4}. %For the irreducible sequence $I$ of Table
%3 the cubic form cannot in general be given by the norm form of any
%Jordan algebras of degree three \cite{GST2}, and therefore such a
%sequence cannot be seen as the dimensional reduction of a
%$\mathcal{J} $-related $N=2$, $d=5$ symmetric MESGT.
The automorphism group $Aut\left( \mathcal{F}\left( \mathcal{J}\right)
\right) $ of $\mathcal{F}\left( \mathcal{J}\right) $ is nothing but the $U$%
-duality group $G\equiv G_{4}$ of the $\mathcal{J}$-related MESGT in $d=4$%
\footnote{$\mathbb{R}\oplus \mathbf{\Gamma }_{n+2}$ denotes the generic
family of reducible Euclidean Jordan algebras of degree 3. $\mathbb{R}$
denotes the one dimensional Jordan algebra and $\mathbf{\Gamma }_{n+2}$
denotes the Jordan algebra of degree 2 associated with a quadratic form of
Lorentzian signature. This signature is required for the Jordan algebra $%
\mathcal{J}$ to be Euclidean, which is defined as a Jordan algebra in which
the equation $x^2+y^2=0 $ implies that $x=y=0$ for all elements $x,y$ of $%
\mathcal{J}$ .}:
\begin{equation}
\begin{array}{l}
II:Aut\left( \mathcal{F}\left( \mathbb{R}\oplus \mathbf{\Gamma }%
_{n+2}\right) \right) =SU(1,1)\otimes SO(2,2+n); \\
\\
III:Aut\left( \mathcal{F}\left( J_{3}^{\mathbb{O}}\right) \right)
=E_{7(-25)}; \\
\\
IV:Aut\left( \mathcal{F}\left( J_{3}^{\mathbb{H}}\right) \right) =SO^{\ast
}\left( 12\right) ; \\
\\
V:Aut\left( \mathcal{F}\left( J_{3}^{\mathbb{C}}\right) \right) =SU(3,3); \\
\\
VI:Aut\left( \mathcal{F}\left( J_{3}^{\mathbb{R}}\right) \right) =Sp(6,%
\mathbb{R}).
\end{array}
\label{Aut-Aut}
\end{equation}
Thus, the homogeneous symmetric special K\"{a}hler scalar manifold of $%
\mathcal{J}$-related symmetric $N=2$, $d=4$ MESGTs can be generally written
as
\begin{equation}
\frac{G_{4}}{H_{4}}=\frac{Aut\left( \mathcal{F}\left( \mathcal{J}\right)
\right) }{m.c.s.\left( Aut\left( \mathcal{F}\left( \mathcal{J}\right)
\right) \right) }=\frac{Aut\left( \mathcal{F}\left( \mathcal{J}\right)
\right) }{H_{0}\otimes U(1)},
\end{equation}
and the following chain of strict group inclusions hold:
\begin{equation}
H_{5}=Aut\left( \mathcal{J}\right) =m.c.s.\left( G_{5}\right) \varsubsetneq
G_{5}=Str_{0}\left( \mathcal{J}\right) \varsubsetneq Str\left( \mathcal{J}%
\right) \varsubsetneq Aut\left( \mathcal{F}\left( \mathcal{J}\right) \right)
=G\equiv G_{4}.  \label{GG}
\end{equation}
\bigskip

The global action of $Aut\left( \mathcal{F}\left( \mathcal{J}\right) \right)
$ on $\mathcal{F}\left( \mathcal{J}\right) $ is generated by the structure
group $Str\left( \mathcal{J}\right) $ of $\mathcal{J}$ $S_{\lambda }\left(
\cdot \right) $ ($\lambda \in \mathbb{R}$), by a discrete transformation $%
\tau $ and by two additional $\mathcal{J}$-parameterized transformations $%
\phi (C)$ and $\psi (D)$. The $\lambda $-parameterized action of $Str(%
\mathcal{J})$ reads
\begin{equation}
Str\left( \mathcal{J}\right) :\left( \alpha ,\beta ,x,y\right) \longmapsto
\left( \frac{\alpha }{\lambda },\lambda \beta ,S_{\lambda }\left( x\right) ,%
\widetilde{S}_{\lambda }(y)\right) ,  \label{Str-Str}
\end{equation}
where $S_{\lambda }$ and and $\tilde{S}_{\lambda }$ respectively denote the
action of $Str(\mathcal{J})$ on $\mathcal{J}$ and its adjoint. The
transformation $\tau $ acts as
\begin{equation}
\tau :\left( \alpha ,\beta ,x,y\right) \longmapsto \left( -\beta ,\alpha
,-y,x\right) .  \label{def-tau}
\end{equation}
The additional transformations (parameterized by elements of $\mathcal{J}$)
act as :
\begin{eqnarray}
&&
\begin{array}{l}
\phi \left( C\right) :\left( \alpha ,\beta ,x,y\right) \longmapsto \left(
\alpha ^{\prime },\beta ,x^{\prime },y^{\prime }\right) , \\
\\
\left\{
\begin{array}{l}
\alpha ^{\prime }\equiv \alpha +\beta I_{3}\left( C\right) +T\left(
x,C\wedge C\right) +T\left( y,C\right) ; \\
\\
x^{\prime }\equiv x+\beta C; \\
\\
y^{\prime }\equiv y+2x\wedge C+\beta C\wedge C; \\
\\
C\in \mathcal{J};
\end{array}
\right.
\end{array}
\label{def-phi} \\
&&  \notag
\end{eqnarray}
\begin{eqnarray}
&&
\begin{array}{l}
\psi \left( D\right) :\left( \alpha ,\beta ,x,y\right) \longmapsto \left(
\alpha ,\beta ^{\prime \prime },x^{\prime \prime },y^{\prime \prime }\right)
, \\
\\
\left\{
\begin{array}{l}
\beta ^{\prime \prime }\equiv \beta +\alpha I_{3}\left( D\right) +T\left(
y,D\wedge D\right) +T\left( x,D\right) ; \\
\\
x^{\prime \prime }\equiv x+\alpha D\wedge D+2y\wedge D; \\
\\
y^{\prime \prime }\equiv y+\alpha D; \\
\\
D\in \mathcal{J}.
\end{array}
\right.
\end{array}
\label{def-psi} \\
&&  \notag
\end{eqnarray}
We should note that \cite{Jacobson,McCrimmon}

\begin{equation}
I_{3}\left( S(x)\right) =\lambda I_{3}(x)
\end{equation}
and
\begin{equation}
\widetilde{S}\left( x\wedge y\right) =S(x)\wedge S(y).
\end{equation}
Furthermore, $Str(\mathcal{J})$ can be written as the direct product of the
reduced structure group $Str_{0}(\mathcal{J})$ times the dilatation group $%
\mathcal{D}$:
\begin{equation}
Str(\mathcal{J})=Str_{0}(\mathcal{J})\otimes \mathcal{D}.
\end{equation}

Now, as it was shown by Krutelevich \cite{Krutelevich}, by suitably acting
with $Aut\left( \mathcal{F}\left( \mathcal{J}\right) \right) $ one can bring
a generic element of $\mathcal{F}\left( \mathcal{J}\right) $ to the
noteworthy form
\begin{equation}
\left( 1,\beta ,x,0\right) \in \mathcal{F}\left( \mathcal{J}\right) ,~\beta
\in \mathbb{R},x\in \mathcal{J}.  \label{Krut-Krut}
\end{equation}
Note that $Str_{0}\left( \mathcal{J}\right) $ preserves such a simplified
form; indeed, Eq. (\ref{Str-Str}) with $\lambda =1$ yields
\begin{equation}
Str_{0}\left( \mathcal{J}\right) :\left( 1,\beta ,x,0\right) \longmapsto
\left( 1,\beta ,S_{0}\left( x\right) ,0\right) .
\end{equation}

Moreover, by suitably acting with $Aut\left( \mathcal{J}\right) =H_{5}$, one
can bring a generic element $x\in \mathcal{J}$ to the diagonal form:
\begin{equation}
x\overset{Aut\left( \mathcal{J}\right) }{\longmapsto }\lambda
_{1}E_{1}+\lambda _{2}E_{2}+\lambda _{3}E_{3},  \label{diag-diag}
\end{equation}
where $\lambda _{1}$, $\lambda _{2}$, $\lambda _{3}\in \mathbb{R}$ are the
real ``eigenvalues''\footnote{%
For the simple Euclidean Jordan algebras of degree three they are the usual
eigenvalues of the $3\times 3$ Hermitian matrix representing $x$, while for
the generic Jordan family $\mathbb{R}+\mathbf{\Gamma }_{n+2}$ they
correspond to the values when a given element $x$ is brought to the form (%
\ref{diag-diag}).} of $x$ and $E_{1}$, $E_{2}$, $E_{3}$ are the three
irreducible idempotents of $\mathcal{J}$ (see further below).

Thus, by using Eqs. (\ref{Krut-Krut}) and (\ref{diag-diag}) and by acting
with the transformation $\phi \left( C=\varepsilon E_{1}\right) $ followed
by the transformation $\psi \left( D=-\varepsilon \left( \lambda
_{3}E_{2}+\lambda _{2}E_{3}\right) \right) $ (recall definitions (\ref
{def-phi}) and (\ref{def-psi})), it can be shown that a generic element of $%
\mathcal{F}\left( \mathcal{J}\right) $ can be brought to the form \cite
{Krutelevich}
\begin{equation}
\begin{array}{l}
\left( 1,\beta -2\lambda _{2}\lambda _{3}\varepsilon ,\lambda _{1}^{\prime
}E_{1}+\lambda _{2}E_{2}+\lambda _{3}E_{3},0\right) \in \mathcal{F}\left(
\mathcal{J}\right) , \\
\\
\lambda _{1}^{\prime }\equiv -\lambda _{2}\lambda _{3}\varepsilon ^{2}+\beta
\varepsilon +\lambda _{1}.
\end{array}
\label{simplified}
\end{equation}
Consequently, by choosing
\begin{equation}
\mathbb{R}_{0}\ni \varepsilon :2\lambda _{2}\lambda _{3}\varepsilon =\beta ,
\end{equation}
one can further simplify the form (\ref{simplified}) into
\begin{equation}
\begin{array}{l}
\left( 1,0,\lambda _{1}^{\prime \prime }E_{1}+\lambda _{2}E_{2}+\lambda
_{3}E_{3},0\right) \in \mathcal{F}\left( \mathcal{J}\right) , \\
\\
\lambda _{1}^{\prime \prime }\equiv \lambda _{2}\lambda _{3}\varepsilon
^{2}+\lambda _{1}.
\end{array}
\label{simplified2}
\end{equation}

Eq. (\ref{simplified2}) shows that the sign of $\lambda _{1}^{\prime \prime
}\in \mathbb{R}$ might differ from the sign of $\lambda _{1}$, depending on
the sign of $\lambda _{2}\lambda _{3}$ and on the relative magnitude of $%
\lambda _{1}$ \textit{versus} $\lambda _{2}\lambda _{3}\varepsilon ^{2}$. In
other words, the sign of the ``eigenvalues'' of the elements of $\mathcal{J}$
in the FTS $\mathcal{F}\left( \mathcal{J}\right) $ could be changed by the
action of $Aut\left( \mathcal{F}\left( \mathcal{J}\right) \right) $.

Now, by a suitable action of $Str_{0}\left( \mathcal{J}\right) $ one can
bring a generic element $x\in \mathcal{J}$ to the diagonal form ($\nu \equiv
\lambda _{1}\lambda _{2}\lambda _{3}$) \cite{FG2}
\begin{equation}
x_{diag,1st}=E_{1}+E_{2}+\nu E_{3}
\end{equation}
if at least two ``eigenvalues'' of $x$ are positive, or to the diagonal form
\begin{equation}
x_{diag,2nd}=-E_{1}-E_{2}+\nu E_{3}
\end{equation}
if at least two ``eigenvalues'' of $x$ are negative (clearly, up to
permutations of the indices $1$, $2$, $3$).\smallskip

For the case of $\mathcal{F}\left( J_{3}^{\mathbb{O}}\right) $, it has been
shown in \cite{Shukuzawa} that for $\nu >0$ the element
\begin{equation}
\left( 1,0,E_{1}+E_{2}+\nu E_{3},0\right) \in \mathcal{F}\left( J_{3}^{%
\mathbb{O}}\right)
\end{equation}
(having all three ``eigenvalues'' $>0$) can be transformed into the element
\begin{equation}
\left( 1,0,-E_{1}-E_{2}+\nu E_{3},0\right) \in \mathcal{F}\left( J_{3}^{%
\mathbb{O}}\right)
\end{equation}
(having two ``eigenvalues'' $<0$ and one ``eigenvalue'' $>0$) - and \textit{%
viceversa} - by a suitable action of $Aut\left( \mathcal{F}\left( J_{3}^{%
\mathbb{O}}\right) \right) =E_{7(-25)}$ (recall Eq. (\ref{Aut-Aut}));
instead, for $\nu <0$ it was proven that such two elements of $\mathcal{F}%
\left( J_{3}^{\mathbb{O}}\right) $ cannot be transformed one into each other
by the action of $E_{7(-25)}$:
\begin{equation}
\begin{array}{l}
\nu >0:\left( 1,0,E_{1}+E_{2}+\left| \nu \right| E_{3},0\right) \overset{%
E_{7(-25)}}{\longleftrightarrow }\left( 1,0,-E_{1}-E_{2}+\left| \nu \right|
E_{3},0\right) ; \\
\\
\nu <0:\left( 1,0,E_{1}+E_{2}-\left| \nu \right| E_{3},0\right) \overset{%
E_{7(-25)}}{\nleftrightarrow }\left( 1,0,-E_{1}-E_{2}-\left| \nu \right|
E_{3},0\right) .
\end{array}
\end{equation}
Since the proof of \cite{Shukuzawa} involves the idempotent elements $E_{1}$%
, $E_{2}$, $E_{3}\in \mathcal{J}$, it is easy to show that such a proof
actually holds true for \textit{all} $\mathcal{F}\left( \mathcal{J}\right) $%
, where $\mathcal{J}$ is a Euclidean Jordan algebra of degree
three.\smallskip \medskip

We shall use $Str_{0}\left( \mathcal{J}\right) \varsubsetneq Aut\left(
\mathcal{F}\left( \mathcal{J}\right) \right) $ to determine the stabilizer
(also called \textit{little group} or \textit{isotropy group}) of the
various classes of charge orbits. Starting from a generic element of $%
\mathcal{F}\left( \mathcal{J}\right) $ with at least two positive
``eigenvalues'' and using $Str_{0}\left( \mathcal{J}\right) $, one can
recast it in the form
\begin{equation}
\left( 1,0,\left| \nu \right| ^{1/3}\left( E_{1}+E_{2}+sgn\left( \nu \right)
E_{3}\right) ,0\right)
\end{equation}
which in turn, by applying the scale transformation $\tau $, can be shown to
become
\begin{equation}
\kappa \left( 1,0,\left( E_{1}+E_{2}+sgn\left( \nu \right) E_{3}\right)
,0\right) ,  \label{ref-v1}
\end{equation}
where $\kappa \equiv \left| \nu \right| ^{1/4}\in \mathbb{R}_{0}^{+}$. On
the other hand, when at least two of the ``eigenvalues'' are negative one
can bring the considered element of $\mathcal{F}\left( \mathcal{J}\right) $
to the form
\begin{equation}
\left( 1,0,\left| \nu \right| ^{1/3}\left( -E_{1}-E_{2}+sgn\left( \nu
\right) E_{3}\right) ,0\right)
\end{equation}
using $Str_{0}\left( \mathcal{J}\right) $, and then, through the scale
transformation $\tau $, to the form
\begin{equation}
\kappa \left( 1,0,\left( -E_{1}-E_{2}+sgn\left( \nu \right) E_{3}\right)
,0\right) .  \label{ref-v2}
\end{equation}
We shall refer to the vector notations (\ref{ref-v1}) and (\ref{ref-v2}) of
elements of $\mathcal{F}\left( \mathcal{J}\right) $ as \textit{reference
vectors}.

In order to determine the little groups of the reference vectors belonging
to each class of charge orbits, we shall use the action of the Lie algebra $%
\mathfrak{Aut}\left( \mathcal{F}\left( \mathcal{J}\right) \right) $ of $%
Aut\left( \mathcal{F}\left( \mathcal{J}\right) \right) $. At the level of $%
\mathfrak{Aut}\left( \mathcal{F}\left( \mathcal{J}\right) \right) $, the
generators of the little groups of the various classes of charge orbits
annihilate the corresponding reference vectors. The subgroups of $%
Str_{0}\left( \mathcal{J}\right) $ that leave the above reference vectors
invariant are known from \cite{FG,FG2}.Therefore, in order to determine the
stability group of a class of orbits under the action of $Aut\left( \mathcal{%
F}\left( \mathcal{J}\right) \right) $, we need only to determine the
generators outside the Lie algebra $\mathfrak{Str}\left( \mathcal{J}\right) $
of $Str\left( \mathcal{J}\right) $ (also called the \textit{structure algebra%
} of $\mathcal{J}$) that annihilate the considered reference vector.

The group actions defined by Eqs. (\ref{def-phi})-(\ref{def-psi}) can be
shown to lead to the following actions by the elements of $\mathfrak{Aut}%
\left( \mathcal{F}\left( \mathcal{J}\right) \right) $ outside $\mathfrak{Str}%
\left( \mathcal{J}\right) $ (i.e. by the infinitesimal generators of $%
Aut\left( \mathcal{F}\left( \mathcal{J}\right) \right) $ which are not
generators of $Str\left( \mathcal{J}\right) \varsubsetneq Aut\left( \mathcal{%
F}\left( \mathcal{J}\right) \right) $) \cite
{Freudenthal1,Freudenthal2,Brown,Seligman}:
\begin{equation}
\begin{array}{l}
\mathcal{T}\left( A,B\right) \left( \alpha ,\beta ,x,y\right) =\left( \alpha
^{\prime },\beta ^{\prime },x^{\prime },y^{\prime }\right) , \\
\\
\left\{
\begin{array}{l}
\alpha ^{\prime }\equiv T\left( A,y\right) ; \\
\\
\beta ^{\prime }\equiv T\left( B,x\right) ; \\
\\
x^{\prime }\equiv 2B\wedge y+\beta A; \\
\\
y^{\prime }\equiv 2A\wedge x+\alpha B; \\
\\
A,B\in \mathcal{J.}
\end{array}
\right.
\end{array}
\label{little-little}
\end{equation}
\medskip The Hermitian conjugate of $\mathcal{T}\left( A,B\right) $ is
defined as
\begin{equation}
\mathcal{T}^{\dagger }\left( A,B\right) \equiv \mathcal{T}\left( B,A\right) ,
\end{equation}
and the non-compact generators are Hermitian, while the compact ones are
anti-Hermitian.\bigskip

We recall that for the simple Euclidean Jordan algebras of degree three $%
J_{3}^{\mathbb{A}}$ a generic element $A$ can be realized as a $3\times 3$
Hermitian matrix over the underlying division algebras $\mathbb{A}=\mathbb{R}%
,\mathbb{C},\mathbb{H},\mathbb{O}$ as follows:

\begin{equation}
A=\left(
\begin{array}{ccccc}
\alpha _{1} &  & a_{3} &  & a_{2}^{\ast } \\
&  &  &  &  \\
a_{3}^{\ast } &  & \alpha _{2} &  & a_{1} \\
&  &  &  &  \\
a_{3} &  & a_{1}^{\ast } &  & \alpha _{3}
\end{array}
\right) ,
\end{equation}
where $\alpha _{i}\in \mathbb{R}$, $a_{i}\in \mathbb{A}$ ($i=1,2,3$), and ``$%
\ast $'' denotes conjugation in $\mathbb{A}$. The Jordan product $\circ $ of
two such elements $A$ and $B$ is simply
\begin{equation}
A\circ B\equiv \frac{1}{2}(A\times B+B\times A),
\end{equation}
where ``$\times $'' stands for the standard ``row-column'' matrix product.
The irreducible idempotents of a Jordan algebra are those elements $E$ such
that
\begin{equation}
E^{2}\equiv E\circ E=E
\end{equation}
and
\begin{equation}
Tr\left( E\right) =1.
\end{equation}
We shall denote the irreducible idempotents as $E_{i}$ ($i=1,2,3$), such
that
\begin{equation}
Tr(A\circ E_{i})=\alpha _{i}.
\end{equation}
The cubic norm $I_{3}$ of a simple Euclidean Jordan algebra of degree three
is simply the determinant of its $3\times 3$ Hermitian matrix realization

\begin{equation}
I_{3}(A)\equiv \alpha _{1}\alpha _{2}\alpha _{3}-\alpha
_{1}|a_{1}|^{2}-\alpha _{2}|a_{2}|^{2}-\alpha
_{3}|a_{3}|^{2}+2Re(a_{1}a_{2}a_{3}),  \label{dett}
\end{equation}
\medskip where $Re(a)\equiv \frac{1}{2}(a+a^{\ast })$ denotes the real part
of $a\in \mathbb{A}$, and
\begin{equation}
Re(a_{1}a_{2}a_{3})=Re[a_{1}(a_{2}a_{3})]=Re[(a_{1}a_{2})a_{3}]
\end{equation}
holds true for all elements of $\mathbb{A}$.

Concerning the generic reducible Jordan family $\mathcal{J}=\mathbb{R}\oplus
\mathbf{\Gamma }_{n+2}$ ($n\in \mathbb{N\cup }\left\{ 0\right\} $), a
generic element $\mathcal{X}$ can be represented in the form
\begin{equation}
(\zeta ;\xi _{0}\mathbf{1}+\overrightarrow{\xi }\cdot \overrightarrow{\sigma
})
\end{equation}
where $\zeta \in \mathbb{R}$, $\xi _{0}\in \mathbb{R}$ , $\overrightarrow{%
\xi }\in \mathbb{R}^{n+1}$ and $\sigma _{i}$ are the gamma matrices in $%
\mathbb{R}^{n+1}$:
\begin{equation}
\{\sigma _{i},\sigma _{j}\}=2\delta _{ij},\text{ }i,j=1,...,n+1.
\end{equation}
The cubic norm of such an element is given by \cite{GST1,GP2}
\begin{equation}
I_{3}\left( \mathcal{X}\right) \equiv \sqrt{2}\zeta \left[ \left( \xi
_{0}\right) ^{2}-\overrightarrow{\xi }\cdot \overrightarrow{\xi }\right] .
\end{equation}
For brevity we shall denote the element $(\zeta ;\xi _{0}\mathbf{1}+%
\overrightarrow{\xi }\cdot \overrightarrow{\sigma })$ as
\begin{equation}
\mathcal{X}=(\zeta ;\xi _{0},\overrightarrow{\xi }).
\end{equation}

The three irreducible idempotents of $\mathbb{R}\oplus \mathbf{\Gamma }%
_{n+2} $ are
\begin{equation}
\begin{array}{l}
E_{1}\equiv (1;0,\overrightarrow{0}); \\
\\
E_{2}\equiv (0;\frac{1}{2},\frac{1}{2},0,...,0); \\
\\
E_{3}\equiv (0;\frac{1}{2},-\frac{1}{2},0,..,0),
\end{array}
\end{equation}
with the identity element $\mathfrak{1}$ $\in \mathbb{R}\oplus \mathbf{%
\Gamma }_{n+2}$ given by
\begin{equation}
\mathfrak{1}=E_{1}+E_{2}+E_{3}=(1;1,\overrightarrow{0}).
\end{equation}
The automorphism group $Aut\left( \mathbb{R}\oplus \mathbf{\Gamma }%
_{n+2}\right) $ is $SO(n+1)$ and the reduced structure group $Str_{0}\left(
\mathbb{R}\oplus \mathbf{\Gamma }_{n+2}\right) $ is $SO(n+1,1)\otimes
SO(1,1) $. The identity element $\mathfrak{1}$ is manifestly invariant under
$Aut\left( \mathbb{R}\oplus \mathbf{\Gamma }_{n+2}\right) $, and the little
group of the element $\mathfrak{b}\equiv -E_{1}-E_{2}+E_{3}=(-1;0,-1,0,..,0)$
is $SO(n,1)$.\bigskip

Let us now determine the explicit form of the generators $\mathcal{T}\left(
A,B\right) $ (\ref{little-little}) annihilating the corresponding reference
vector in each class of charge orbits:\medskip

\textbf{1}. For
\begin{equation}
\left( \alpha ,\beta ,x,y\right) =\kappa \left(
1,0,E_{1}+E_{2}+E_{3},0\right)  \label{ast}
\end{equation}
%\textbf{(corresponding, by Eq.
%(\ref {diag-diag}), to }$\lambda _{1}=\lambda _{2}=\lambda
%$_{3}=1$\textbf{)},
the annihilation condition
\begin{equation}
\mathcal{T}(A,B)\kappa \left( 1,0,E_{1}+E_{2}+E_{3},0\right) =0
\end{equation}
requires
\begin{equation}
\left\{
\begin{array}{l}
\alpha ^{\prime }=0; \\
\\
\beta =0\Rightarrow \beta ^{\prime }=0=Tr\left( B\right) ; \\
\\
x^{\prime }=0; \\
\\
y=0\Rightarrow y^{\prime }=0=-A+B,
\end{array}
\right.  \label{First}
\end{equation}
implying that
\begin{equation}
A=B,~~~Tr(A)=0.
\end{equation}
Consequently, the generators $\mathcal{T}\left( A,A\right) $ with $Tr(A)=0$
annihilate the reference vector (\ref{ast}). They correspond to $dim\left(
\mathcal{J}\right) -1=n_{V}-1=\widehat{n}=dim_{\mathbb{R}}\left( \frac{G_{5}%
}{H_{5}}\right) $ non-compact (Hermitian) generators outside $\mathfrak{Str}%
\left( \mathcal{J}\right) $.

The subgroup of $Str_{0}(\mathcal{J})$ that leaves the reference vector (\ref
{ast}) invariant is $Aut\left( \mathcal{J}\right) =H_{5}$, which is compact.
Hence, the stability group of the reference vector (\ref{ast}) is
non-compact and isomorphic (not identical) to the reduced structure group $%
Str_{0}(\mathcal{J})$; we shall denote it as $Str_{0}^{\ast }(\mathcal{J})$.

Therefore, the corresponding non-degenerate charge orbits, listed in Table
4, are of the form

\begin{equation}
\frac{Aut\left( \mathcal{F}\left( \mathcal{J}\right) \right) }{Str_{0}^{\ast
}(\mathcal{J})}=\frac{G_{4}\equiv G}{\widehat{H}},
\end{equation}
with $m.c.s.\left( Str_{0}^{\ast }(\mathcal{J})\right) =Aut\left( \mathcal{J}%
\right) =\widehat{h}=H_{5}$.

Now, the quartic invariant of a vector of the form $(\alpha ,0,(\lambda
_{1}E_{1}+\lambda _{2}E_{2}+\lambda _{3}E_{3}),0)\in \mathcal{F}\left(
\mathcal{J}\right) $ is simply
\begin{equation}
I_{4}=-\alpha \lambda _{1}\lambda _{2}\lambda _{3}\equiv \lambda _{0}\lambda
_{1}\lambda _{2}\lambda _{3}
\end{equation}
where we identify $-\alpha \equiv \lambda _{0}$. Furthermore, using the
dilatation symmetry one can rescale $\alpha $ relative to the eigenvalues $%
\lambda _{i}$, and hence one can assume $\alpha =1$ without any loss of
generality.

Thus, the orbits $\frac{Aut\left( \mathcal{F}\left( \mathcal{J}\right)
\right) }{Str_{0}^{\ast }(\mathcal{J})}$ have negative quartic invariant
\begin{equation}
I_{4}=-\kappa ^{4}=-\lambda _{1}\lambda _{2}\lambda _{3}<0,
\end{equation}
and a single negative ``eigenvalue'' $\lambda _{0}=-1$.

As it can be seen at a glance, such orbits can be interpreted as the
non-BPS, $Z\neq 0$ ones (the right column of Table 4 coincides with the
column of $\mathcal{O}_{non-BPS,Z\neq 0}$ orbits in Table 3).
\begin{table}[t]
\begin{center}
\begin{tabular}{|c||c|c|}
\hline
$
\begin{array}{c}
\\
~ \\
~
\end{array}
$ & $\mathcal{J}$ & $\frac{Aut\left( \mathcal{F}\left( \mathcal{J}\right)
\right) }{Str_{0}^{\ast }(\mathcal{J})}$ \\ \hline\hline
$
\begin{array}{c}
\\
II \\
~
\end{array}
$ & $\mathbb{R}\oplus \mathbf{\Gamma }_{n+2}$ & $\frac{SU(1,1)\otimes
SO(2,2+n)}{SO(1,1)\otimes SO(1,1+n)}$ \\ \hline
$
\begin{array}{c}
\\
III \\
~
\end{array}
$ & $J_{3}^{\mathbb{O}}$ & $~\frac{E_{7(-25)}}{E_{6(-26)}}$ \\ \hline
$
\begin{array}{c}
\\
IV \\
~
\end{array}
$ & $J_{3}^{\mathbb{H}}$ & $\frac{SO^{\ast }(12)}{SU^{\ast }(6)}~$ \\ \hline
$
\begin{array}{c}
\\
V \\
~
\end{array}
$ & $J_{3}^{\mathbb{C}}$ & $\frac{SU(3,3)}{SL(3,\mathbb{C})}$ \\ \hline
$
\begin{array}{c}
\\
VI \\
~
\end{array}
$ & $J_{3}^{\mathbb{R}}$ & $\frac{Sp(6,\mathbb{R})}{SL(3,\mathbb{R})}$ \\
\hline
\end{tabular}
\end{center}
\caption{\textbf{Class} $\frac{Aut\left( \mathcal{F}\left( \mathcal{J}%
\right) \right) }{Str_{0}^{\ast }(\mathcal{J})}$ \textbf{of non-degenerate
charge orbits with a negative quartic invariant and a single negative
``eigenvalue'' for all }$N=2$\textbf{, }$d=4$\textbf{\ MESGTs defined by
Euclidean Jordan algebras of degree three.}}
\end{table}

\medskip

\textbf{2}. For
\begin{equation}
\left( \alpha ,\beta ,x,y\right) =\kappa \left(
1,0,-E_{1}-E_{2}-E_{3},0\right)  \label{ast-ast}
\end{equation}
%$ \textbf{(corresponding, by Eq. (\ref
%{diag-diag}), to }$\lambda _{1}=\lambda _{2}=\lambda _{3}=-1$\textbf{)}, one
%gets:
the annihilation condition
\begin{equation}
\mathcal{T}(A,B)\kappa \left( 1,0,-E_{1}-E_{2}-E_{3},0\right) =0
\end{equation}
requires
\begin{equation}
\left\{
\begin{array}{l}
\alpha ^{\prime }=0; \\
\\
\beta =0\Rightarrow \beta ^{\prime }=0=-Tr\left( B\right) ; \\
\\
x^{\prime }=0; \\
\\
y=0\Rightarrow y^{\prime }=0=A+B,
\end{array}
\right.  \label{Second}
\end{equation}
implying that
\begin{equation}
A=-B,~~~Tr(A)=0.
\end{equation}
Consequently, the generators $\mathcal{T}\left( A,-A\right) $ with $Tr(A)=0$
annihilate the reference vector (\ref{ast-ast}). They correspond to $%
dim\left( \mathcal{J}\right) -1$ compact (anti-Hermitian) generators outside
$\mathfrak{Str}\left( \mathcal{J}\right) $.

The subgroup of $Str_{0}(\mathcal{J})$ that leaves the reference vector (\ref
{ast-ast}) invariant is, as in case 1, the compact automorphism group $%
Aut\left( \mathcal{J}\right) $. Thus, the stability group of the reference
vector (\ref{ast-ast}) is the compact form of $Str_{0}(\mathcal{J})$, which
we denote as $Ktr_{0}(\mathcal{J})$. Therefore, the corresponding
non-degenerate charge orbits are of the form

\begin{equation}
\frac{Aut\left( \mathcal{F}\left( \mathcal{J}\right) \right) }{Ktr_{0}(%
\mathcal{J})}=\frac{G_{4}\equiv G}{H_{0}},
\end{equation}
with $Ktr_{0}(\mathcal{J})=\frac{m.c.s.\left( Aut\left( \mathcal{F}\left(
\mathcal{J}\right) \right) \right) }{U(1)}=H_{0}$. They have positive
quartic invariant
\begin{equation}
I_{4}=\kappa ^{4}=\left| \lambda _{1}\lambda _{2}\lambda _{3}\right| >0
\end{equation}
and all four``eigenvalues'' negative.

These orbits correspond to the $N=2$, $\frac{1}{2}$-BPS orbits, and they are
listed in Table 5 (whose right column coincides with the column of $\mathcal{%
O}_{\frac{1}{2}-BPS}$ orbits in Table 3).
\begin{table}[t]
\begin{center}
\begin{tabular}{|c||c|c|}
\hline
$
\begin{array}{c}
\\
~ \\
~
\end{array}
$ & $\mathcal{J}$ & $\frac{Aut\left( \mathcal{F}\left( \mathcal{J}\right)
\right) }{Ktr_{0}(\mathcal{J})}$ \\ \hline\hline
$
\begin{array}{c}
\\
II \\
~
\end{array}
$ & $\mathbb{R}\oplus \mathbf{\Gamma }_{n+2}$ & $\frac{SU(1,1)\otimes
SO(2,2+n)}{SO(2)\otimes SO(2+n)}$ \\ \hline
$
\begin{array}{c}
\\
III \\
~
\end{array}
$ & $J_{3}^{\mathbb{O}}$ & $~\frac{E_{7(-25)}}{E_{6}}$ \\ \hline
$
\begin{array}{c}
\\
IV \\
~
\end{array}
$ & $J_{3}^{\mathbb{H}}$ & $\frac{SO^{\ast }(12)}{SU(6)}~$ \\ \hline
$
\begin{array}{c}
\\
V \\
~
\end{array}
$ & $J_{3}^{\mathbb{C}}$ & $\frac{SU(3,3)}{SU(3)\otimes SU(3)}$ \\ \hline
$
\begin{array}{c}
\\
VI \\
~
\end{array}
$ & $J_{3}^{\mathbb{R}}$ & $\frac{Sp(6,\mathbb{R})}{SU(3)}$ \\ \hline
\end{tabular}
\end{center}
\caption{\textbf{Class }$\frac{Aut\left( \mathcal{F}\left( \mathcal{J}%
\right) \right) }{Ktr_{0}(\mathcal{J})}$\textbf{\ of non-degenerate charge
orbits with positive quartic invariant and all negative ``eigenvalues'' for
all }$N=2$\textbf{, }$d=4$\textbf{\ symmetric MESGTs defined by Euclidean
Jordan algebras of degree three.}}
\end{table}
\medskip

\textbf{3}. For
\begin{equation}
\left( \alpha ,\beta ,x,y\right) =\kappa \left(
1,0,-E_{1}-E_{2}+E_{3},0\right)  \label{ast-ast-ast}
\end{equation}
%\textbf{(corresponding, by Eq. (\ref {diag-diag}), to }$\lambda
%_{1}=\lambda _{2}=-\lambda _{3}=-1$\textbf{)}, one gets:
the annihilation condition
\begin{equation}
\mathcal{T}(A,B)\kappa \left( 1,0,-E_{1}-E_{2}+E_{3},0\right) =0
\end{equation}
requires $\alpha ^{\prime }=0$, $x^{\prime }=0$ and
\begin{equation}
\left\{
\begin{array}{l}
\beta =0\Rightarrow \beta ^{\prime }=0=T\left( B,x\right) =Tr\left( B\circ
\left( -E_{1}-E_{2}+E_{3}\right) \right) ; \\
\\
y=0\Rightarrow y^{\prime }=0=2A\wedge x+B=2A\wedge \left(
-E_{1}-E_{2}+E_{3}\right) +B.
\end{array}
\right.  \label{Third}
\end{equation}

For generic elements $A$, $B$ belonging to the simple Euclidean Jordan
algebra of degree three $J_{3}^{\mathbb{A}}$ (where $\mathbb{A}=\mathbb{O}$,
$\mathbb{H}$, $\mathbb{C}$, $\mathbb{R}$), the general solution of Eqs. (\ref
{Third}) yields non-compact (Hermitian) generators of the form $\mathcal{T}%
(A,A)$ with (the $3\times 3$ Hermitian matrix representation of $A$ reading)
\begin{equation}
A=\left(
\begin{array}{ccccc}
\alpha _{1} &  & a_{3} &  & 0 \\
&  &  &  &  \\
a_{3}^{\ast } &  & \alpha _{2} &  & 0 \\
&  &  &  &  \\
0 &  & 0 &  & \alpha _{1}+\alpha _{2}
\end{array}
\right) ,~~dim_{\mathbb{R}}\left( A\right) =2+dim_{\mathbb{R}}\left( \mathbb{%
A}\right)  \label{repr-AA}
\end{equation}
and compact (anti-Hermitian) generators of the form $\mathcal{T}(B,-B)$ with
(the $3\times 3$ Hermitian matrix representation of $B$ reading)
\begin{equation}
B=\left(
\begin{array}{ccccc}
0 &  & 0 &  & a_{2}^{\ast } \\
&  &  &  &  \\
0 &  & 0 &  & a_{1} \\
&  &  &  &  \\
a_{2} &  & a_{1}^{\ast } &  & 0
\end{array}
\right) ,~~dim_{\mathbb{R}}\left( B\right) =2dim_{\mathbb{R}}\left( \mathbb{A%
}\right) .  \label{repr-BB}
\end{equation}
$a_{1}$, $a_{2}$ and $a_{3}$ are elements of the division algebra $\mathbb{A}
$, whereas $\alpha _{1}$ and $\alpha _{2}$ are real numbers.

The subgroup of $Str_{0}\left( \mathcal{J}\right) $ that leaves the
reference vector (\ref{ast-ast-ast}) invariant is a (non-compact) real form
of $Aut\left( \mathcal{J}\right) $, which we will denote as $\breve{H}\left(
\mathcal{J}\right) $.

The non-compact generators $\mathcal{T}(A,A)$ and the compact generators $%
\mathcal{T}(B,-B)$ (with $A$ and $B$ respectively given by Eqs. (\ref
{repr-AA}) and (\ref{repr-BB})) form a representation of $\breve{H}\left(
\mathcal{J}\right) \varsubsetneq Str_{0}\left( \mathcal{J}\right) $ of
dimension
\begin{equation}
dim_{\mathbb{R}}\left( A\right) +dim_{\mathbb{R}}\left( B\right) =2+3dim_{%
\mathbb{R}}\left( \mathbb{A}\right) =dim\left( \mathcal{J}\right) -1.
\end{equation}

Furthermore, the non-compact generators $\mathcal{T}(A,A)$, the compact
generators $\mathcal{T}(B,-B)$ (with $A$ and $B$ respectively given by Eqs. (%
\ref{repr-AA}) and (\ref{repr-BB})) and the generators of $\breve{H}\left(
\mathcal{J}\right) $ all together generate the non-compact stability group
of the reference vector (\ref{ast-ast-ast}), which we will denote as $\Sigma
_{0}\left( \mathcal{J}\right) $.

A similar treatment can be given also for the reducible case $\mathcal{J}=%
\mathbb{R}\oplus \mathbf{\Gamma }_{n+2}$ . By also using the results of \cite
{FG,FG2}, in Table 6 we list the groups $\breve{H}\left( \mathcal{J}\right) $
and $\Sigma _{0}\left( \mathcal{J}\right) $ for all the symmetric $N=2$, $%
d=4 $ MESGTs related to an Euclidean Jordan algebra of degree three $%
\mathcal{J}$.
\begin{table}[t]
\begin{center}
\begin{tabular}{|c||c|c|c|}
\hline
$
\begin{array}{c}
\\
~ \\
~
\end{array}
$ & $\mathcal{J}$ & $\breve{H}\left( \mathcal{J}\right) $ & $\Sigma
_{0}\left( \mathcal{J}\right) $ \\ \hline\hline
$
\begin{array}{c}
\\
II \\
~
\end{array}
$ & $\mathbb{R}\oplus \mathbf{\Gamma }_{n+2}$ & $SO(n,1)$ & $SO(n+1,1)\times
SO(1,1)$ \\ \hline
$
\begin{array}{c}
\\
III \\
~
\end{array}
$ & $J_{3}^{\mathbb{O}}$ & $F_{4(-20)}$ & $E_{6(-26)} $ \\ \hline
$
\begin{array}{c}
\\
IV \\
~
\end{array}
$ & $J_{3}^{\mathbb{H}}$ & $USp(4,2)$ & $SU^*(6)$ \\ \hline
$
\begin{array}{c}
\\
V \\
~
\end{array}
$ & $J_{3}^{\mathbb{C}}$ & $SU(2,1)$ & $SL(3,\mathbb{C})$ \\ \hline
$
\begin{array}{c}
\\
VI \\
~
\end{array}
$ & $J_{3}^{\mathbb{R}}$ & $SL(2,\mathbb{R})$ & $SL(3,\mathbb{R})$ \\ \hline
\end{tabular}
\end{center}
\caption{\textbf{Non-compact group }$\breve{H}\left( \mathcal{J}\right)
\varsubsetneq Str_{0}\left( \mathcal{J}\right) $\textbf{\ and non-compact
stability group }$\Sigma _{0}\left( \mathcal{J}\right) \sim Str_{0}^{\ast
}\left( \mathcal{J}\right) $\textbf{\ of the reference vector }$\protect%
\kappa \left( 1,0,-E_{1}-E_{2}+E_{3},0\right) $\textbf{\ of the FTSs
corresponding to all }$N=2$\textbf{, }$d=4$\textbf{\ MESGTs defined by
Euclidean Jordan algebras of degree three. }}
\end{table}

Since $\Sigma _{0}\left( \mathcal{J}\right) \sim Str_{0}^{\ast }\left(
\mathcal{J}\right) $, the charge orbits of case 1 and the charge orbits $%
\frac{Aut\left( \mathcal{F}\left( \mathcal{J}\right) \right) }{\Sigma
_{0}\left( \mathcal{J}\right) }$ of the present case overlap:
\begin{equation}
\frac{G_{4}\equiv G}{\widehat{H}}=\frac{Aut\left( \mathcal{F}\left( \mathcal{%
J}\right) \right) }{Str_{0}^{\ast }\left( \mathcal{J}\right) }\sim \frac{%
Aut\left( \mathcal{F}\left( \mathcal{J}\right) \right) }{\Sigma _{0}\left(
\mathcal{J}\right) }.
\end{equation}
This is consistent with the result that the corresponding reference vectors (%
\ref{ast}) and (\ref{ast-ast-ast}) can be mapped into each other by the
action of $Aut\left( \mathcal{F}\left( \mathcal{J}\right) \right) $, as
discussed above ($\kappa _{1},\kappa _{3}\in \mathbb{R}_{0}^{+}$):
\begin{equation}
\kappa _{1}\left( 1,0,E_{1}+E_{2}+E_{3},0\right) \overset{Aut\left( \mathcal{%
F}\left( \mathcal{J}\right) \right) }{\longleftrightarrow }\kappa _{3}\left(
1,0,-E_{1}-E_{2}+E_{3},0\right) .
\end{equation}

The orbits $\frac{Aut\left( \mathcal{F}\left( \mathcal{J}\right) \right) }{%
\Sigma _{0}\left( \mathcal{J}\right) }$ can be interpreted as the non-BPS, $%
Z\neq 0$ ones with a negative quartic invariant
\begin{equation}
I_{4}=-4\kappa ^{4}=-4\lambda _{1}\lambda _{2}\lambda _{3}<0,
\end{equation}
and with three negative ``eigenvalues''.\medskip

\textbf{4}. For
\begin{equation}
\left( \alpha ,\beta ,x,y\right) =\kappa \left(
1,0,E_{1}+E_{2}-E_{3},0\right)  \label{ast-ast-ast-ast}
\end{equation}
% \textbf{(corresponding, by Eq.
%(\ref {diag-diag}), to }$\lambda _{1}=\lambda _{2}=-\lambda
%_{3}=1$\textbf{)}, one gets:
the annihilation condition
\begin{equation}
\mathcal{T}(A,B)\kappa \left( 1,0,E_{1}+E_{2}-E_{3},0\right) =0
\end{equation}
requires $\alpha ^{\prime }=0$, $x^{\prime }=0$ and
\begin{equation}
\left\{
\begin{array}{l}
\beta =0\Rightarrow \beta ^{\prime }=0=T\left( B,x\right) =-Tr\left( B\circ
\left( -E_{1}-E_{2}+E_{3}\right) \right) ; \\
\\
y=0\Rightarrow y^{\prime }=0=2A\wedge x+B=-2A\wedge \left(
-E_{1}-E_{2}+E_{3}\right) +B.
\end{array}
\right.  \label{Fourth}
\end{equation}
For generic elements $A,B\in J_{3}^{\mathbb{A}}$ ($\mathbb{A}=\mathbb{O}$, $%
\mathbb{H}$, $\mathbb{C}$, $\mathbb{R}$), the general solution of Eqs. (\ref
{Fourth}) yields compact (anti-Hermitian) generators of the form $\mathcal{T}%
(A,-A)$ and non-compact (Hermitian) generators of the form $\mathcal{T}(B,B)$%
, with (the $3\times 3$ Hermitian matrix representations of) $A$ and $B$
respectively given by Eqs. (\ref{repr-AA}) and (\ref{repr-BB}).

The subgroup of $Str_{0}\left( \mathcal{J}\right) $ that leaves the
reference vector (\ref{ast-ast-ast-ast}) invariant is $\breve{H}\left(
\mathcal{J}\right) $ as in case 3. However, in the present case the number
of compact and non-compact generators outside $\mathfrak{Str}\left( \mathcal{%
J}\right) $ is respectively equal to $2+dim_{\mathbb{R}}\left( \mathbb{A}%
\right) $ and $2dim_{\mathbb{R}}\left( \mathbb{A}\right) $, i.e. the
opposite of what happens in case 3.

Furthermore, the compact generators $\mathcal{T}(A,-A)$, the non-compact
generators $\mathcal{T}(B,B)$ (with $A$ and $B$ respectively given by Eqs. (%
\ref{repr-AA}) and (\ref{repr-BB})) and the generators of $\breve{H}\left(
\mathcal{J}\right) $ all together generate the non-compact stability group
of the reference vector (\ref{ast-ast-ast-ast}), which we will denote as $%
\Delta _{0}\left( \mathcal{J}\right) $.

A similar treatment can be given also for the reducible case $\mathcal{J}=%
\mathbb{R}\oplus \mathbf{\Gamma }_{n+2}$ using results of \cite{FG,FG2}.

Therefore, the corresponding non-degenerate charge orbits are of the form

\begin{equation}
\frac{Aut\left( \mathcal{F}\left( \mathcal{J}\right) \right) }{\Delta
_{0}\left( \mathcal{J}\right) }=\frac{G_{4}\equiv G}{\widetilde{H}}.
\end{equation}
They have positive quartic invariant
\begin{equation}
I_{4}=4\kappa ^{4}=4\left| \lambda _{1}\lambda _{2}\lambda _{3}\right| >0
\end{equation}
and two negative ``eigenvalues''.

By also using the results of \cite{FG,FG2}, in Table 7 we list the groups $%
\mathcal{\breve{H}}\left( \mathcal{J}\right) $ and the orbits $\frac{%
Aut\left( \mathcal{F}\left( \mathcal{J}\right) \right) }{\Delta _{0}\left(
\mathcal{J}\right) }$ for all the symmetric $N=2$, $d=4$ MESGTs defined by
Euclidean Jordan algebras of degree three $\mathcal{J}$. The orbits $\frac{%
Aut\left( \mathcal{F}\left( \mathcal{J}\right) \right) }{\Delta _{0}\left(
\mathcal{J}\right) }$ correspond to the non-BPS, $Z=0$ ones (the right
column of Table 7 coincides with the column of $\mathcal{O}_{non-BPS,Z=0}$
orbits in Table 3).

\begin{table}[t]
\begin{center}
\begin{tabular}{|c||c|c|c|}
\hline
$
\begin{array}{c}
\\
~ \\
~
\end{array}
$ & $\mathcal{J}$ & $\breve{\mathcal{H}}\left( \mathcal{J}\right) $ & $\frac{%
Aut\left( \mathcal{F}\left( \mathcal{J}\right) \right) }{\Delta _{0}\left(
\mathcal{J}\right) }$ \\ \hline\hline
$
\begin{array}{c}
\\
II \\
~
\end{array}
$ & $\mathbb{R}\oplus \mathbf{\Gamma }_{n+2}$ & $SO(n,1)$ & $\frac{%
SU(1,1)\otimes SO(2,2+n)}{SO(2)\otimes SO(2,n)}$ \\ \hline
$
\begin{array}{c}
\\
III \\
~
\end{array}
$ & $J_{3}^{\mathbb{O}}$ & $F_{4(-20)}$ & $\frac{E_{7(-25)}}{E_{6(-14)}}$ \\
\hline
$
\begin{array}{c}
\\
IV \\
~
\end{array}
$ & $J_{3}^{\mathbb{H}}$ & $USp(4,2)$ & $\frac{SO^{\ast }(12)}{SU(4,2)}$ \\
\hline
$
\begin{array}{c}
\\
V \\
~
\end{array}
$ & $J_{3}^{\mathbb{C}}$ & $SU(2,1)$ & $\frac{SU(3,3)}{SU(2,1)\otimes SU(1,2)%
}$ \\ \hline
$
\begin{array}{c}
\\
VI \\
~
\end{array}
$ & $J_{3}^{\mathbb{R}}$ & $SL(2,\mathbb{R})$ & $\frac{Sp(6,\mathbb{R})}{%
SU(2,1)}$ \\ \hline
\end{tabular}
\end{center}
\caption{\textbf{Non-compact group }$\breve{H}\left( \mathcal{J}\right) $%
\textbf{\ and class }$\frac{Aut\left( \mathcal{F}\left( \mathcal{J}\right)
\right) }{\Delta _{0}\left( \mathcal{J}\right) }$\textbf{\ of non-degenerate
charge orbits with a positive quartic norm and two negative ``eigenvalues''
for all }$N=2$\textbf{, }$d=4$\textbf{\ symmetric MESGTs defined by
Euclidean Jordan algebras of degree three. }}
\end{table}

Notice that, consistently with the previous treatment, the reference vectors
(\ref{ast-ast}) and (\ref{ast-ast-ast-ast}) (both having $\lambda
_{1}\lambda _{2}\lambda _{3}<0$) cannot be mapped into each other by $%
Aut\left( \mathcal{F}\left( \mathcal{J}\right) \right) $, and therefore the
corresponding non-degenerate charge orbits do not overlap ($\kappa
_{2},\kappa _{4}\in \mathbb{R}_{0}^{+}$):
\begin{gather}
\kappa _{2}\left( 1,0,-E_{1}-E_{2}-E_{3},0\right) \overset{Aut\left(
\mathcal{F}\left( \mathcal{J}\right) \right) }{\nleftrightarrow }\kappa
_{4}\left( 1,0,E_{1}+E_{2}-E_{3},0\right) \\
\Downarrow  \notag \\
\mathcal{O}_{\frac{1}{2}-BPS}=\frac{G_{4}\equiv G}{H_{0}}=\frac{Aut\left(
\mathcal{F}\left( \mathcal{J}\right) \right) }{Ktr_{0}(\mathcal{J})}\neq
\frac{Aut\left( \mathcal{F}\left( \mathcal{J}\right) \right) }{\Delta
_{0}\left( \mathcal{J}\right) }=\frac{G_{4}\equiv G}{\widetilde{H}}=\mathcal{%
O}_{non-BPS,Z=0}.
\end{gather}
\ \medskip

\section{\label{N=2-Attractors}Classification of Attractors}

The three classes of orbits in Table 3 correspond to the three distinct
classes of solutions of the $N=2$, $d=4$ extremal black hole attractor
equations (\ref{AEs1}) and (\ref{AEs2}). \setcounter{equation}0
\def\theequation{4.\arabic{subsection}.\arabic{equation}}
\subsection{\label{N=2-Attractors-BPS}$\frac{1}{2}$-BPS solutions}

As already mentioned, the class of $\frac{1}{2}$-BPS orbits corresponds to
the solution (\ref{BPS}) determining $N=2$, $\frac{1}{2}$-BPS critical
points of $V_{BH}$. Such a solution yields the following value of the black
hole scalar potential at the considered attractor point(s) \cite{FGK}:
\begin{equation}
V_{BH,\frac{1}{2}-BPS}=\left| Z\right| _{\frac{1}{2}-BPS}^{2}+\left[ G^{i%
\overline{i}}D_{i}Z\overline{D}_{\overline{i}}\overline{Z}\right] _{\frac{1}{%
2}-BPS}=\left| Z\right| _{\frac{1}{2}-BPS}^{2}.
\end{equation}
\textit{The overall symmetry group at }$N=2$\textit{\ }$\frac{1}{2}$\textit{%
-BPS critical point(s) is }$H_{0}$\textit{, stabilizer of }$~\mathcal{O}_{%
\frac{1}{2}-BPS}=\frac{G}{H_{0}}$. The \textit{symmetry enhancement} is
given by Eq. (\ref{symm-en-BPS}). For such a class of orbits
\begin{equation}
I_{4,\frac{1}{2}-BPS}=\left| Z\right| _{\frac{1}{2}-BPS}^{4}>0.
\end{equation}
\setcounter{equation}0
\def\theequation{4.\arabic{subsection}.\arabic{equation}}
\subsection{\label{N=2-Attractors-non-BPS}Non-BPS solutions}

The two classes of $N=2$ non-BPS non-degenerate charge orbits respectively
correspond to the following solutions of $N=2$ attractor eqs. (\ref{AEs1}):
\begin{equation}
\text{non-BPS, }Z\neq 0\text{:}\left\{
\begin{array}{l}
Z\neq 0, \\
\\
D_{i}Z\neq 0\text{ for some }i\in \left\{ 1,...,n_{V}\right\} , \\
\\
I_{4,non-BPS,Z\neq 0}=-\left( \left| Z\right| _{non-BPS,Z\neq 0}^{2}+\left(
G^{i\overline{j}}D_{i}Z\overline{D}_{\overline{j}}\overline{Z}\right)
_{non-BPS,Z\neq 0}\right) ^{2}= \\
\\
=-16\left| Z\right| _{non-BPS,Z\neq 0}^{4}<0;
\end{array}
\right.
\end{equation}
\begin{equation}
\text{non-BPS, }Z=0\text{:}\left\{
\begin{array}{l}
Z=0, \\
\\
D_{i}Z\neq 0\text{ for some }i\in \left\{ 1,...,n_{V}\right\} , \\
\\
I_{4,non-BPS,Z=0}=\left( G^{i\overline{j}}D_{i}Z\overline{D}_{\overline{j}}%
\overline{Z}\right) _{non-BPS,Z=0}^{2}>0.
\end{array}
\right.
\end{equation}

In Subsubsects. \ref{N=2-Attractors-non-BPS-1} and \ref
{N=2-Attractors-non-BPS-2} we will show how the general solutions of Eqs. (%
\ref{AEs1}), respectively determining the two aforementioned classes of $N=2$
non-BPS extremal black hole attractors, can be easily given by using
``flat'' local $I$-coordinates in the scalar manifold.

Otherwise speaking, we will consider the ``flatted'' attractor eqs. (\ref
{AEs2}), which can be specialized in the regular non-BPS cases as follows:
\begin{eqnarray}
&&
\begin{array}{l}
\text{non-BPS, }Z\neq 0\text{:~~~}2\overline{Z}D_{I}Z=-iC_{IJK}\delta ^{J%
\overline{J}}\delta ^{K\overline{K}}\overline{D}_{\overline{J}}\overline{Z}%
\overline{D}_{\overline{K}}\overline{Z};
\end{array}
\notag \\
&&  \label{AEs-non-BPS1-flat} \\
&&
\begin{array}{l}
\text{non-BPS, }Z=0\text{:~~~}C_{IJK}\delta ^{J\overline{J}}\delta ^{K%
\overline{K}}\overline{D}_{\overline{J}}\overline{Z}\overline{D}_{\overline{K%
}}\overline{Z}=0.
\end{array}
\notag \\
&&  \label{AEs-non-BPS2-flat}
\end{eqnarray}
Thus, by respectively denoting with $\widehat{H}$ ($\widetilde{H}$) the
stabilizer of the $N=2$, non-BPS, $Z\neq 0$ ($Z=0$) classes of orbits listed
in Table 3, our claim is the following: \textit{the general solution of Eqs.
}(\ref{AEs-non-BPS1-flat})\textit{\ (}(\ref{AEs-non-BPS2-flat})\textit{) is
obtained by retaining a complex charge vector }$\left( Z,D_{I}Z\right) $%
\textit{\ which is invariant under }$\widehat{h}$\textit{\ (}$\frac{%
\widetilde{h}}{U(1)}$\textit{), where }$\widehat{h}$\textit{\ (}$\widetilde{h%
}$\textit{) is the m.c.s.}\footnote{%
Indeed, while $H_{0}$ is a proper compact subgroup of $G$, the groups $%
\widehat{H}$, $\widetilde{H}$ are real (non-compact) forms of $H_{0}$, as it
can be seen from Table 3 (see also\textbf{\ }\cite{Gilmore,Helgason}).
Therefore in general they admit a m.c.s. $\widehat{h}$, $\widetilde{h}$,
which in turn is a (non-maximal) compact subgroup of $G$ and a proper
compact subgroup of $H_{0}$.
\par
It is interesting to notice that in all cases (listed in Table 3) $G$ always
admits only 2 real (non-compact) forms $\widehat{H}$, $\widetilde{H}$ of $%
H_{0}$ as proper subgroups (consistent with the required dimension of
orbits). The inclusion of $\widehat{H}$, $\widetilde{H}$ in $G$ is such that
in all cases $\widehat{H}\otimes SO(1,1)$ and $\widetilde{H}\otimes U(1)$
are different maximal non-compact subgroups of $G$.}\textit{\ of }$\widehat{H%
}$\textit{\ (}$\widetilde{H}$\textit{)}.

As a consequence, \textit{the overall symmetry group of the }$N=2$\textit{,
non-BPS, }$Z\neq 0$\textit{\ (}$Z=0$\textit{) critical point(s) is }$%
\widehat{h}$\textit{\ (}$\frac{\widetilde{h}}{U(1)}$\textit{). }Thus, at $%
N=2 $, non-BPS, $Z\neq 0$\ ($Z=0$) critical point(s) the following \textit{%
enhancement of symmetry} holds:
\begin{equation}
\begin{array}{l}
N=2\text{, non-BPS, }Z\neq 0:\mathcal{S}\longrightarrow \widehat{h}\
=m.c.s.\left( \widehat{H}\right) ; \\
\\
N=2\text{, non-BPS, }Z=0:\mathcal{S}\longrightarrow \ \frac{\widetilde{h}}{%
U(1)}=\frac{m.c.s.\left( \widetilde{H}\right) }{U(1)}.
\end{array}
\label{symm-en-non-BPS}
\end{equation}

It is worth remarking that the non-compact group $\widehat{H}$ stabilizing
the non-BPS, $Z\neq 0$ class of orbits of $N=2$, $d=4$ symmetric MESGTs,
beside being a real (non-compact) form of $H_{0}$, \ is isomorphic to the
duality group $G_{5}$ of $N=2$, $d=5$ symmetric MESGTs and was denoted as $%
Str_{0}^{\ast }(\mathcal{J})$ in Sect. 3 (see Table 4) \footnote{%
Such a feature is missing in the $N=2$, $d=4$ symmetric MESGTs whose scalar
manifolds belong to the sequence $I$, simply because such theories do not
have a class of non-BPS, $Z\neq 0$ orbits.}. Consequently, in the cases $II$-%
$VI$ of Table 3, Eq. (\ref{GG}) can be completed as follows: \textbf{
\begin{equation}
H_{5}=Aut\left( \mathcal{J}\right) =m.c.s.\left( G_{5}\right) \varsubsetneq
\widehat{H}=Str_{0}^{\ast }\left( \mathcal{J}\right) \varsubsetneq Aut\left(
\mathcal{F}\left( \mathcal{J}\right) \right) =G\equiv G_{4}.
\end{equation}
}

Since the scalar manifolds of $N=2$, $d=5$ symmetric MESGTs are endowed with
a real special geometry (see Footnote 10) \cite{GST1,GST2,GST3}, the complex
representation $R_{H_{0}}$ of $H_{0}$ decomposes in a pair of irreducible
real representations $\left( R_{\widehat{h}}+\mathbf{1}\right) _{\mathbb{R}}$%
's of $\widehat{h}=m.c.s.\left( \widehat{H}\right) \varsubsetneq $ $H_{0}$
(see Subsubsect. \ref{N=2-Attractors-non-BPS-1}, and in particular Eq. (\ref
{KR})). As we will see in Subsubsects. \ref{N=2-Attractors-non-BPS-1} and
\ref{N=2-Attractors-non-BPS-2}, such a fact crucially distinguishes the
non-BPS, $Z\neq 0$ and $Z=0$ cases.

The stabilizers (and the corresponding m.c.s.s) of the non-BPS, $Z\neq 0$
and $Z=0$ classes of orbits of $N=2$, $d=4$ symmetric MESGTs are given in
Table 8.

\begin{table}[t]
\begin{center}
\begin{tabular}{|c||c|c|c|c|c|}
\hline
& $
\begin{array}{c}
\\
H_{0} \\
~
\end{array}
$ & $
\begin{array}{c}
\\
\widehat{H} \\
~
\end{array}
$ & $
\begin{array}{c}
\\
\widetilde{H} \\
~
\end{array}
$ & $
\begin{array}{c}
\\
\widehat{h}\equiv m.c.s.\left( \widehat{H}\right)  \\
~
\end{array}
$ & $
\begin{array}{c}
\\
\widetilde{h}^{\prime }\equiv \frac{m.c.s.\left( \widetilde{H}\right) }{U(1)}
\\
~
\end{array}
$ \\ \hline\hline
$I$ & $
\begin{array}{c}
\\
SU(n+1) \\
~
\end{array}
$ & $-$ & $SU(1,n)$ & $-$ & $SU(n)$ \\ \hline
$II$ & $
\begin{array}{c}
SO(2) \\
\otimes  \\
SO(2+n)
\end{array}
$ & $
\begin{array}{c}
SO(1,1) \\
\otimes  \\
SO(1,1+n)
\end{array}
$ & $
\begin{array}{c}
SO(2) \\
\otimes  \\
SO(2,n)
\end{array}
$ & $SO(1+n)$ & $
\begin{array}{c}
SO(2) \\
\otimes  \\
SO(n)
\end{array}
$ \\ \hline
$III$ & $
\begin{array}{c}
\\
E_{6}\equiv E_{6(-78)} \\
~
\end{array}
$ & $E_{6(-26)}$ & $E_{6(-14)}$ & $F_{4}\equiv F_{4(-52)}$ & $SO(10)$ \\
\hline
$IV$ & $SU(6)$ & $SU^{\ast }(6)$ & $SU(4,2)$ & $USp(6)$ & $
\begin{array}{c}
SU(4) \\
\otimes  \\
SU(2)
\end{array}
$ \\ \hline
$V$ & $SU(3)\otimes SU(3)$ & $SL(3,\mathbb{C})$ & $
\begin{array}{c}
SU(2,1) \\
\otimes  \\
SU(1,2)
\end{array}
$ & $SU(3)$ & $
\begin{array}{c}
SU(2) \\
\otimes  \\
SU(2)\otimes U(1)
\end{array}
$ \\ \hline
$VI$ & $
\begin{array}{c}
\\
SU(3) \\
~
\end{array}
$ & $SL(3,\mathbb{R})$ & $SU(2,1)$ & $SO(3)$ & $SU(2)$ \\ \hline
\end{tabular}
\end{center}
\caption{\textbf{Stabilizers and corresponding m.c.s.s of the non-degenerate
classes of orbits of }$N=2$\textbf{, }$d=4$\textbf{\ symmetric MESGTs}. {$%
\widehat{H}$ and $\widetilde{H}$ are real (non-compact) forms of $H_{0}$,
the stabilizer of $\frac{1}{2}$-BPS orbits.}}
\end{table}
\setcounter{equation}0
\def\theequation{4.2.\arabic{subsubsection}.\arabic{equation}}
\subsubsection{\label{N=2-Attractors-non-BPS-1}Non-BPS, $Z\neq 0$\ solutions}

Let us start by considering the class of non-BPS, $Z\neq 0$ orbits of $N=2$,
$d=4$ symmetric MESGTs.

As mentioned, the ``flatted matter charges'' $D_{I}Z$ are a vector of $%
R_{H_{0}}$. In general, $R_{H_{0}}$ decomposes under the m.c.s. $\widehat{h}%
\subset \widehat{H}$ as follows:
\begin{equation}
R_{H_{0}}\longrightarrow \left( R_{\widehat{h}}+\mathbf{1}\right) _{\mathbb{C%
}}\mathbf{,}  \label{KR}
\end{equation}
where the r.h.s. is made of the complex singlet representation of $\widehat{h%
}$ and by another non-singlet real representation of $\widehat{h}$, denoted
above with $R_{\widehat{h}}$. As previously mentioned, despite being
complex, $\left( R_{\widehat{h}}+\mathbf{1}\right) _{\mathbb{C}}$ is not
charged with respect to $U(1)$ symmetry because, due to the 5-dimensional
origin of the non-compact stabilizer $\widehat{H}$ whose m.c.s. is $\widehat{%
h}$, actually $\left( R_{\widehat{h}}+\mathbf{1}\right) _{\mathbb{C}}$ is
nothing but the complexification of its real counterpart$\left( R_{\widehat{h%
}}+\mathbf{1}\right) _{\mathbb{R}}$. The decomposition (\ref{KR})yields the
following splitting of ``flatted matter charges'':
\begin{equation}
D_{I}Z\longrightarrow \left( D_{\widehat{I}}Z,D_{\widehat{I}_{0}}Z\right) ,
\end{equation}
where $\widehat{I}$ are the indices along the representation $R_{\widehat{h}%
} $, and $\widehat{I}_{0}$ is the $\widehat{h}$-singlet index.

By considering the related attractor eqs., it should be noticed that the
rank-3 symmetric tensor $C_{IJK}$ in Eqs.\textit{\ }(\ref{AEs-non-BPS1-flat}%
) corresponds to a cubic $H_{0}$-invariant coupling $\left( R_{H_{0}}\right)
^{3}$. By decomposing $\left( R_{H_{0}}\right) ^{3}$ in terms of
representations of $\widehat{h}$, one finds
\begin{equation}
\left( R_{H_{0}}\right) ^{3}\longrightarrow \left( R_{\widehat{h}}\right)
^{3}+\left( R_{\widehat{h}}\right) ^{2}\mathbf{1}_{\mathbb{C}}+\left(
\mathbf{1}_{\mathbb{C}}\right) ^{3}\mathbf{.}  \label{decomp-non-BPS-Z<>0}
\end{equation}
Notice that a term $R_{\widehat{h}}\left( \mathbf{1}_{\mathbb{C}}\right)
^{2} $ cannot be in such a representation decomposition, since it is not $%
\widehat{h}$-invariant, and thus not $H_{0}$-invariant. This implies that
components of the form $C_{\widehat{I}\widehat{I}_{0}\widehat{I}_{0}}$
cannot exist. Also, a term like $\left( \mathbf{1}_{\mathbb{C}}\right) ^{3}$
can appear in the r.h.s. of the decomposition (\ref{AEs-non-BPS1-flat})
since as we said the $\widehat{h}$-singlet $\mathbf{1}_{\mathbb{C}}$,
despite being complex, is \textit{not} $U(1)$-charged.

It is then immediate to conclude that the solution of $N=2$, $d=4$ non-BPS, $%
Z\neq 0$ extremal black hole attractor eqs. in ``flat'' indices (\ref
{AEs-non-BPS1-flat}) corresponds to keep the ``flatted matter charges'' $%
D_{I}Z$ $\widehat{h}$-invariant. By virtue of decomposition (\ref
{decomp-non-BPS-Z<>0}), this is obtained by putting
\begin{equation}
D_{\widehat{I}}Z=0,D_{\widehat{I}_{0}}Z\neq 0,  \label{sol-non-BPS-Z<>0}
\end{equation}
i.e. by putting all ``flatted matter charges'' to zero, except the one along
the $\widehat{h}$-singlet (and thus $\widehat{h}$-invariant) direction in
scalar manifold. By substituting the solution (\ref{sol-non-BPS-Z<>0}) in
Eqs. (\ref{AEs-non-BPS1-flat}), one obtains
\begin{gather}
2\overline{Z}D_{\widehat{I}_{0}}Z=-iC_{\widehat{I}_{0}\widehat{I}_{0}%
\widehat{I}_{0}}\left( \overline{D}_{\overline{\widehat{I}_{0}}}\overline{Z}%
\right) ^{2}\overset{Z\neq 0}{\Leftrightarrow }D_{\widehat{I}_{0}}Z=-\frac{i%
}{2}\frac{C_{\widehat{I}_{0}\widehat{I}_{0}\widehat{I}_{0}}}{\overline{Z}}%
\left( \overline{D}_{\overline{\widehat{I}_{0}}}\overline{Z}\right) ^{2} \\
\Downarrow  \notag \\
\left| D_{\widehat{I}_{0}}Z\right| ^{2}\left( 1-\frac{1}{4}\frac{\left| C_{%
\widehat{I}_{0}\widehat{I}_{0}\widehat{I}_{0}}\right| ^{2}}{\left| Z\right|
^{2}}\left| D_{\widehat{I}_{0}}Z\right| ^{2}\right) =0  \notag \\
\Updownarrow  \notag \\
\left| D_{\widehat{I}_{0}}Z\right| ^{2}=4\frac{\left| Z\right| ^{2}}{\left|
C_{\widehat{I}_{0}\widehat{I}_{0}\widehat{I}_{0}}\right| ^{2}};
\label{sol-non-BPS-Z<>0-2}
\end{gather}
this is nothing but the general criticality condition of $V_{BH}$ for the
1-modulus case in the locally ``flat'' coordinate $\widehat{I}_{0}$, which
in this case corresponds to the $\widehat{h}$-singlet direction in the
scalar manifold. Such a case has been thoroughly studied in non-flat $i$%
-coordinates in \cite{BFM}.\smallskip

All $N=2$, $d=4$ symmetric MESGTs (disregarding the sequence $I$ having $%
C_{ijk}=0$) have a cubic prepotential ($F=\frac{1}{3!}d_{ijk}t^{i}t^{j}t^{k}$
in special coordinates), and thus in special coordinates it holds that $%
C_{ijk}=e^{K}d_{ijk}$, with $K$ and $d_{ijk}$ respectively denoting the K%
\"{a}hler potential and the completely symmetric rank-3 constant tensor that
is determined by the norm form of the underlying Jordan algebra of degree
three \cite{GST2}. In the cubic $n_{V}=1$-modulus case, by using Eq. (\ref
{CERN2}) it follows that
\begin{equation}
\left( G^{1\overline{_{s}1}_{s}}\right) ^{3}\left|
C_{1_{s}1_{s}1_{s}}\right| ^{2}=\left| C_{1_{f}1_{f}1_{f}}\right| ^{2}=\frac{%
4}{3},  \label{ress1}
\end{equation}
where the subscripts ``$s$'' and ``$f$'' respectively stand for ``special''
and ``flat'', denoting the kind of coordinate system being considered. By
substituting Eq. (\ref{ress1}) in Eq. (\ref{sol-non-BPS-Z<>0-2}) one obtains
the result
\begin{equation}
\left| D_{\widehat{I}_{0}}Z\right| ^{2}=3\left| Z\right| ^{2}.  \label{ress2}
\end{equation}
Another way of proving Eq. (\ref{ress2})\ is by computing the quartic
invariant along the $\widehat{h}$-singlet direction, then yielding
\begin{equation}
I_{4,non-BPS,Z\neq 0}=-16\left| Z\right| _{non-BPS,Z\neq 0}^{2}.
\end{equation}

The considered solution (\ref{sol-non-BPS-Z<>0})-(\ref{sol-non-BPS-Z<>0-2}),
(\ref{ress2}) is the $N=2$ analogue of the $N=8$, $d=4$ non-BPS regular
solution discussed in \cite{FKlast}, and it yields the following value of
the black hole scalar potential at the considered attractor point(s) \cite
{BFM,FKlast}:
\begin{equation}
V_{BH,non-BPS,Z\neq 0}=4\left| Z\right| _{non-BPS,Z\neq 0}^{2}.
\label{ress3}
\end{equation}
Once again, as for the non-BPS $N=8$ regular solutions (see Eq. (\ref
{S-non-BPS})), we find the extra factor $4$.

From the above considerations, \textit{the overall symmetry group at }$N=2$%
\textit{\ non-BPS, }$Z\neq 0$\textit{\ critical point(s) is }$\widehat{h}$%
\textit{, m.c.s. of the non-compact stabilizer }$\widehat{H}$\textit{\ of }$%
\mathcal{O}_{non-BPS,Z\neq 0}$\textit{.} \setcounter{equation}0
\def\theequation{4.2.\arabic{subsubsection}.\arabic{equation}}
\subsubsection{\label{N=2-Attractors-non-BPS-2}Non-BPS, $Z=0$\ solutions}

Let us now move to consider the other class of non-BPS orbits of $N=2$, $d=4$
symmetric MESGTs.

It has $Z=0$ and it was not considered in \cite{FG} (see also Footnote 7).
We will show that the solution of the $N=2$, $d=4$, non-BPS, $Z=0$ extremal
black hole attractor eqs. (\ref{AEs-non-BPS2-flat}) are the ``flatted matter
charges'' $D_{I}Z$ which are invariant under $\frac{\widetilde{h}}{U(1)}$,
where $\widetilde{h}$ is the m.c.s. of $\widetilde{H}$, the stabilizer of
the class $\mathcal{O}_{non-BPS,Z=0}=\frac{G}{\widetilde{H}}$.

Differently from the non-BPS, $Z\neq 0$ case, in the considered non-BPS, $%
Z=0 $ case there is always a $U(1)$ symmetry acting, since the scalar
manifolds of $N=2$, $d=4$ symmetric MESGTs \textit{all} have the group $%
\widetilde{h}$ of the form
\begin{equation}
\widetilde{h}=\widetilde{h}^{\prime }\otimes U(1),\text{ \ \ }\widetilde{h}%
^{\prime }\equiv \frac{\widetilde{h}}{U(1)}.  \label{gen-struct}
\end{equation}

The compact subgroups $\widetilde{h}^{\prime }$ for all $N=2$, $d=4$
symmetric MESGTs are listed in Table 8. In the case at hand, we thence have
to consider the decomposition of the previously introduced complex
representation $R_{H_{0}}$ under the compact subgroup $\widetilde{h}^{\prime
}\varsubsetneq H_{0}$. In general, $R_{H_{0}}$ decomposes under $\widetilde{h%
}^{\prime }\varsubsetneq \widetilde{H}$ as follows:
\begin{equation}
R_{H_{0}}\longrightarrow \left( \mathcal{W}_{\widetilde{h}^{\prime }}+%
\mathcal{Y}_{\widetilde{h}^{\prime }}+\mathbf{1}\right) _{\mathbb{C}}\mathbf{%
,}  \label{KR-KR-KR}
\end{equation}
where in the r.h.s. the complex singlet representation of $\widetilde{h}%
^{\prime }$ and two complex non-singlet representations $\mathcal{W}_{%
\widetilde{h}^{\prime }}$ and $\mathcal{Y}_{\widetilde{h}^{\prime }}$ of $%
\widetilde{h}^{\prime }$ appear. In general, $\mathcal{W}_{\widetilde{h}%
^{\prime }}$, $\mathcal{Y}_{\widetilde{h}^{\prime }}$ and $\mathbf{1}_{%
\mathbb{C}}$ are charged (and thus not invariant) with respect to the $U(1)$
explicit factor appearing in (\ref{gen-struct}). The decomposition (\ref
{KR-KR-KR}) yields the following splitting of ``flatted matter charges'':
\begin{equation}
D_{I}Z\longrightarrow \left( D_{\widetilde{I}_{\mathcal{W}}^{\prime }}Z,D_{%
\widetilde{I}_{\mathcal{Y}}^{\prime }}Z,D_{\widetilde{I}_{0}^{\prime
}}Z\right) ,
\end{equation}
where $\widetilde{I}_{\mathcal{W}}^{\prime }$ and $\widetilde{I}_{\mathcal{Y}%
}^{\prime }$ respectively denote the indices along the complex
representations $\mathcal{W}_{\widetilde{h}^{\prime }}$ and $\mathcal{Y}_{%
\widetilde{h}^{\prime }}$, and $\widetilde{I}_{0}^{\prime }$ is the $%
\widetilde{h}^{\prime }$-singlet index.

Once again, the related $N=2$, $d=4$ non-BPS, $Z=0$ extremal black hole
attractor eqs. (\ref{AEs-non-BPS2-flat}) contain the rank-3 symmetric tensor
$C_{IJK}$, corresponding to a cubic $H_{0}$-invariant coupling $\left(
R_{H_{0}}\right) ^{3}$. The decomposition of $\left( R_{H_{0}}\right) ^{3}$
in terms of representations of $\widetilde{h}^{\prime }$ yields
\begin{equation}
\left( R_{H_{0}}\right) ^{3}\longrightarrow \left( \mathcal{W}_{\widetilde{h}%
^{\prime }}\right) ^{2}\mathcal{Y}_{\widetilde{h}^{\prime }}+\left( \mathcal{%
Y}_{\widetilde{h}^{\prime }}\right) ^{2}\mathbf{1}_{\mathbb{C}}\mathbf{.}
\label{decomp-non-BPS-Z=0}
\end{equation}
When decomposed under $\widetilde{h}^{\prime }$, $\left( R_{H_{0}}\right)
^{3}$ must be nevertheless $\widetilde{h}$-invariant, and therefore, beside
the $\widetilde{h}^{\prime }$-invariance, one has to consider the invariance
under the $U(1)$ factor, too. Thus, terms of the form $\left( \mathcal{W}_{%
\widetilde{h}^{\prime }}\right) ^{3}$, $\left( \mathcal{Y}_{\widetilde{h}%
^{\prime }}\right) ^{3}$, $\mathcal{W}_{\widetilde{h}^{\prime }}\left(
\mathbf{1}_{\mathbb{C}}\right) ^{2}$, $\mathcal{Y}_{\widetilde{h}^{\prime
}}\left( \mathbf{1}_{\mathbb{C}}\right) ^{2}$ and $\left( \mathbf{1}_{%
\mathbb{C}}\right) ^{3}$ cannot exist in the $\widetilde{h}$-invariant
r.h.s. of decomposition (\ref{decomp-non-BPS-Z=0}).

Notice also that the structure of the decomposition (\ref{decomp-non-BPS-Z=0}%
) implies that components of the cubic coupling of the form $C_{\widetilde{I}%
_{\mathcal{W}}^{\prime }\widetilde{I}_{0}^{\prime }\widetilde{I}_{0}^{\prime
}}$, $C_{\widetilde{I}_{\mathcal{Y}}^{\prime }\widetilde{I}_{0}^{\prime }%
\widetilde{I}_{0}^{\prime }}$ and $C_{\widetilde{I}_{0}^{\prime }\widetilde{I%
}_{0}^{\prime }\widetilde{I}_{0}^{\prime }}$ cannot exist. For such a
reason, it is immediate to conclude that the solution of $N=2$, $d=4$
non-BPS, $Z=0$ extremal black hole attractor eqs. in ``flat'' indices (\ref
{AEs-non-BPS2-flat}) corresponds to keep the ``flatted matter charges'' $%
D_{I}Z$ $\widetilde{h}^{\prime }$-invariant. By virtue of decomposition (\ref
{decomp-non-BPS-Z=0}), this is obtained by putting
\begin{equation}
D_{\widetilde{I}_{\mathcal{W}}^{\prime }}Z=0=D_{\widetilde{I}_{\mathcal{Y}%
}^{\prime }}Z,~~~~D_{\widetilde{I}_{0}^{\prime }}Z\neq 0,
\label{sol-non-BPS-Z=0}
\end{equation}
i.e. by putting all ``flatted matter charges'' to zero, except the one along
the $\widetilde{h}^{\prime }$-singlet (and thus $\widetilde{h}^{\prime }$%
-invariant, but not $U(1)$-invariant and therefore not $\widetilde{h}$%
-invariant) direction in the scalar manifold.\medskip

The considered solution (\ref{sol-non-BPS-Z=0}) does not have any analogue
in $N=8$, $d=4$ SUGRA, and it yields the following value of the black hole
scalar potential at the considered attractor point(s):
\begin{eqnarray}
V_{BH,non-BPS,Z=0} &=&\left| Z\right| _{non-BPS,Z=0}^{2}+\left[ G^{i%
\overline{i}}D_{i}Z\overline{D}_{\overline{i}}\overline{Z}\right]
_{non-BPS,Z=0}=  \notag \\
&&  \notag \\
&=&\left| D_{\widetilde{I}_{0}^{\prime }}Z\right| _{non-BPS,Z=0}^{2}.
\end{eqnarray}
It is here worth remarking that in the $stu$ model it can be explicitly
computed that \cite{to-appear}
\begin{equation}
V_{BH,non-BPS,Z=0}=\left| D_{\widetilde{I}_{0}^{\prime }}Z\right|
_{non-BPS,Z=0}^{2}=\left| Z\right| _{\frac{1}{2}-BPS}^{2}=V_{BH,\frac{1}{2}%
-BPS}.  \label{stu-stu}
\end{equation}

From above considerations, \textit{the overall symmetry group at }$N=2$%
\textit{\ non-BPS, }$Z=0$\textit{\ critical point(s) is }$\widetilde{h}%
^{\prime }=\frac{\widetilde{h}}{U(1)}$\textit{, }$\widetilde{h}$\textit{\
being the m.c.s. of the non-compact stabilizer }$\widetilde{H}$\textit{\ of }%
$\mathcal{O}_{non-BPS,Z=0}$\textit{.\medskip }

The general analysis carried out above holds for all $N=2$, $d=4$ symmetric
``magical'' MESGTs, namely for the irreducible cases $III$-$VI$ listed in
Tables 2 and 3. The cases of irreducible sequence $I$ and of generic Jordan
family $II$ deserve suitable, slightly different treatments, respectively
given in Appendices I and II. \setcounter{equation}0
\def\theequation{4.\arabic{subsection}.\arabic{equation}}
\subsection{\label{Orb-Attr-O-H}Orbits and Attractors of $J_{3}^{\mathbb{O}}$
and $J_{3}^{\mathbb{H}}$}

\ Let us now apply the above analysis to the $N=2$, $d=4$ ``magical''
MESGTs, based on the symmetric special K\"{a}hler manifolds
\begin{equation}
\frac{E_{7(-25)}}{E_{6}\otimes U(1)},\text{ \ \ \ }\frac{SO^{\ast }(12)}{U(6)%
}.
\end{equation}
defined by simple Jordan algebras $J_{3}^{\mathbb{O}}$and $J_{3}^{\mathbb{H}%
} $ of Hermitian $3\times 3$ matrices over octonions $\mathbb{O}$ and
quaternions $\mathbb{H}$, respectively. \setcounter{equation}0
\def\theequation{4.3.\arabic{subsubsection}.\arabic{equation}}
\subsubsection{\label{dual-2-6}$N=2,6$ SUGRAs and the dual role of \ $\frac{%
SO^{\ast }(12)}{U(6)}$}

Before proceeding further, also in order to stress the relevance of such two
``magical'' MESGTs, it is here worth pointing out some similarities and
differences with respect to $N=8$, $d=4$ SUGRA, based on $\frac{E_{7(7)}}{%
SU(8)}$ and treated in Sect. \ref{Intro}.

Since the duality groups of $N=8$, $d=4$ SUGRA and $N=2$, $d=4$ ``magical'' $%
\frac{E_{7(-25)}}{E_{6}\otimes U(1)}$-based MESGT are two different real
(non-compact) forms of the Lie exceptional group $E_{7}\equiv E_{7(-133)}$ ($%
E_{7(7)}$ and $E_{7(-25)}$, respectively), both their ``charge vectors'' are
in the real symplectic representation\footnote{%
As mentioned in Sect. \ref{Intro}, for $E_{7}$ the real symplectic and the
fundamental representations coincide, but this is not generally the case.} $%
\mathbf{56}$ of $E_{7}$.

Nevertheless, while the real (non-compact) form $E_{7(7)}$ contains only two
different forms of $E_{6}$, namely the real (non-compact) forms $E_{6(2)}$
and $E_{6(6)}$, the group $E_{7(-25)}$ contains three different forms of $%
E_{6}$: $E_{6}\equiv E_{6(-78)}$ and its real (non-compact) forms $%
E_{6(-14)} $ and $E_{6(-26)}$. This fact is of course related to the
fundamental difference that in $N=8$, $d=4$ $\frac{E_{7(7)}}{SU(8)}$-based
SUGRA there are only two classes of 55-dim. non-degenerate orbits, whereas
all $N=2$, $d=4$ MESGTs with symmetric scalar manifolds that originate from
5 dimensions , in particular the ``exceptional'' MESGT with the scalar
manifold $\frac{E_{7(-25)}}{E_{6}\otimes U(1)}$, admit three distinct
classes of such orbits.

By recalling the criticality conditions (\ref{AEs1}) and (\ref{AEs2}) for $%
V_{BH}$, also previously referred to as the $N=2$, $d=4$ extremal black hole
attractor eqs., let us stress once again that the aforementioned three
classes of $N=2$ non-degenerate orbits correspond to the following three
classes of attractor solutions:\medskip

$\frac{1}{2}$\textbf{-BPS} : $Z\neq 0$, $D_{i}Z=0,\forall i=1,...,n_{V}$%
;\medskip

$\left( \text{\textbf{non-BPS}}\right) _{\mathbf{1}}$ : $Z\neq 0$, $%
D_{i}Z\neq 0$, for some $i\in \left\{ 1,...,n_{V}\right\} $;\medskip

$\left( \text{\textbf{non-BPS}}\right) _{\mathbf{2}}$ : $Z=0$, $D_{i}Z\neq 0$%
, for some $i\in \left\{ 1,...,n_{V}\right\} $.\medskip

Notice that the class $\left( \text{non-BPS}\right) _{\mathbf{2}}$,
corresponding to non-degenerate orbits of the class $\mathcal{O}%
_{non-BPS,Z=0}$, does not exist in $N=8$, $d=4$ $\frac{E_{7(7)}}{SU(8)}$%
-based SUGRA.

As it can be explicitly computed in the manageable yet interestingly rich
case of the $stu$ model \cite{to-appear}, the classes $\frac{1}{2}$-BPS and $%
\left( \text{non-BPS}\right) _{\mathbf{2}}$ of solutions, and
correspondingly the classes $\mathcal{O}_{\frac{1}{2}-BPS}$ and $\mathcal{O}%
_{non-BPS,Z=0}$ of orbits, have a strictly positive quartic $E_{7}$%
-invariant $I_{4}>0$. On the other hand, the class $\left( \text{non-BPS}%
\right) _{\mathbf{1}}$ of solutions, and thus the class $\mathcal{O}%
_{non-BPS,Z\neq 0}$ of orbits, have a strictly negative quartic $E_{7}$%
-invariant $I_{4}<0$.\medskip

At this point, we notice that both real (non-compact) forms \ of $E_{7}$ ,
namely $E_{7(7)}$ of Cremmer and Julia \cite{CJ} and $E_{7(-25)}$ of GST
\cite{GST1} contain the factorized group $SO^{\ast }(12)\otimes SU(2)$ as a
subgroup, and indeed both manifolds $\frac{E_{7(-25)}}{E_{6}\otimes U(1)}$ ($%
dim_{\mathbb{R}}=54$) and $\frac{E_{7(7)}}{SU(8)}$ ($dim_{\mathbb{R}}=70$)
contain as a submanifold the coset space $\frac{SO^{\ast }(12)}{U(6)}$,
which is the symmetric special K\"{a}hler manifold of the $N=2$, $d=4$
``magical'' MESGT defined by the Jordan algebra $J_{3}^{\mathbb{H}}$ \cite
{GST1}.\medskip

Such an observation reveals the \textit{dual} role of the manifold $\frac{%
SO^{\ast }(12)}{U(6)}$: it is at the same time the $\sigma $-model part of
an $N=6$ SUGRA (if one starts from $N=8$, i.e. from $E_{7(7)}$) and of an $%
N=2$ MESGT (if one starts from $N=2$, i.e. from $E_{7(-25)}$).\medskip

The truncation from $N=8$ starts from the following decomposition of the
real representation $R_{H}=\mathbf{70}$ of the $\mathcal{R}$-symmetry group $%
H=SU(8)$ in terms of representations of $SU(6)\otimes SU(2)$ \cite
{GST1,ADFFT}:
\begin{equation}
\mathbf{70}\longrightarrow \left( \mathbf{15},\mathbf{1}\right) +\left(
\overline{\mathbf{15}},\overline{\mathbf{1}}\right) +(\mathbf{20},\mathbf{2})%
\text{ }.  \label{decompo1}
\end{equation}
Under such a splitting the $70$ $N=8$ real scalars decompose in $15$ complex
scalars belonging to the submanifold $\frac{SO^{\ast }(12)}{U(6)}%
\varsubsetneq \frac{E_{7(7)}}{SU(8)}$ and in $20$ (half-)hypermultiplet
scalars belonging to the submanifold $\frac{E_{6(2)}}{SU(6)\otimes SU(2)}%
\varsubsetneq \frac{E_{7(7)}}{SU(8)}$. Consequently, the $\frac{SO^{\ast
}(12)}{U(6)}$-based model is then reached by disregarding all the $SU(2)$%
-non-singlet terms in r.h.s. of decomposition (\ref{decompo1}), namely $(%
\mathbf{20},\mathbf{2})$. Such a procedure eliminates two gravitinos from
the SUGRA multiplet, and thus an $N=6$, $d=4$ $\frac{SO^{\ast }(12)}{U(6)}$%
-based SUGRA is obtained.

On the other hand, the truncation from the exceptional $N=2$ MESGT yields
the following decomposition of the complex representation $R_{H_{0}}=\mathbf{%
27}$ of $H_{0}=E_{6}$ in terms of representations of its (non-maximal)
compact subgroup $SU(6)\otimes SU(2)$:
\begin{equation}
\mathbf{27}\longrightarrow \left( \mathbf{15},\mathbf{1}\right) +(\mathbf{6},%
\mathbf{2}).  \label{decompo2}
\end{equation}
Once again, the $\frac{SO^{\ast }(12)}{U(6)}$-based model is reached by
disregarding all the $SU(2)$-non-singlet terms in r.h.s. of decomposition (%
\ref{decompo2}), namely $(\mathbf{6},\mathbf{2})$. In this case, the
elimination of $SU(2)$-non-singlets does not touch the two gravitinos of the
SUGRA multiplet, and thus an $N=2$, $d=4$ , $\frac{SO^{\ast }(12)}{U(6)}$%
-based SUGRA is obtained, namely the ``magical'' MESGT defined by $J_{3}^{%
\mathbb{H}}$ \cite{GST1}.\medskip

Thus, the $N=2$ and $N=6$ $d=4$ SUGRAs with scalar manifold $\frac{SO^{\ast
}(12)}{U(6)}$ have \textit{indistinguishable} full bosonic sectors, and
therefore their charge orbits are the same and their attractor Eqs. have the
same solutions. Since we discussed above that an $N=2$, $d=4$ MESGT
generally yield three distinct classes of non-degenerate orbits, the same
must hold for the considered $N=6$ SUGRA.\medskip

This actually holds true, because the ``charge vector'' $\mathcal{C}$ of the
$N=6$, $d=4$ $\frac{SO^{\ast }(12)}{U(6)}$-based theory, although it is a
pure SUGRA, contains a vector which is a singlet under $SU(6)$, $U(6)$ being
the $\mathcal{R}$-symmetry \cite{ADF}. Thus, the whole $N=6$ ``charge
vector'' can be written as
\begin{equation}
\mathcal{C}=\left( X,\widehat{Z}_{AB}\right) ,  \label{C-N=6}
\end{equation}
where $X$ is the $SU(6)$-singlet ``charge vector'' and $\widehat{Z}_{AB}$ ($%
A=1,...,N=6$) is the complex $6\times 6$ antisymmetric central charge matrix.

Correspondingly, $N=6$, $d=4$ $\frac{SO^{\ast }(12)}{U(6)}$-based SUGRA has
three distinct classes of extremal black hole solutions with finite,
non-vanishing entropy, corresponding to the following structures of $%
\mathcal{C}$ \cite{ADF}:\medskip

$X=0$, $\widehat{Z}_{AB}\neq 0$ (precisely $\widehat{Z}_{12}\neq 0$, $%
\widehat{Z}_{34}=0$, $\widehat{Z}_{56}=0$ in the ``normal'' frame).

This solution is a $\frac{1}{6}$-BPS $N=6$ solution, and it has
\begin{equation}
I_{4,\frac{1}{6}-BPS}=\left| Z\right| _{\frac{1}{6}-BPS}^{4}>0.
\end{equation}
\smallskip

$X\neq 0$, $\widehat{Z}_{AB}=0$.

Such a case corresponds to an $N=6$ non-BPS solution (since $\widehat{Z}%
_{AB}=0$), but, due to the contribution of $X$, it has
\begin{equation}
I_{4,non-BPS,\widehat{Z}_{AB}=0}=\left| X\right| _{non-BPS,\widehat{Z}%
_{AB}=0}^{4}>0\medskip .
\end{equation}
\smallskip

$X\neq 0$, $\widehat{Z}_{AB}\neq 0$ (precisely $\left| X\right| =\left|
\widehat{Z}_{12}\right| =\left| \widehat{Z}_{34}\right| =\left| \widehat{Z}%
_{56}\right| $ in the ``normal'' frame).

Such a case corresponds to an $N=6$ non-BPS solution with
\begin{equation}
I_{4,non-BPS,\widehat{Z}_{AB}\neq 0}=-16\left| X\right| _{non-BPS,\widehat{Z}%
_{AB}\neq 0}^{4}<0\medskip .
\end{equation}
\medskip

At a generic point of the scalar manifold $\frac{SO^{\ast }(12)}{U(6)}$ the
overall symmetry of the moduli-dependent matrix $\widehat{Z}_{AB}$, which
can be put in the skew-diagonal form $\widehat{Z}_{AB,normal}$, is $(SU(2))^{%
\frac{N}{2}=3}$.\medskip\ Thus, by considering the aforementioned structures
of $\widehat{Z}_{AB}$ at regular solutions, one can determine the overall
symmetry and the coset expression of the corresponding classes of
non-degenerate orbits:\medskip

$X=0$\textbf{, }$\frac{1}{6}$\textbf{-BPS solution: }overall symmetry $%
SU(2)\otimes SU(4)$.\textbf{\smallskip\ }Consequently, the unique choice for
the coset expression of the class of non-degenerate $N=6$ $\frac{1}{6}$-BPS
orbits is ($SU(2)\otimes SU(4)\otimes U(1)=m.c.s.\left( SU(4,2)\right) $)
\begin{equation}
\mathcal{O}_{\frac{1}{6}-BPS}=\frac{SO^{\ast }(12)}{SU(4,2)}.
\end{equation}

$X\neq 0$\textbf{, non-BPS, }$\widehat{Z}_{AB}=0$\textbf{\
solution:\smallskip\ }overall symmetry $SU(6)$:
\begin{equation}
\mathcal{O}_{non-BPS,\widehat{Z}_{AB}=0}=\frac{SO^{\ast }(12)}{SU(6)}.
\end{equation}
Notice that such a class of $N=6$ solutions does not have an $N=8$ analogue;
it is peculiar to $N=6$ SUGRA, due to the particular form (\ref{C-N=6}) of
the ``charge vector'' $\mathcal{C}$.\medskip

$X\neq 0$\textbf{, non-BPS, }$\widehat{Z}_{AB}\neq 0$\textbf{\
solution:\smallskip } overall symmetry is $USp(6)$ and the corresponding
orbit reads ($USp(6)=m.c.s.\left( SU^{\ast }(6)\right) $)
\begin{equation}
\mathcal{O}_{non-BPS,\widehat{Z}_{AB}\neq 0}=\frac{SO^{\ast }(12)}{SU^{\ast
}(6)}.
\end{equation}
\smallskip \bigskip

It is now immediate to notice that the classes of non-degenerate orbits of
the considered $N=6$, $d=4$ case are the same of those of the $N=2$, $d=4$
``magical'' MESGT based on $\frac{SO^{\ast }(12)}{U(6)}$, \textit{but with
different BPS features}. The $N$-dependent BPS-interpretations of the
classes of non-degenerate orbits of the irreducible homogeneous symmetric
special K\"{a}hler manifold $\frac{SO^{\ast }(12)}{U(6)}$ are summarized in
Table 9.

\begin{table}[t]
\begin{center}
\begin{tabular}{|c||c|c|}
\hline
$
\begin{array}{c}
\\
\text{Orbit} \\
~
\end{array}
$ & $N=6$ & $N=2$ \\ \hline\hline
$
\begin{array}{c}
\\
\frac{SO^{\ast }(12)}{SU(6)} \\
~
\end{array}
$ & $\mathcal{O}_{non-BPS,\widehat{Z}_{AB}=0}:\left\{
\begin{array}{l}
X\neq 0, \\
\widehat{Z}_{AB}=0, \\
\forall \left( A,B\right) \in \left\{ 1,...,6\right\} ^{2}
\end{array}
\right. $ & $\mathcal{O}_{\frac{1}{2}-BPS}:\left\{
\begin{array}{l}
Z\neq 0, \\
D_{i}Z=0, \\
\forall i=1,...,n_{V}=15
\end{array}
\right. $ \\ \hline
$
\begin{array}{c}
\\
\frac{SO^{\ast }(12)}{SU(4,2)} \\
~
\end{array}
$ & $\mathcal{O}_{\frac{1}{6}-BPS}:\left\{
\begin{array}{c}
X=0 \\
\widehat{Z}_{AB}\neq 0
\end{array}
\right. $ & $\mathcal{O}_{non-BPS,Z=0}:\left\{
\begin{array}{c}
Z=0 \\
D_{i}Z\neq 0
\end{array}
\right. $ \\ \hline
$
\begin{array}{c}
\\
\frac{SO^{\ast }(12)}{SU^{\ast }(6)} \\
~
\end{array}
$ & $\mathcal{O}_{non-BPS,\widehat{Z}_{AB}\neq 0}$ $:\left\{
\begin{array}{c}
X\neq 0 \\
\widehat{Z}_{AB}\neq 0
\end{array}
\right. $ & $\mathcal{O}_{non-BPS,Z\neq 0}:\left\{
\begin{array}{c}
Z\neq 0 \\
D_{i}Z\neq 0
\end{array}
\right. $ \\ \hline
\end{tabular}
\end{center}
\caption{\textbf{$N$-dependent BPS-interpretations of the classes of
non-degenerate orbits of the homogeneous symmetric special K\"{a}hler
manifold $\frac{SO^{\ast }(12)}{U(6)}$.} $\widehat{{Z}}${$_{AB}\neq 0$ and $%
D_{i}Z\neq 0$ are generally understood for some values of $\left( A,B\right)
\in \left\{ 1,...,6\right\} ^{2}$ and of $i\in \left\{
1,...,n_{V}=15\right\} $, respectively.}}
\end{table}

As it can be seen from Table 9, the role of regular BPS orbits and non-BPS
orbits with (all) central charge(s) vanishing is \textit{flipped} under the
\textit{exchange} $N=2\longleftrightarrow N=6$. Such a kind of \textit{%
``cross-symmetry''} is easily understood when noticing that the $N=2$
central charge $Z$ corresponds to the $SU(6)$-singlet component $X$ of the $%
N=6$ ``charge vector'' $\mathcal{C}$, and that the $15$ complex $N=2$
``matter charges'' $D_{i}Z$ correspond to the $15$ independent complex
elements of the $6\times 6$ antisymmetric $N=6$ central charge matrix $%
\widehat{Z}_{AB}$.

Also, Table 9 immediately yields another consequence of the \textit{``}$%
N=2\longleftrightarrow N=6$\textit{\ cross-symmetry''}: while in all $N=2$, $%
d=4$ irreducible symmetric MESGTs the class of regular BPS critical points
is more symmetric than both classes of regular non-BPS critical points, a
different result holds for the $N=6$\textbf{, }$d=4$\textbf{\ }$\frac{%
SO^{\ast }(12)}{U(6)}$-based SUGRA: \textit{the most symmetric regular
solutions are the non-BPS ones with }$\widehat{Z}_{AB}=0$\textit{\ and }$%
X\neq 0$\textit{, and not the }$\frac{1}{6}$\textit{-BPS ones}.
\setcounter{equation}0
\def\theequation{4.3.\arabic{subsubsection}.\arabic{equation}}
\subsubsection{\label{Orb-Attr-O}Orbits and Attractors of $J_{3}^{\mathbb{O}%
} $}

Let us now apply the general analysis performed in Sect. \ref{N=2-Attractors}
to the case of the exceptional $N=2$, $d=4$ MESGT with scalar manifold $%
\frac{E_{7(-25)}}{E_{6}\otimes U(1)}$, defined by the exceptional Jordan
algebra $J_{3}^{\mathbb{O}}$. As already mentioned above, this is the MESGT
with the largest scalar manifold (apart from the two sequences $I$ and $II$%
), and it is the only ``magical'' MESGT which is not a consistent truncation
of $N=8$, $d=4$ $\frac{E_{7(7)}}{SU(8)}$-based SUGRA \cite{GST1}. In this
case
\begin{equation}
G=E_{7(-25)},H_{0}=E_{6}=\frac{m.c.s.\left( E_{7(-25)}\right) }{U(1)}%
=Ktr_0(J_3^{\mathbb{O}}).
\end{equation}
\medskip

$\frac{1}{2}$\textbf{-BPS solutions }$\mathcal{O}_{\frac{1}{2}-BPS,J_{3}^{%
\mathbb{O}}}=\frac{E_{7(-25)}}{E_{6}}\medskip $

They are given by
\begin{equation}
\begin{array}{l}
Z\neq 0,\text{ }D_{i}Z=0\Leftrightarrow D_{I}Z=0,\forall i,I=1,...,n_{V}=27;
\\
\\
V_{BH,\frac{1}{2}-BPS,J_{3}^{\mathbb{O}}}=\left| Z\right| _{\frac{1}{2}%
-BPS,J_{3}^{\mathbb{O}}}^{2},
\end{array}
\label{BBPS}
\end{equation}
and they are manifestly $E_{6}$-invariant.\medskip

\textbf{Non-BPS, }$Z\neq 0$\textbf{\ solutions }$\mathcal{O}_{non-BPS,Z\neq
0,J_{3}^{\mathbb{O}}}=\frac{E_{7(-25)}}{E_{6(-26)}}\medskip $

The non-compact stabilizer of the orbit is $\widehat{H}%
=E_{6(-26)}=Str_0^*(J_3^{\mathbb{O}})$. The ``flatted matter charges'' $%
D_{I}Z$ sit in the complex representation $R_{H_{0}}=\mathbf{27}$ of $%
H_{0}=E_{6}$. Under $\widehat{h}=F_{4}=m.c.s.(E_{6(-26)})$, $R_{H_{0}}$
decomposes as
\begin{equation}
\mathbf{27}\longrightarrow \mathbf{26}+\mathbf{1,}  \label{27}
\end{equation}
where the r.h.s. is made of the complex representations $\mathbf{1}$ and $%
\mathbf{26}$ of $F_{4}$. Such a decomposition yields the following splitting
of ``flatted matter charges'':
\begin{equation}
D_{I}Z\longrightarrow \left( D_{\mathbf{26}}Z,D_{\mathbf{1}}Z\right) ,
\end{equation}
where the subscripts ``$\mathbf{26}$'' and ``$\mathbf{1}$'' denote the
directions along the $\mathbf{26}$ and the $F_{4}$-singlet direction in $%
\frac{E_{7(-25)}}{E_{6}\otimes U(1)}$, respectively. By decomposing $\left(
R_{H_{0}}\right) ^{3}=\left( \mathbf{27}\right) ^{3}$ in terms of
representations of $F_{4}$, one therefore finds
\begin{equation}
\left( \mathbf{27}\right) ^{3}\longrightarrow \left( \mathbf{26}\right)
^{3}+\left( \mathbf{26}\right) ^{2}\mathbf{1}+\mathbf{1}^{3}\mathbf{%
\Leftrightarrow }C_{IJK}\longrightarrow \left\{ C_{\mathbf{26,26,26}},C_{%
\mathbf{26,26,1}},C_{\mathbf{1,1,1}}\right\} .
\end{equation}

Thus, the solution of $N=2$, $d=4$ non-BPS, $Z\neq 0$ extremal black hole
attractor eqs. in ``flat'' indices (\ref{AEs-non-BPS1-flat}) for the treated
case is obtained by putting
\begin{equation}
D_{\mathbf{26}}Z=0,D_{\mathbf{1}}Z\neq 0,
\end{equation}
constrained by
\begin{equation}
2\overline{Z}D_{\mathbf{1}}Z=-iC_{\mathbf{1,1,1}}\left( \overline{D_{\mathbf{%
1}}Z}\right) ^{2},  \label{constr1}
\end{equation}
where we recall that (see Eq. (\ref{ress1}))
\begin{equation}
\left| C_{\mathbf{1,1,1}}\right| _{non-BPS,Z\neq 0,J_{3}^{\mathbb{O}}}^{2}=%
\frac{4}{3}.  \label{res1}
\end{equation}

The resulting value of the black hole scalar potential at the considered
attractor point(s) is
\begin{eqnarray}
V_{BH,non-BPS,Z\neq 0,J_{3}^{\mathbb{O}}} &=&\left| Z\right| _{non-BPS,Z\neq
0,J_{3}^{\mathbb{O}}}^{2}+\left[ G^{i\overline{i}}D_{i}Z\overline{D}_{%
\overline{i}}\overline{Z}\right] _{non-BPS,Z\neq 0,J_{3}^{\mathbb{O}}}=
\notag \\
&&  \notag \\
&=&\left| Z\right| _{non-BPS,Z\neq 0,J_{3}^{\mathbb{O}}}^{2}+\left| D_{%
\mathbf{1}}Z\right| _{non-BPS,Z\neq 0,J_{3}^{\mathbb{O}}}^{2}=4\left|
Z\right| _{non-BPS,Z\neq 0,J_{3}^{\mathbb{O}}}^{2}.  \notag \\
&&  \label{V-non-BPS1-O}
\end{eqnarray}
\textit{The overall symmetry group at }$N=2$\textit{\ non-BPS, }$Z\neq 0$%
\textit{\ critical point(s) of }$J_{3}^{\mathbb{O}}$\textit{\ is }$\widehat{h%
}=F_{4}=m.c.s.(E_{6(-26)})$\textit{.\medskip }

\textbf{Non-BPS, }$Z=0$\textbf{\ solutions }$\mathcal{O}_{non-BPS,Z=0,J_{3}^{%
\mathbb{O}}}=\frac{E_{7(-25)}}{E_{6(-14)}} \medskip $

The non-compact stabilizer of the orbit is $\widetilde{H}=E_{6(-14)}=%
\Delta_0(J_3^{\mathbb{O}})$. Under $\widetilde{h}^{\prime }=SO(10)=\frac{%
m.c.s.(E_{6(-14)})}{U(1)}$, $R_{H_{0}}=\mathbf{27}$ decomposes as
\begin{equation}
\mathbf{27}\longrightarrow \mathbf{16}_{1}+\mathbf{10}_{-2}+\mathbf{1}_{4}%
\mathbf{,}  \label{decompo3}
\end{equation}
where $\mathcal{W}_{\widetilde{h}^{\prime }}=\mathbf{16}_{1}$, $\mathcal{Y}_{%
\widetilde{h}^{\prime }}=\mathbf{10}_{-2}$ and $\mathbf{1}_{4}$ are complex (%
$U(1)$-charged) representations of $SO(10)$, and here the numeric subscripts
denote the charges with respect to the explicit factor $U(1)$ in $\widetilde{%
h}=SO(10)\otimes U(1)=m.c.s.(E_{6(-14)})$.

The decomposition (\ref{decompo3}) yields the following splitting of
``flatted matter charges'':
\begin{equation}
D_{I}Z\longrightarrow \left( D_{\mathbf{16}}Z,D_{\mathbf{10}}Z,D_{\mathbf{1}%
}Z\right) .
\end{equation}
Thence, by decomposing the $E_{6}$-invariant representation $\left(
R_{H_{0}}\right) ^{3}=\left( \mathbf{27}\right) ^{3}$ in terms of
representations of $SO(10)$, one finds
\begin{equation}
\left( \mathbf{27}\right) ^{3}\longrightarrow \mathbf{16}_{1}\mathbf{16}_{1}%
\mathbf{10}_{-2}+\mathbf{10}_{-2}\mathbf{10}_{-2}\mathbf{1}_{4}\mathbf{%
\Leftrightarrow }C_{IJK}\longrightarrow \left\{ C_{\mathbf{16,16,10}},C_{%
\mathbf{10,10,1}}\right\} .  \label{27cube}
\end{equation}

Thus, the solution of $N=2$, $d=4$ non-BPS, $Z=0$ extremal black hole
attractor eqs. in ``flat'' indices (\ref{AEs-non-BPS2-flat}) for the treated
case is obtained by putting
\begin{equation}
D_{\mathbf{16}}Z=0=D_{\mathbf{10}}Z,D_{\mathbf{1}}Z\neq 0.  \label{sol-sol}
\end{equation}

The resulting value of the black hole scalar potential at the considered
attractor point(s) is
\begin{equation}
V_{BH,non-BPS,Z=0,J_{3}^{\mathbb{O}}}=\left| D_{\mathbf{1}}Z\right|
_{non-BPS,Z=0,J_{3}^{\mathbb{O}}}^{2}.  \label{V-non-BPS2-O}
\end{equation}
Attention should be paid to distinguish between the $D_{\mathbf{1}}Z$'s
appearing in Eqs. (\ref{V-non-BPS1-O}) and (\ref{V-non-BPS2-O}), since they
refer to different directions in $\frac{E_{7(-25)}}{E_{6}\otimes U(1)}$: in
Eq. (\ref{V-non-BPS1-O}) $D_{\mathbf{1}}Z$ denotes the ``flatted matter
charge'' along the complex $F_{4}$-singlet, whereas in (\ref{V-non-BPS2-O}) $%
D_{\mathbf{1}}Z$ stands for the ``flatted matter charge'' along the complex (%
$U(1)$-charged) $SO(10)$-singlet.

\textit{The overall symmetry group at }$N=2$\textit{\ non-BPS, }$Z=0$\textit{%
\ critical point(s) of }$J_{3}^{\mathbb{O}}$ \textit{is }$\widetilde{h}%
^{\prime }=SO(10)=\frac{\widetilde{h}}{U(1)}$\textit{, where }$\widetilde{h}%
=SO(10)\otimes U(1)=m.c.s.$\textit{\ }$\left( E_{6(-14)}\right) $\textit{.}
\setcounter{equation}0
\def\theequation{4.3.\arabic{subsubsection}.\arabic{equation}}
\subsubsection{\label{Orb-Attr-H}Orbits and Attractors of $J_{3}^{\mathbb{H}%
} $}

Let us now move to consider the case of the $N=2$, $d=4$ symmetric
``magical'' MESGT with scalar manifold $\frac{SO^{\ast }(12)}{U(6)}$,
related to $J_{3}^{\mathbb{H}}$.

This is the ``magical'' MESGT with the largest scalar manifold which can be
obtained as a consistent truncation of $N=8$, $d=4$ $\frac{E_{7(7)}}{SU(8)}$%
-based SUGRA. The other two $N=2$, $d=4$ symmetric ``magical'' MESGTs with a
smaller scalar manifold (namely those based on $\frac{SU(3,3)}{SU(3)\otimes
SU(3)\otimes U(1)}$ and $\frac{Sp(6,\mathbb{R})}{U(3)}$: see Tables 2 and 3)
can actually be obtained as consistent truncations of such a $\frac{SO^{\ast
}(12)}{U(6)}$-based ``magical'' MESGT.

Moreover, as pointed out in Subsubsect. \ref{dual-2-6}, it plays a dual
role, since $\frac{SO^{\ast }(12)}{U(6)}$ is also the scalar manifold of $%
N=6 $, $d=4$ SUGRA. It is now worth analyzing the same classes of
non-degenerate charge orbits as in Subsubsect. \ref{dual-2-6} with an $N=2$,
$d=4$ approach based on special K\"{a}hler geometry.

In this case
\begin{equation}
G=SO^{\ast }(12),H_{0}=SU(6)=\frac{m.c.s.\left( SO^{\ast }(12)\right) }{U(1)}%
=Ktr_0(J_3^{\mathbb{H}}).
\end{equation}
\medskip

$\frac{1}{2}$\textbf{-BPS solutions }$\mathcal{O}_{\frac{1}{2}-BPS,J_{3}^{%
\mathbb{H}}}=\frac{SO^{\ast }(12)}{SU(6)}\medskip $

They are given by
\begin{equation}
\begin{array}{l}
Z\neq 0,\text{ }D_{i}Z=0\Leftrightarrow D_{I}Z=0,\forall i,I=1,...,n_{V}=15;
\\
\\
V_{BH,\frac{1}{2}-BPS,J_{3}^{\mathbb{H}}}=\left| Z\right| _{\frac{1}{2}%
-BPS,J_{3}^{\mathbb{H}}}^{2},
\end{array}
\end{equation}
and they are manifestly $SU(6)$-invariant.\medskip

\textbf{Non-BPS, }$Z\neq 0$\textbf{\ solutions }$\mathcal{O}_{non-BPS,Z\neq
0,J_{3}^{\mathbb{H}}}=\frac{SO^{\ast }(12)}{SU^{\ast }(6)}\medskip $

The non-compact stabilizer of the orbit is $\widehat{H}=SU^{\ast
}(6)=Str_0^*(J_3^{\mathbb{H}})$. The ``flatted matter charges'' $D_{I}Z$ sit
in the complex representation $R_{H_{0}}=\mathbf{15}$ of $H_{0}$; under $%
\widehat{h}=USp(6)=m.c.s.(SU^{\ast }(6)),$ $R_{H_{0}}$ decomposes as
\begin{equation}
\mathbf{15}\longrightarrow \mathbf{14}+\mathbf{1,}
\end{equation}
where the r.h.s. is made of the complex representations $\mathbf{1}$ and $%
\mathbf{14}$ of $USp(6)$. Such a decomposition yields the following
splitting of ``flatted matter charges'':
\begin{equation}
D_{I}Z\longrightarrow \left( D_{\mathbf{14}}Z,D_{\mathbf{1}}Z\right) .
\end{equation}
By decomposing $\left( R_{H_{0}}\right) ^{3}=\left( \mathbf{15}\right) ^{3}$
in terms of representations of $USp(6)$, one therefore finds
\begin{equation}
\left( \mathbf{15}\right) ^{3}\longrightarrow \left( \mathbf{14}\right)
^{3}+\left( \mathbf{14}\right) ^{2}\mathbf{1}+\mathbf{1}^{3}\mathbf{%
\Leftrightarrow }C_{IJK}\longrightarrow \left\{ C_{\mathbf{15,15,15}},C_{%
\mathbf{15,15,1}},C_{\mathbf{1,1,1}}\right\} .
\end{equation}

Thus, the solution of $N=2$, $d=4$ non-BPS, $Z\neq 0$ extremal black hole
attractor eqs. in ``flat'' indices (\ref{AEs-non-BPS1-flat}) for the treated
case is obtained by putting
\begin{equation}
D_{\mathbf{14}}Z=0,D_{\mathbf{1}}Z\neq 0,
\end{equation}
constrained by Eq. (\ref{constr1}), once again with $\left| C_{\mathbf{1,1,1}%
}\right| _{non-BPS,Z\neq 0,J_{3}^{\mathbb{H}}}^{2}=\frac{4}{3}$.

The resulting value of the black hole scalar potential at the considered
attractor point(s) is once again
\begin{equation}
\begin{array}{l}
V_{BH,non-BPS,Z\neq 0,J_{3}^{\mathbb{H}}}= \\
\\
=\left| Z\right| _{non-BPS,Z\neq 0,J_{3}^{\mathbb{H}}}^{2}+\left| D_{\mathbf{%
1}}Z\right| _{non-BPS,Z\neq 0,J_{3}^{\mathbb{H}}}^{2}=4\left| Z\right|
_{non-BPS,Z\neq 0,J_{3}^{\mathbb{H}}}^{2}.
\end{array}
\label{V-non-BPS1-H}
\end{equation}
\textit{The overall symmetry group at }$N=2$\textit{\ non-BPS, }$Z\neq 0$%
\textit{\ critical point(s) of }$J_{3}^{\mathbb{H}}$\textit{\ is }$\widehat{h%
}=USp(6)=m.c.s.(SU^{\ast }(6)=Str_0^*(J_3^{\mathbb{H}})$\textit{.\medskip }

\textbf{Non-BPS, }$Z=0$\textbf{\ solutions }$\mathcal{O}_{non-BPS,Z=0,J_{3}^{%
\mathbb{H}}}=\frac{SO^{\ast }(12)}{SU(4,2)}\medskip $

The non-compact stabilizer of the orbit is $\widetilde{H}=SU(4,2)$. Under $%
\widetilde{h}^{\prime }=SU(4)\otimes SU(2)=\frac{m.c.s.(SU(4,2))}{U(1)}$, $%
R_{H_{0}}=\mathbf{15}$ decomposes as
\begin{equation}
\mathbf{15}\longrightarrow \left( \mathbf{4,2}\right) _{-1}+\left( \mathbf{6}%
,\mathbf{1}\right) _{2}+\left( \mathbf{1,1}\right) _{-4}\mathbf{,}
\label{decompo4}
\end{equation}
where $\mathcal{W}_{\widetilde{h}^{\prime }}=\left( \mathbf{4,2}\right)
_{-1} $, $\mathcal{Y}_{\widetilde{h}^{\prime }}=\left( \mathbf{6},\mathbf{1}%
\right) _{2}$ and $\left( \mathbf{1,1}\right) _{-4}$ are complex ($U(1)$%
-charged) representations of $SU(4)\otimes SU(2)$, and as above the numeric
subscripts denote the charges with respect to the explicit factor $U(1)$ in $%
\widetilde{h}=SU(4)\otimes SU(2)\otimes U(1)=m.c.s.(SU(4,2))$.

The decomposition (\ref{decompo4}) yields the following splitting of
``flatted matter charges'':
\begin{equation}
D_{I}Z\longrightarrow \left( D_{\left( \mathbf{4,2}\right) }Z,D_{\left(
\mathbf{6},\mathbf{1}\right) }Z,D_{\left( \mathbf{1,1}\right) }Z\right) .
\end{equation}
Thence, by decomposing the $SU(6)$-invariant representation $\left(
R_{H_{0}}\right) ^{3}=\left( \mathbf{15}\right) ^{3}$ in terms of
representations of $SU(4)\otimes SU(2)$, one finds
\begin{gather}
\left( \mathbf{15}\right) ^{3}\longrightarrow \left( \mathbf{4,2}\right)
_{-1}\left( \mathbf{4,2}\right) _{-1}\left( \mathbf{6},\mathbf{1}\right)
_{2}+\left( \mathbf{6},\mathbf{1}\right) _{2}\left( \mathbf{6},\mathbf{1}%
\right) _{2}\left( \mathbf{1,1}\right) _{-4} \\
\Updownarrow  \notag \\
C_{IJK}\longrightarrow \left\{ C_{\mathbf{\left( \mathbf{4,2}\right) ,}%
\left( \mathbf{4,2}\right) ,\left( \mathbf{6},\mathbf{1}\right) },C_{\left(
\mathbf{6},\mathbf{1}\right) \mathbf{,\left( \mathbf{6},\mathbf{1}\right) ,}%
\left( \mathbf{1,1}\right) }\right\} .
\end{gather}

Thus, the solution of $N=2$, $d=4$ non-BPS, $Z=0$ extremal black hole
attractor eqs. in ``flat'' indices (\ref{AEs-non-BPS2-flat}) for the treated
case is obtained by putting
\begin{equation}
D_{\left( \mathbf{4,2}\right) }Z=0=D_{\left( \mathbf{6},\mathbf{1}\right) }Z,%
\text{ \ }D_{\left( \mathbf{1,1}\right) }Z\neq 0.
\end{equation}

The resulting value of the black hole scalar potential at the considered
attractor point(s) is
\begin{equation}
V_{BH,non-BPS,Z=0,J_{3}^{\mathbb{H}}}=\left| D_{\left( \mathbf{1,1}\right)
}Z\right| _{non-BPS,Z=0,J_{3}^{\mathbb{H}}}^{2}.  \label{V-non-BPS2-H}
\end{equation}
Once again, attention should be paid to distinguish between the $D_{\mathbf{1%
}}Z$'s appearing in Eqs. (\ref{V-non-BPS1-H}) and (\ref{V-non-BPS2-H}),
since they refer to different directions in $\frac{SO^{\ast }(12)}{U(6)}$:
in Eq. (\ref{V-non-BPS1-H}) $D_{\mathbf{1}}Z$ denotes the ``flatted matter
charge'' along the complex $USp(6)$-singlet, whereas in (\ref{V-non-BPS2-H})
$D_{\left( \mathbf{1,1}\right) }Z$ stands for the ``flatted matter charge''
along the complex ($U(1)$-charged) $\left( SU(4)\otimes SU(2)\right) $%
-singlet.

\textit{The overall symmetry group at }$N=2$\textit{\ non-BPS, }$Z=0$\textit{%
\ critical point(s) of }$J_{3}^{\mathbb{H}}$ \textit{is }$\widetilde{h}%
^{\prime }=SU(4)\otimes SU(2)=\frac{\widetilde{h}}{U(1)}$\textit{, where }$%
\widetilde{h}=SU(4)\otimes SU(2)\otimes U(1)=m.c.s.$\textit{\ }$\left(
SU(4,2)\right) $\textit{.\medskip }

A completely analogous analysis may be performed for the cases $V$ ($J_{3}^{%
\mathbb{C}}$) and $VI$ ($J_{3}^{\mathbb{R}}$) of Tables 2 and 3. We leave
such an analysis as an instructive exercise for the reader.
\setcounter{equation}0
\def\theequation{\arabic{section}.\arabic{equation}}
\section{\label{N=2-Spectra}The Mass Spectra at Critical Points}

The black hole scalar potential $V_{BH}$ gives different masses to the
different BPS-phases of the considered symmetric $N=2$, $d=4$ MESGTs. The
fundamental object to be considered in such a framework is the
moduli-dependent $2n_{V}\times 2n_{V}$ Hessian matrix of $V_{BH}$, which in
complex basis reads\footnote{%
The reported formul{\ae } for $\mathcal{M}_{ij}$ and $\mathcal{N}_{i%
\overline{j}}$ hold for any special K\"{a}hler manifold. In the symmetric
case formula (\ref{M}) gets simplified using Eq. (\ref{CERN1}).} \cite{BFM}
\begin{eqnarray}
&&
\begin{array}{c}
\mathbf{H}^{V_{BH}}\equiv \left(
\begin{array}{ccc}
D_{i}D_{j}V_{BH} &  & D_{i}\overline{D}_{\overline{j}}V_{BH} \\
&  &  \\
D_{j}\overline{D}_{\overline{i}}V_{BH} &  & \overline{D}_{\overline{i}}%
\overline{D}_{\overline{j}}V_{BH}
\end{array}
\right) \equiv \left(
\begin{array}{ccc}
\mathcal{M}_{ij} &  & \mathcal{N}_{i\overline{j}} \\
&  &  \\
\overline{\mathcal{N}}_{j\overline{i}} &  & \overline{\mathcal{M}}_{%
\overline{i}\overline{j}}
\end{array}
\right) ;
\end{array}
\\
&&  \notag \\
&&  \notag \\
&&
\begin{array}{l}
\mathcal{M}_{ij}\equiv D_{i}D_{j}V_{BH}=D_{j}D_{i}V_{BH}= \\
\\
=4i\overline{Z}C_{ijk}G^{k\overline{k}}\overline{D}_{\overline{k}}\overline{Z%
}+iG^{k\overline{k}}G^{l\overline{l}}D_{j}C_{ikl}\overline{D}_{\overline{k}}%
\overline{Z}\overline{D}_{\overline{l}}\overline{Z};
\end{array}
\label{M} \\
&&  \notag \\
&&  \notag \\
&&
\begin{array}{l}
\mathcal{N}_{i\overline{j}}\equiv D_{i}\overline{D}_{\overline{j}}V_{BH}=%
\overline{D}_{\overline{j}}D_{i}V_{BH}= \\
\\
=2\left[ G_{i\overline{j}}\left| Z\right| ^{2}+D_{i}Z\overline{D}_{\overline{%
j}}\overline{Z}+G^{l\overline{n}}G^{k\overline{k}}G^{m\overline{m}}C_{ikl}%
\overline{C}_{\overline{j}\overline{m}\overline{n}}\overline{D}_{\overline{k}%
}\overline{Z}D_{m}Z\right] ;
\end{array}
\label{N} \\
&&  \notag \\
&&
\begin{array}{l}
\mathcal{M}^{T}=\mathcal{M},\mathcal{N}^{\dag }=\mathcal{N}.
\end{array}
\end{eqnarray}
By analyzing $\mathbf{H}^{V_{BH}}$ at regular critical points, it is
possible to formulate general conclusions about the mass spectrum of the
corresponding extremal black hole solutions with finite, non-vanishing
entropy, i.e. about the mass spectrum along the related classes of
non-degenerate charge orbits of the symplectic real representation $R_{V}$
of the $d=4$ duality group $G$.\smallskip

Let us start by remarking that, due to its very definition (\ref{VBH-def}),
the $N=2$ black hole scalar potential $V_{BH}$ is positive for any (not
necessarily strictly) positive definite metric $G_{i\overline{i}}$ of the
scalar manifold. Consequently, the \textit{stable} critical points (i.e. the
\textit{attractors} in a strict sense) will necessarily be minima of such a
potential. As already pointed out above and as done also in \cite{BFM,AoB},
the geometry of the scalar manifold is usually assumed to be \textit{%
regular, }i.e. endowed with a metric tensor $G_{i\overline{j}}$ being
strictly positive definite everywhere.\bigskip

$\frac{1}{2}$\textbf{-BPS critical points\medskip }

It is now well known that regular special K\"{a}hler geometry implies that
\textit{all} $N=2$ $\frac{1}{2}$-BPS critical points of \textit{all} $N=2$, $%
d=4$ MESGTs are stable, and therefore they are attractors in a strict sense.
Indeed, the Hessian matrix $\mathbf{H}_{\frac{1}{2}-BPS}^{V_{BH}}$ evaluated
at such points is strictly positive definite \cite{FGK}:
\begin{equation}
\begin{array}{l}
\mathcal{M}_{ij,\frac{1}{2}-BPS}=0, \\
\\
\mathcal{N}_{i\overline{j},\frac{1}{2}-BPS}=2\left. G_{i\overline{j}}\right|
_{\frac{1}{2}-BPS}\left| Z\right| _{\frac{1}{2}-BPS}^{2}>0,
\end{array}
\label{BPS-crit}
\end{equation}
where the notation ``$>0$'' is clearly understood as strict positive
definiteness of the quadratic form related to the square matrix being
considered. Notice that the Hermiticity and strict positive definiteness of $%
\mathbf{H}_{\frac{1}{2}-BPS}^{V_{BH}}$ are respectively due to the
Hermiticity and strict positive definiteness of the K\"{a}hler metric $G_{i%
\overline{j}}$ of the scalar manifold.

By switching from the non-flat $i$-coordinates to the ``flat'' local $I$%
-coordinates by using the (inverse) Vielbein $e_{I}^{i}$ of the scalar
manifold, Eqs. (\ref{BPS-crit}) can be rewritten as
\begin{equation}
\begin{array}{l}
\mathcal{M}_{IJ,\frac{1}{2}-BPS}=0, \\
\\
\mathcal{N}_{I\overline{J},\frac{1}{2}-BPS}=2\delta _{I\overline{J}}\left|
Z\right| _{\frac{1}{2}-BPS}^{2}>0.
\end{array}
\end{equation}
Thus, one obtains that in \textit{all} $N=2$, $d=4$ MESGTs the $\frac{1}{2}$%
-BPS mass spectrum in ``flat'' coordinates is \textit{monochromatic}, i.e.
that all ``particles'' (i.e. the ``modes'' related to the degrees of freedom
described by the ``flat'' local $I$-coordinates) acquire \textit{the same}
mass at $\frac{1}{2}$-BPS critical points of $V_{BH}$.\medskip

\textbf{Non-BPS, }$Z\neq 0$\textbf{\ critical points}

In this case the result of Tripathy and Trivedi \cite{TT} should apply,
namely the Hessian matrix $\mathbf{H}_{non-BPS,Z\neq 0}^{V_{BH}}$ should
have $n_{V}+1$ strictly positive and $n_{V}-1$ vanishing real eigenvalues.

By recalling the analysis performed in Sect. \ref{N=2-Attractors}, it is
thence clear that such massive and massless non-BPS, $Z\neq 0$ ``modes'' fit
distinct real representations of $\widehat{h}=m.c.s.\left( \widehat{H}%
\right) $, where $\widehat{H}$ is the non-compact stabilizer of the class $%
\mathcal{O}_{non-BPS,Z\neq 0}=\frac{G}{\widehat{H}}$ of non-BPS, $Z\neq 0$
non-degenerate charge orbits.

This is perfectly consistent with the decomposition (\ref{KR}) of the
complex representation $R_{H_{0}}$ ($dim_{\mathbb{R}}R_{H_{0}}=2n_{V}$) of $%
H_{0}$ in terms of representations of $\widehat{h}$:
\begin{equation}
R_{H_{0}}\longrightarrow \left( R_{\widehat{h}}+\mathbf{1}\right) _{\mathbb{C%
}}=\left( R_{\widehat{h}}+\mathbf{1}+R_{\widehat{h}}+\mathbf{1}\right) _{%
\mathbb{R}}\mathbf{,}\text{ \ }dim_{\mathbb{R}}\left( R_{\widehat{h}}\right)
_{\mathbb{R}}=n_{V}-1.  \label{KR-KR}
\end{equation}
As yielded by the treatment given in Subsubsect. \ref
{N=2-Attractors-non-BPS-1}, the notation ``$\left( R_{\widehat{h}}+\mathbf{1}%
\right) _{\mathbb{C}}=\left( R_{\widehat{h}}+\mathbf{1}+R_{\widehat{h}}+%
\mathbf{1}\right) _{\mathbb{R}}$'' denotes nothing but the
decomplexification of $\left( R_{\widehat{h}}+\mathbf{1}\right) _{\mathbb{C}%
} $, which is actually composed by a pair of real irreducible
representations $\left( R_{\widehat{h}}+\mathbf{1}\right) _{\mathbb{R}}$ of $%
\widehat{h}$.

Therefore, Tripathy and Trivedi's result can be understood in terms of real
representations of the m.c.s. of the non-compact stabilizer of $\mathcal{O}%
_{non-BPS,Z\neq 0}$: the $n_{V}-1$ massless non-BPS, $Z\neq 0$ ``modes'' are
in one of the two real $R_{\widehat{h}}$'s of $\widehat{h}$ in the r.h.s. of
Eq. (\ref{KR-KR}), say the first one, whereas the $n_{V}+1$ massive non-BPS,
$Z\neq 0$ ``modes'' are split in the remaining real $R_{\widehat{h}}$ of $%
\widehat{h}$ and in the two real $\widehat{h}$-singlets. The resulting
interpretation of the decomposition (\ref{KR-KR}) is
\begin{equation}
R_{H_{0}}\longrightarrow \left(
\begin{array}{c}
\left( R_{\widehat{h}}\right) _{\mathbb{R}} \\
\\
n_{V}-1~~\text{\textit{massless} }
\end{array}
\right) +\left(
\begin{array}{c}
\left( R_{\widehat{h}}\right) _{\mathbb{R}}+\mathbf{1}_{\mathbb{R}}+\mathbf{1%
}_{\mathbb{R}} \\
\\
n_{V}+1~~\text{\textit{massive}}
\end{array}
\right) .
\end{equation}
It is interesting to notice once again that there is no $U(1)$ symmetry
relating the two real $R_{\widehat{h}}$'s (and thus potentially relating the
splitting of ``modes'' along $\mathcal{O}_{non-BPS,Z\neq 0}$), since in
\textit{all} symmetric $N=2$, $d=4$ MESGTs $\widehat{h}$ \textit{never}
contains an explicit factor $U(1)$ (as instead it \textit{always} happens
for $\widetilde{h}$!); this can be related to the fact that the non-compact
stabilizer is $\widehat{H}=Str_{0}^{\ast }(\mathcal{J})$ whose $\widehat{h}$
is the m.c.s..\medskip

For further elucidation, let us consider the explicit example of the $J_{3}^{%
\mathbb{O}}$-related symmetric $N=2$, $d=4$ ``magical'' MESGT with scalar
manifold $\frac{E_{7(-25)}}{E_{6}\otimes U(1)}$ ($n_{V}=27$), treated in
Subsubsect. \ref{Orb-Attr-O}. In the sense of aforementioned
decomplexification, the decomposition (\ref{27}) of the complex
representation $R_{H_{0}}=\mathbf{27}$ under $\widehat{h}=F_{4}$ can also be
written as

\begin{equation}
\mathbf{27}\longrightarrow \mathbf{26}+\mathbf{1}=\mathbf{26}_{\mathbb{R}}+%
\mathbf{1}_{\mathbb{R}}\mathbf{+\mathbf{26}}_{\mathbb{R}}\mathbf{+\mathbf{1}}%
_{\mathbb{R}}\mathbf{.}  \label{27-27}
\end{equation}
It is then clear that the mass spectrum of $\mathcal{O}_{non-BPS,Z\neq
0,J_{3}^{\mathbb{O}}}$ splits under $F_{4}$ as follows: the $26$ massless
non-BPS, $Z\neq 0$ ``modes'' are in one of the two $\mathbf{26}_{\mathbb{R}}$%
's of $F_{4}$ in the r.h.s. of Eq. (\ref{27-27}), say the first one, whereas
the $28$ massive non-BPS, $Z\neq 0$ ``modes'' are split in the remaining $%
\mathbf{26}_{\mathbb{R}}$ of $F_{4}$ and in the two real $F_{4}$-singlets.
The resulting interpretation of the decomposition (\ref{27-27}) is
\begin{equation}
\mathbf{27}\longrightarrow \left(
\begin{array}{c}
\mathbf{26}_{\mathbb{R}} \\
\\
26~~\text{\textit{massless} }
\end{array}
\right) +\left(
\begin{array}{c}
\mathbf{\mathbf{26}}_{\mathbb{R}}+\mathbf{1}_{\mathbb{R}}+\mathbf{1}_{%
\mathbb{R}} \\
\\
28~~\text{\textit{massive}}
\end{array}
\right) .
\end{equation}

\textbf{Non-BPS, }$Z=0$\textbf{\ critical points\smallskip }

For the class $\mathcal{O}_{non-BPS,Z=0}$ of non-degenerate non-BPS, $Z=0$
orbits the situation changes, and Tripathy and Trivedi's result no longer
holds true, due to the local vanishing of the $N=2$ central charge.\medskip

In order to illustrate the consequences of $Z=0$ along $\mathcal{O}%
_{non-BPS,Z=0}$ on the related mass spectrum, let us consider again the
symmetric $N=2$, $d=4$ ``magical'' MESGT with scalar manifold $\frac{%
E_{7(-25)}}{E_{6}\otimes U(1)}$.

In such a case, $\mathcal{O}_{non-BPS,Z=0,J_{3}^{\mathbb{O}}}=\frac{%
E_{7(-25)}}{E_{6(-14)}}$, and therefore the non-compact stabilizer is $%
\widetilde{H}=E_{6(-14)}$. The complex $\mathbf{27}$ and the $E_{6}$%
-invariant $\left( \mathbf{27}\right) ^{3}$ of $E_{6}$ decompose under $%
\widetilde{h}^{\prime }=SO(10)=\frac{m.c.s.(E_{6(-14)})}{U(1)}$ as given by
Eqs. (\ref{decompo3}) and (\ref{27cube}), respectively:
\begin{equation}
\mathbf{27}\longrightarrow \mathbf{16}_{1}+\mathbf{10}_{-2}+\mathbf{1}_{4}%
\mathbf{;}  \label{27-non-BPS-Z=0}
\end{equation}
\begin{equation}
\left( \mathbf{27}\right) ^{3}\longrightarrow \mathbf{16}_{1}\mathbf{16}_{1}%
\mathbf{10}_{-2}+\mathbf{10}_{-2}\mathbf{10}_{-2}\mathbf{1}_{4}\mathbf{.}
\label{27cube2}
\end{equation}
Consequently, the rank-3 $E_{6}$-invariant tensor coupling $C_{IJK}$
decomposes in its non-vanishing components as follows:
\begin{equation}
C_{IJK}\longrightarrow \left\{ C_{\mathbf{16,16,10}},C_{\mathbf{10,10,1}%
}\right\} .  \label{decomp-C}
\end{equation}

In Subsubsect. \ref{Orb-Attr-O} the solution of $N=2$, $d=4$ non-BPS, $Z=0$
extremal black hole attractor eqs. in ``flat'' indices (\ref
{AEs-non-BPS2-flat}) for the case at hand was found to be given by Eq. (\ref
{sol-sol}):
\begin{equation}
D_{\mathbf{16}}Z=0=D_{\mathbf{10}}Z,~D_{\mathbf{1}}Z\neq 0.  \label{sol-sol2}
\end{equation}

By using Eqs. (\ref{decomp-C}) and (\ref{sol-sol2}), the block matrix
components of the $54\times 54$ critical Hessian $\mathbf{H}%
_{non-BPS,Z=0,J_{3}^{\mathbb{O}}}^{V_{BH}}$ in ``flat'' coordinates can be
computed to be:
\begin{equation}
\begin{array}{l}
\begin{array}{l}
\mathcal{M}_{IJ,non-BPS,Z=0,J_{3}^{\mathbb{O}}}=0;
\end{array}
\\
\\
\\
\begin{array}{l}
\mathcal{N}_{I\overline{J},non-BPS,Z=0,J_{3}^{\mathbb{O}}}= \\
\\
=2\left[ D_{I}Z\overline{D}_{\overline{J}}\overline{Z}+\delta ^{L\overline{N}%
}\delta ^{K\overline{K}}\delta ^{M\overline{M}}C_{IKL}\overline{C}_{%
\overline{J}\overline{M}\overline{N}}\overline{D}_{\overline{K}}\overline{Z}%
D_{M}Z\right] _{_{non-BPS,Z=0,J_{3}^{\mathbb{O}}}}.
\end{array}
\end{array}
\end{equation}
\newline
The only non-vanishing elements of the $27\times 27$ critical (diagonal and
real) matrix $\mathcal{N}_{I\overline{J},non-BPS,Z=0,J_{3}^{\mathbb{O}}}$
are the following ones:
\begin{equation}
\begin{array}{l}
\mathcal{N}_{\mathbf{1}\overline{\mathbf{1}},non-BPS,Z=0,J_{3}^{\mathbb{O}%
}}=2\left| D_{\mathbf{1}}Z\right| _{non-BPS,Z=0,J_{3}^{\mathbb{O}}}^{2}; \\
\\
\mathcal{N}_{\mathbf{10}\overline{\mathbf{10}},non-BPS,Z=0,J_{3}^{\mathbb{O}%
}}=2\left| C_{\mathbf{10,10,1}}\right| _{non-BPS,Z=0,J_{3}^{\mathbb{O}%
}}^{2}\left| D_{\mathbf{1}}Z\right| _{non-BPS,Z=0,J_{3}^{\mathbb{O}}}^{2},
\end{array}
\end{equation}
and therefore one gets (the subscripts denote the matrix dimension)
\begin{eqnarray}
&&
\begin{array}{l}
\mathbf{H}_{non-BPS,Z=0,J_{3}^{\mathbb{O}}}^{V_{BH}}=2\left| D_{\mathbf{1}%
}Z\right| _{non-BPS,Z=0,J_{3}^{\mathbb{O}}}^{2}\cdot \\
\\
\cdot \left(
\begin{array}{ccc}
0_{27\times 27} &  &
\begin{array}{ccc}
1_{1\times 1} &  &  \\
& \mathbf{C\cdot }1_{10\times 10} &  \\
&  & 0_{16\times 16}
\end{array}
\\
&  &  \\
\begin{array}{ccc}
1_{1\times 1} &  &  \\
& \mathbf{C\cdot }1_{10\times 10} &  \\
&  & 0_{16\times 16}
\end{array}
&  & 0_{27\times 27}
\end{array}
\right) ,
\end{array}
\notag \\
&&
\end{eqnarray}
where $\mathbf{C}\equiv \left| C_{\mathbf{10,10,1}}\right|
_{non-BPS,Z=0,J_{3}^{\mathbb{O}}}^{2}$ $\in \mathbb{R}_{0}^{+}$. The real
form \cite{BFM} of the non-BPS $Z=0$ critical Hessian finally reads
\begin{eqnarray}
&&
\begin{array}{l}
\mathbf{H}_{non-BPS,Z=0,J_{3}^{\mathbb{O}},real}^{V_{BH}}=\left| D_{\mathbf{1%
}}Z\right| _{non-BPS,Z=0,J_{3}^{\mathbb{O}}}^{2}\cdot \\
\\
\cdot \left(
\begin{array}{ccc}
\begin{array}{ccc}
1_{1\times 1} &  &  \\
& \mathbf{C\cdot }1_{10\times 10} &  \\
&  & 0_{16\times 16}
\end{array}
&  & 0_{27\times 27} \\
&  &  \\
0_{27\times 27} &  &
\begin{array}{ccc}
1_{1\times 1} &  &  \\
& \mathbf{C\cdot }1_{10\times 10} &  \\
&  & 0_{16\times 16}
\end{array}
\end{array}
\right) ;
\end{array}
\notag \\
&&
\end{eqnarray}
thus, there are $32$ vanishing and $22$ strictly positive real eigenvalues.

It is then clear that the mass spectrum of $\mathcal{O}_{non-BPS,Z=0,J_{3}^{%
\mathbb{O}}}$ splits under $SO(10)=\frac{m.c.s.(E_{6(-14)})}{U(1)}$ as
follows: there are $32$ massless non-BPS, $Z=0$ ``modes'' fitting the $32$
real degrees of freedom corresponding to the complex $\mathbf{16}_{1}$ of $%
SO(10)$ in the r.h.s. of decomposition (\ref{27-non-BPS-Z=0}), and $20+2=22$
massive non-BPS, $Z=0$ ``modes'' fitting the remaining real degrees of
freedom corresponding to the complex $\mathbf{10}_{-2}$ and $\mathbf{1}_{4}$
of $SO(10)$. The resulting interpretation of the decomposition (\ref
{27-non-BPS-Z=0}) is
\begin{equation}
\mathbf{27}\longrightarrow \left(
\begin{array}{c}
\mathbf{16}_{1} \\
\\
32~~\text{\textit{massless}}
\end{array}
\right) +\left(
\begin{array}{c}
\mathbf{10}_{-2}+\mathbf{1}_{4} \\
\\
20+2~~\text{\textit{massive}}
\end{array}
\right) \mathbf{.}  \label{decccomp}
\end{equation}

By looking at the form of the non-BPS, $Z=0$ solution (\ref{sol-sol2}) in
``flat'' coordinates, it is then easy to realize that the $\mathbf{16}_{1}$
of $SO(10)$ remains massless at regular non-BPS, $Z=0$ critical points
\textit{because it does not couple to the }$SO(10)$\textit{-singlet} in the
representation decomposition (\ref{27cube2}).\medskip

The results obtained above for the $\frac{E_{7(-25)}}{E_{6}\otimes U(1)}$%
-based ``magical'' $N=2$, $d=4$ symmetric MESGT can be immediately extended
to all the other three ``magical'' $N=2$, $d=4$ symmetric MESGTs
(corresponding to cases $IV$, $V$ and $VI$ of Tables 2 and 3) as follows.

In all ``magical'' $N=2$, $d=4$ MESGTs the complex representation $R_{H_{0}}$
of $H_{0}$ decomposes under $\widetilde{h}^{\prime }=\frac{m.c.s.\left(
\widetilde{H}\right) }{U(1)}$ in the following way (see Eq. (\ref{KR-KR-KR}%
)):
\begin{equation}
R_{H_{0}}\longrightarrow \mathcal{W}_{\widetilde{h}^{\prime }}+\mathcal{Y}_{%
\widetilde{h}^{\prime }}+\mathbf{1}_{\mathbb{C}}\mathbf{,}  \label{deccomp}
\end{equation}
where in the r.h.s. the complex $\widetilde{h}^{\prime }$-singlet and the
complex non-singlet representations $\mathcal{W}_{\widetilde{h}^{\prime }}$
and $\mathcal{Y}_{\widetilde{h}^{\prime }}$ of $\widetilde{h}^{\prime }$
appear. Correspondingly, the decomposition of the $H_{0}$-invariant
representation $\left( R_{H_{0}}\right) ^{3}$ in terms of representations of
$\widetilde{h}^{\prime }$ reads (see Eq. (\ref{decomp-non-BPS-Z=0}))
\begin{equation}
\left( R_{H_{0}}\right) ^{3}\longrightarrow \left( \mathcal{W}_{\widetilde{h}%
^{\prime }}\right) ^{2}\mathcal{Y}_{\widetilde{h}^{\prime }}+\left( \mathcal{%
Y}_{\widetilde{h}^{\prime }}\right) ^{2}\mathbf{1}_{\mathbb{C}}\mathbf{.}
\label{decomp-decomp}
\end{equation}
Let us now recall that $dim_{\mathbb{R}}R_{H_{0}}=2n_{V}$ and $dim_{\mathbb{R%
}}\mathbf{1}_{\mathbb{C}}=2$, and let us define
\begin{equation}
\left.
\begin{array}{r}
dim_{\mathbb{R}}\mathcal{W}_{\widetilde{h}^{\prime }}\equiv \mathbf{W}_{%
\widetilde{h}^{\prime }}; \\
\\
dim_{\mathbb{R}}\mathcal{Y}_{\widetilde{h}^{\prime }}\equiv \mathbf{Y}_{%
\widetilde{h}^{\prime }};
\end{array}
\right\} :\mathbf{W}_{\widetilde{h}^{\prime }}+\mathbf{Y}_{\widetilde{h}%
^{\prime }}+2=2n_{V}.  \label{one}
\end{equation}

Thus, it can generally be stated that the mass spectrum along $\mathcal{O}%
_{non-BPS,Z=0}$ of all ``magical'' $N=2$, $d=4$ symmetric MESGTs splits
under $\widetilde{h}^{\prime }=\frac{m.c.s.(\widetilde{H})}{U(1)}$ as
follows:

$\mathbf{-}$\textbf{\ }the mass ``modes'' fitting the $\mathbf{W}_{%
\widetilde{h}^{\prime }}$ real degrees of freedom corresponding to the
complex ($U(1)$-charged) non-$\widetilde{h}^{\prime }$-singlet
representation $\mathcal{W}_{\widetilde{h}^{\prime }}$ (which does \textit{%
not }couple to the complex $\widetilde{h}^{\prime }$-singlet in the $H_{0}$%
-invariant decomposition (\ref{decomp-decomp})) remain \textit{massless};

$\mathbf{-}$\textbf{\ }the mass ``modes'' fitting the $\mathbf{Y}_{%
\widetilde{h}^{\prime }}+2$ real degrees of freedom corresponding to the
complex ($U(1)$-charged) non-$\widetilde{h}^{\prime }$-singlet
representation $\mathcal{Y}_{\widetilde{h}^{\prime }}$ and to the ($U(1)$%
-charged) $\widetilde{h}^{\prime }$-singlet $\mathbf{1}_{\mathbb{C}}$
\textit{all} become \textit{massive}.

The resulting interpretation of the decomposition (\ref{deccomp}) is
\begin{equation}
R_{H_{0}}\longrightarrow \left(
\begin{array}{c}
\mathcal{W}_{\widetilde{h}^{\prime }} \\
\\
\mathbf{W}_{\widetilde{h}^{\prime }}~~\text{\textit{massless}}
\end{array}
\right) +\left(
\begin{array}{c}
\mathcal{Y}_{\widetilde{h}^{\prime }}+\mathbf{1}_{\mathbb{C}} \\
\\
\mathbf{Y}_{\widetilde{h}^{\prime }}+2~~\text{\textit{massive}}
\end{array}
\right) \mathbf{.}  \label{two}
\end{equation}

The interpretations (\ref{decccomp}) and (\ref{two}) show that, even though
the complex representations $\mathcal{W}_{\widetilde{h}^{\prime }}$, $%
\mathcal{Y}_{\widetilde{h}^{\prime }}$ and $\mathbf{1}_{\mathbb{C}}$ of $%
\widetilde{h}^{\prime }$ are charged with respect to the explicit factor $%
U(1)$ \textit{always} appearing in $\widetilde{h}$, this fact does \textit{%
not} affect in any way the splitting of the non-BPS, $Z=0$ mass
``modes''.\medskip

It is worth pointing out that in the $N=2$, $6$, $d=4$ SUGRAs based on $%
\frac{SO^{\ast }(12)}{U(6)}$ (see Subsubsects. \ref{dual-2-6} and \ref
{Orb-Attr-H}) the splitting of the critical non-BPS, $Z=0$ mass spectrum
obtained by the above general analysis is in agreement with the results of
the $N=6$ analysis of the $\frac{1}{6}$-BPS solutions, yielding $7$ massive
vector multiplets and $4$ massless hypermultiplets \cite{ADF}.\bigskip\

The critical mass spectra of the irreducible sequence $\frac{SU(1,1+n)}{%
U(1)\otimes SU(1+n)}$ and of the reducible sequence $\frac{SU(1,1)}{U(1)}%
\otimes \frac{SO(2,2+n)}{SO(2)\otimes SO(2+n)}$ are treated in Appendices I
and II, respectively.\bigskip

Generally, the Hessian $\mathbf{H}^{V_{BH}}$ at regular $N=2$, non-BPS
critical points of $V_{BH}$ exhibits the following features: it does not
have \textit{``repelling''} directions (i.e. strictly \textit{negative} real
eigenvalues), it has a certain number of \textit{``attracting''} directions
(related to strictly \textit{positive} real eigenvalues), but it is also
characterized by some \textit{``flat''} directions, i.e. by some \textit{%
vanishing} eigenvalues, corresponding to massless non-BPS ``modes''.

In order to establish whether the considered $N=2$, non-BPS citical points
of $V_{BH}$ are actually \textit{attractors} in a strict sense, i.e. whether
they actually are \textit{stable minima} of $V_{BH}$ in the scalar manifold,
one has to proceed further with K\"{a}hler-covariant differentiation of $%
V_{BH}$, dealing (at least) with third and higher-order derivatives.
Tripathy and Trivedi \cite{TT} presented a charge-dependent statement about
the stability of $N=2$, non-BPS critical points in the case $Z\neq 0$ for
cubic prepotentials, which would be interesting to check and interpret for $%
N=2$, $d=4$ symmetric MESGTs in the \textit{``representation decomposition
approach''} exploited above.

We leave the detailed analysis of the issue of stability of both classes of
regular non-BPS critical points ($Z\neq 0$ and $Z=0$) of $V_{BH}$ in $N=2$, $%
d=4$ (symmetric) MESGTs for future work.\textbf{\bigskip }
\setcounter{equation}0
\def\theequation{\arabic{section}.\arabic{equation}}
\section{\label{Conclusion}Conclusion}

In the present work we have classified the regular solutions of the $N=2$, $%
d=4$ extremal black hole attractor equations for homogeneous symmetric
special K\"{a}hler geometries. For rank-3 symmetric manifolds $\frac{G}{%
H=H_{0}\otimes U(1)}$ of MESGTs defined by Jordan algebras of degree three
such solutions exist in three distinct classes, one $\frac{1}{2}$-BPS and
the other two non-BPS, one of which corresponds to vanishing central charge $%
Z=0$.

We have also shown that these three classes of solutions are in one to one
correspondence with the non-degenerate charge orbits of the actions of the
duality groups $G$ on the corresponding charge spaces that were classified
in Sect. 3.

The non-BPS, $Z=0$ class of regular solutions has no analogue in $d=5$,
where a similar classification has been recently given \cite{FG2}.

For the considered rank-3 symmetric manifolds the tensor $C_{ijk}$ of
special K\"{a}hler geometry is covariantly constant, and in ``flat''
coordinates it is proportional to the numeric symmetric tensor $d_{IJK}$ of
the real special geometry of $N=2$, $d=5$ MESGTs \cite{GST1,GST2,GST3,GST4}.

The rank-1 family of symmetric manifolds (case $I$ of Table 3) has $%
C_{ijk}=0 $. It has only two classes of regular solutions to the attractor
equations: one $\frac{1}{2}$-BPS and one non-BPS with $Z=0$.

For all the rank-3 symmetric spaces the classical black hole entropy is
given by the Bekenstein-Hawking entropy-area formula \cite{BH1}
\begin{equation}
S_{BH}=\frac{A_{H}}{4}=\pi \sqrt{\left| I_{4}\right| },
\end{equation}
where
\begin{equation}
\left. V_{BH}\right| _{\partial V_{BH}=0}=\sqrt{\left| I_{4}\right| }
\end{equation}
and $I_{4}$ is the quartic invariant of the charge vector in the considered
non-degenerate charge orbit. For the $\frac{1}{2}$-BPS and non-BPS $Z=0$
classes $I_{4}>0$, while the non-BPS $Z\neq 0$ class has $I_{4}<0$.

For the family of rank-1 homogeneous symmetric manifolds the invariant is
instead quadratic; it is positive for $\frac{1}{2}$-BPS orbits and negative
for the non-BPS ones (see Appendix I).

We also investigated the (splittings of the) mass spectra of the theory
along the three species of classes of non-degenerate charge orbits, finding
an agreement with a result recently obtained by Tripathy and Trivedi \cite
{TT}, holding true for generic cubic holomorphic prepotentials and for
non-BPS $Z\neq 0$ critical points of $V_{BH}$.\smallskip

In order to proceed further, it would be interesting to extend the analysis
performed in the present work to non-symmetric special K\"{a}hler spaces
with cubic (holomorphic) prepotentials. A proper subclass of such manifolds
is given by the homogeneous non-symmetric spaces described in \cite{dWVVP}.
We should stress that all special K\"{a}hler cubic geometries have an
uplifting to five dimensions.

In \cite{GZ} it was shown that the $N=2$, $d=5$ ``magical'' MESGTs defined
by $J_{3}^{\mathbb{C}},J_{3}^{\mathbb{H}}$ and $J_{3}^{\mathbb{O}}$ are
simply the ``lowest'' members of three infinite families of unified $N=2$, $%
d=5$ MESGTs defined by Lorentzian Jordan algebras of degree $>3$. The scalar
manifolds of such theories are not homogeneous except for the ``lowest''
members. It would be interesting to extend the analysis of \cite{FG2} and of
the present work to these theories in five dimensions and to their
descendants in four dimensions, respectively.

Another direction for further investigations could be the analysis of
non-cubic (holomorphic) prepotentials, corresponding to more general special
K\"{a}hler geometries, such as the ones of the moduli spaces of Calabi-Yau
compactifications.

Finally, another possible extension of the present work might deal with
higher $N(=3,4)$ SUGRAs coupled to matter multiplets, which are known to
have BPS attractor solutions \cite{FKS,FK1,FK2}.\bigskip

\textbf{Acknowledgements}

S. F. would like to thank L. Andrianopoli and R. D'Auria for discussions. A.
M. would like to thank the Department of Physics, Theory Unit Group at CERN
for its kind hospitality during the completion of the present paper.

The work of S.B. has been supported in part by the European Community Human
Potential Program under contract MRTN-CT-2004-005104 ``Constituents,
fundamental forces and symmetries of the universe''.

The work of S.F.~has been supported in part by the European Community Human
Potential Program under contract MRTN-CT-2004-005104 ``Constituents,
fundamental forces and symmetries of the universe'', in association with
INFN Frascati National Laboratories and by D.O.E.~grant DE-FG03-91ER40662,
Task C.

The work of M.G was supported in part by the National Science Foundation
under grant number PHY-0245337 and PHY-0555605. Any opinions, findings and
conclusions or recommendations expressed in this material are those of the
authors and do not necessarily reflect the views of the National Science
Foundation.

The work of A.M. has been supported by a Junior Grant of the ``Enrico
Fermi'' Center, Rome, in association with INFN Frascati National
Laboratories. \appendix
\setcounter{equation}0
\def\theequation{AI.\arabic{equation}}
\section{\label{App1}Appendix I\newline
~The sequence $\frac{SU(1,1+n)}{U(1)\otimes SU(1+n)}$}

The $N=2$, $d=4$ symmetric MESGTs with scalar manifolds belonging to the
infinite sequence of irreducible rank-1 homogeneous symmetric special
K\"{a}hler manifolds $\frac{SU(1,1+n)}{U(1)\otimes SU(1+n)}$ feature another
fundamental property: they \textit{all} have a quadratic prepotential, and
thus globally
\begin{equation}
C_{ijk}=0.~  \label{TWO}
\end{equation}

Consequently, in such a case the attractor eqs. (\ref{AEs1}) acquire the
trivial form
\begin{equation}
\overline{Z}D_{i}Z=0,\forall i=1,...,n_{V},
\end{equation}
and therefore they admit only two classes of solutions:

\textbf{1. }$\frac{1}{2}$\textbf{-BPS solutions}:
\begin{equation}
\begin{array}{l}
Z\neq 0,\text{ \ }D_{i}Z=0,\forall i=1,...,n_{V}; \\
\\
V_{BH,\frac{1}{2}-BPS,I}=\left| Z\right| _{\frac{1}{2}-BPS,I}^{2}.
\end{array}
\label{THREE}
\end{equation}

\textbf{2. non-BPS, }$Z=0$\textbf{\ solutions}:
\begin{equation}
\begin{array}{l}
Z=0,\text{ \ }D_{i}Z\neq 0,\text{ for some }i\in \left\{ 1,...,n_{V}\right\}
; \\
\\
V_{BH,non-BPS,Z=0,I}=\left[ G^{i\overline{i}}D_{i}Z\overline{D}_{\overline{i}%
}\overline{Z}\right] _{non-BPS,Z=0,I}.
\end{array}
\label{FOUR}
\end{equation}

No non-BPS, $Z\neq 0$ solutions exist at all.

Recall that in this case $n_{V}=n+1$. Since the little group of the $SU(n+1)$%
-vector $D_{i}Z$ is $SU(n)$, it is then clear that the non-BPS, $Z=0$
solutions (\ref{FOUR}) are $SU(n)$-invariant. On the other hand, since in $%
\frac{1}{2}$-BPS solutions (\ref{THREE}) the vector $D_{i}Z$ vanishes, the
symmetry of the solution gets enhanced to $SU(n+1)$. Consequently, the two
classes of non-degenerate charge orbits of the case at hand respectively
reads (see Table 3):
\begin{equation}
\begin{array}{l}
\mathcal{O}_{\frac{1}{2}-BPS,I}=\frac{SU(1,n+1)}{SU(n+1)}; \\
\\
\mathcal{O}_{non-BPS,Z=0,I}=\frac{SU(1,n+1)}{SU(1,n)}~.
\end{array}
\end{equation}
Such two species of orbits are classified by a quadratic invariant, which in
``flat'' coordinates reads
\begin{equation}
\mathbf{I}_{2}=\left| Z\right| ^{2}-\sum_{I=1}^{n_{V}}\left| D_{I}Z\right|
^{2}.  \label{I2}
\end{equation}
One thus gets that
\begin{equation}
\begin{array}{l}
V_{BH,\frac{1}{2}-BPS,I}=\left| Z\right| _{\frac{1}{2}-BPS,I}^{2}=\mathbf{I}%
_{2,\frac{1}{2}-BPS}>0; \\
\\
V_{BH,non-BPS,Z=0,I}=\sum_{I=1}^{n_{V}}\left| D_{I}Z\right| _{\frac{1}{2}%
-BPS,I}^{2}=-\mathbf{I}_{2,non-BPS,Z=0}>0~.
\end{array}
\end{equation}
\medskip\

For what concerns the critical mass spectrum, $\frac{1}{2}$-BPS case is well
known to be always stable for all $N=2$, $d=4$ MESGTs, and it has been
treated in Sect. \ref{N=2-Spectra}. Considering the non-BPS, $Z=0$ critical
mass spectrum and using Eq. (\ref{TWO}), Eqs. (\ref{M}) and (\ref{N})
immediately yield the following elements of the $\left( 2n+2\right) \times
\left( 2n+2\right) $ critical Hessian $\mathbf{H}_{non-BPS,Z=0,I}^{V_{BH}}$:
\begin{equation}
\begin{array}{l}
\mathcal{M}_{IJ,non-BPS,Z=0,I}=0; \\
\\
\mathcal{N}_{I\overline{J}}=2\left[ D_{I}Z\overline{D}_{\overline{J}}%
\overline{Z}\right] _{non-BPS,Z=0,I}.
\end{array}
\end{equation}
Since the $\left( n+1\right) \times \left( n+1\right) $ Hermitian matrix $%
\left[ D_{I}Z\overline{D}_{\overline{J}}\overline{Z}\right] _{non-BPS,Z=0,I}$
has rank $1$, one immediately gets that the $\left( 2n+2\right) \times
\left( 2n+2\right) $ Hermitian matrix $\mathbf{H}_{non-BPS,Z=0,I}^{V_{BH}}$
has rank $2$: it has $2$ strictly positive and $2n$ vanishing real
eigenvalues. This is the general splitting pattern of the mass spectrum
along $\mathcal{O}_{non-BPS,Z=0}$ in the irreducible $N=2$, $d=4$ symmetric $%
\frac{SU(1,1+n)}{U(1)\otimes SU(1+n)}$-based MESGTs. \setcounter{equation}0
\def\theequation{AII.\arabic{equation}}
\section{\label{App2}Appendix II\newline
~The sequence $\frac{SU(1,1)}{U(1)}\otimes \frac{SO(2,2+n)}{SO(2)\otimes
SO(2+n)}$}

The case $II$ of Tables 2 and 3 is given by the sequence of $N=2$, $d=4$
symmetric MESGTs based on rank-3 reducible homogeneous symmetric special
K\"{a}hler manifolds $\frac{SU(1,1)}{U(1)}\otimes \frac{SO(2,2+n)}{%
SO(2)\otimes SO(2+n)}$ ($n\in \mathbb{N}\cup \left\{ 0\right\} $) with real
dimension $2(n+3)$. The main difference with respect to the four ``magical''
$N=2$, $d=4$ symmetric MESGTs (cases $III$-$VI$) and the sequence $\frac{%
SU(1,1+n)}{U(1)\otimes SU(1+n)}$ (case $I$) treated above is the fact that
for case $II$ the scalar manifold is \textit{reducible}, being the direct
product of two distinct manifolds.

In this case $G=SU(1,1)\otimes SO(2,2+n)$ and $H_{0}=SO(2)\otimes SO(2+n)$.
However, the holomorphic prepotential $F$ has at most an $SO(n+1)$ manifest
compact symmetry. For instance, in a suitable basis of special coordinates, $%
F$ may be written as follows:
\begin{equation}
F(t)=t^{1}\left[ \left( t^{2}\right) ^{2}-\left( t^{3}\right)
^{2}-\sum_{i=4}^{n+3}\left( t^{i}\right) ^{2}\right] ,  \label{F}
\end{equation}
which can be identified with the norm form of the underlying generic family
of non-simple Jordan algebras (see Sect. 3).

Beside the manifest overall symmetry $SO(1,n+1)$ (which is clearly related
to its 5-dimensional origin, since $SO(1,n+1)=\frac{\widehat{H}=G_{5}}{%
SO(1,1)}$), it should be remarked the full factorization of the holomorphic
cubic function (\ref{F}) in linear and quadratic components, due to the
reducibility of the scalar manifold, which follows from the fact that the
underlying Jordan algebras are not simple.\medskip

A particular, noteworthy case is given by the $n=0$ element $II_{0}$ which
is the $N=2$, $d=4$ symmetric MESGT with $n_{V}=3$-moduli usually called $%
stu $ model \cite{BKRSW}. Its scalar manifold and prepotential are given by
Eq. (\ref{stu}).\medskip

At this point, it is worth recalling that there are actually two other
typologies of $N=2$, $d=4$ symmetric MESGTs, corresponding to the following
cosets and prepotentials (in suitable systems of special coordinates):
\begin{eqnarray}
&&
\begin{array}{l}
VII:\frac{SU(1,1)}{U(1)},r=1,F(t)=\frac{1}{3}t^{3};
\end{array}
\\
&&  \notag \\
&&
\begin{array}{l}
VIII:\frac{SU(1,1)}{U(1)}\otimes \frac{SO(2,1)}{SO(2)}=\left( \frac{SU(1,1)}{%
U(1)}\right) ^{2},r=2,F(s,t)=st^{2}.
\end{array}
\end{eqnarray}
Notice that, even though they share the same coset structure, the cases $%
I_{0}$ and $VII$ correspond to different prepotential functions: while $%
I_{0} $ is the $n=0$, 1-modulus element of the sequence $I$ with quadratic
prepotential, the 1-modulus case $VII$ can actually be obtained from the $%
stu $ model $II_{0}$ by putting $s=t=u$ (and rescaling everything by $\sqrt[3%
]{3} $). On the other hand, the 2-moduli case $VIII$ can also be obtained
from the $stu$ model $II_{0}$, e.g. by putting $t=u$, or also by putting $%
n=-1$ in sequence $II$.

Summarizing, the cases $VII$ and $VIII$ can both be obtained as consistent
truncations of the $stu$ model, simply by identifying some or all moduli,
and consequently destroying the related \textit{triality symmetry}.

While the $n_{V}=2$-moduli case $VIII$ has the usual three classes of $N=2$
critical solutions, the $n_{V}=1$-moduli case $VII$ has only $\frac{1}{2}$%
-BPS and non-BPS, $Z\neq 0$ solutions, as it can be easily seen by
considering the structure of the attractor eqs. (\ref{AEs1}) for such cases.
Models $II_{0}$ ($stu$), $VII$ and $VIII$ are important also because in such
highly symmetric cases with a few moduli one can actually manage to
analytically solve the attractor eqs. (\ref{AEs1}) for the purely
charge-dependent critical values of the moduli \cite{to-appear}.\medskip

In order to solve the $N=2$ attractor eqs. (\ref{AEs1}) for the $\frac{%
SU(1,1)}{U(1)}\otimes \frac{SO(2,2+n)}{SO(2)\otimes SO(2+n)}$-based $N=2$, $%
d=4$ symmetric MESGTs, it is once again crucially convenient to switch to
``flat'' $I$-coordinates, in which the $N=2$ attractor eqs. are given by
Eqs. (\ref{AEs2}):
\begin{equation}
\begin{array}{l}
\partial _{I}V_{BH}=0\Leftrightarrow 2\overline{Z}D_{I}Z=-iC_{IJK}\delta ^{J%
\overline{J}}\delta ^{K\overline{K}}\overline{D}_{\overline{J}}\overline{Z}%
\overline{D}_{\overline{K}}\overline{Z}=0, \\
\\
\forall I=0,1,...,n_{V}-1=n+2.
\end{array}
\label{AESS}
\end{equation}
In such a ``flat'' $I$-coordinate system, even though it cannot be written
as the third partial derivative tensor of the prepotential (since no $F$
exists at all: see Footnote 6), the rank-3 symmetric tensor $C_{IJK}$
becomes very simple, since its unique non-vanishing components are simply
determined by the norm forms of the underlying reducible family of Jordan
algebras of degree three:
\begin{equation}
C_{IJK}=C_{0\widehat{J}\widehat{K}}=C_{0}\delta _{\widehat{J}\widehat{K}},%
\text{ \ }C_{0}\in \mathbb{C}_{0},
\end{equation}
where the ``hatted'' ``flat'' indices' range is $\left\{ 1,...,n+2\right\} $
and $\delta _{\widehat{J}\widehat{K}}$ is the $SO(n+2)$-invariant Euclidean
metric. Consequently, Eqs. (\ref{AESS}) split as follows:
\begin{equation}
\left\{
\begin{array}{l}
2\overline{Z}D_{0}Z=-iC_{0}\delta ^{\overline{\widehat{J}}\overline{\widehat{%
K}}}\overline{D}_{\overline{\widehat{J}}}\overline{Z}\overline{D}_{\overline{%
\widehat{K}}}\overline{Z}=-iC_{0}\sum_{\widehat{J}=1}^{n+2}\left( \overline{D%
}_{\overline{\widehat{J}}}\overline{Z}\right) ^{2}=0; \\
\\
2\overline{Z}D_{\widehat{I}}Z=-iC_{0}\overline{D}_{\overline{0}}\overline{Z}%
\overline{D}_{\overline{\widehat{I}}}\overline{Z}=0.
\end{array}
\right.  \label{AESSS}
\end{equation}
\smallskip

Thus, it is immediate to obtain the three classes of regular solutions of $%
N=2$ extremal black hole attractor eqs. (\ref{AESSS}) in ``flat'' $I$%
-coordinates for the $\frac{SU(1,1)}{U(1)}\otimes \frac{SO(2,2+n)}{%
SO(2)\otimes SO(2+n)}$-based $N=2$, $d=4$ symmetric MESGTs:\smallskip
\medskip

$\frac{1}{2}$\textbf{-BPS class}:
\begin{equation}
\begin{array}{l}
Z\neq 0,\text{ \ }D_{0}Z=0=D_{\widehat{I}}Z,\text{ \ }\forall \widehat{I}%
=1,...,n+2; \\
\\
V_{BH,\frac{1}{2}-BPS,II}=\left| Z\right| _{\frac{1}{2}-BPS,II}^{2}.
\end{array}
\label{II-BPS}
\end{equation}
\smallskip \smallskip

\textbf{Non-BPS, }$Z\neq 0$\textbf{\ class} (recall Eq. ($\sim $) in
Footnote 15):
\begin{equation}
\begin{array}{l}
Z\neq 0,\text{ \ }D_{0}Z\neq 0,\text{ \ }D_{1}Z\neq 0,~~D_{\breve{I}}Z=0,%
\text{ }\breve{I}\in \left\{ 2,...,n+2\right\} ; \\
\\
\\
V_{BH,non-BPS,Z\neq 0,II}=\left| Z\right| _{non-BPS,Z\neq
0,II}^{2}+\sum_{I=0}^{n_{V}-1=n+2}\left| D_{I}Z\right| _{non-BPS,Z\neq
0,II}^{2}= \\
\\
=\left| Z\right| _{non-BPS,Z\neq 0,II}^{2}+\left| D_{0}Z\right|
_{non-BPS,Z\neq 0,II}^{2}+\left| D_{1}Z\right| _{non-BPS,Z\neq 0,II}^{2}= \\
\\
=4\left| Z\right| _{non-BPS,Z\neq 0,II}^{2}.
\end{array}
\label{II-non-BPS-Z<>0}
\end{equation}
\medskip

\textbf{Non-BPS, }$Z=0$\textbf{\ classes}.

In this case, there are two distinct sets of independent regular non-BPS, $%
Z=0$ solutions to $N=2$ extremal black hole attractor eqs. (\ref{AESSS});
they read:
\begin{equation}
\left( \text{non-BPS}\right) _{1}:\left\{
\begin{array}{l}
Z=0,\text{ \ }D_{0}Z=0,\text{ \ }D_{1}Z=\pm iD_{2}Z,~D_{\widehat{I}}Z=0,%
\text{ \ }\forall \widehat{I}=3,...,n+2; \\
\\
V_{BH,non-BPS,Z=0,II,1}=2\left| D_{1}Z\right| _{non-BPS,Z=0,II,1}^{2}.
\end{array}
\right.  \label{Gunayy}
\end{equation}
\smallskip
\begin{equation}
\left( \text{non-BPS}\right) _{2}:\left\{
\begin{array}{l}
Z=0,\text{ \ }D_{0}Z\neq 0,\text{ \ }D_{\widehat{I}}Z=0,\text{ \ }\forall
\widehat{I}=1,...,n+2; \\
\\
V_{BH,non-BPS,Z=0,II,2}=\left| D_{0}Z\right| _{non-BPS,Z=0,II,2}^{2}.
\end{array}
\right.  \label{ours}
\end{equation}
\medskip

Let us analyze the overall symmetry of the solutions (\ref{II-BPS})-(\ref
{ours}).

For the $\frac{1}{2}$-BPS case $D_{0}Z=0=D_{\widehat{I}}Z$, thus the compact
symmetry is $SO(2)\otimes SO(2+n)$. This is also the case of the type $2$ of
non-BPS $Z=0$\ class (see Eq. (\ref{ours}). For the type $1$ of non-BPS $Z=0$%
\ class $\left( D_{1}Z\right) ^{2}+\left( D_{2}Z\right) ^{2}=0$ and $%
D_{0}Z=0 $, therefore the compact symmetry is $SO(2)\otimes SO(2)\otimes
SO(n)$. Finally, for the non-BPS $Z\neq 0$\ class\ $D_{0}Z\neq 0$ and $%
D_{1}Z\neq 0$, thence the compact symmetry is $SO(1+n)$.\smallskip

The cases $\frac{1}{2}$-BPS, non-BPS $Z\neq 0$ and non-BPS $Z=0$ type $1$
correspond to the three classes of non-degenerate orbits of $N=2$, $d=4$
symmetric $\frac{SU(1,1)}{U(1)}\otimes \frac{SO(2,2+n)}{SO(2)\otimes SO(2+n)}
$-based MESGTs given in Table 3.\smallskip

The case non-BPS $Z=0$ type $2$ corresponds instead to the class of orbits $%
\frac{SU(1,1)\otimes SO(2,2+n)}{SO(2)\otimes SO(2+n)}$, isomorphic to (%
\textit{but physically distinct from}) the $\frac{1}{2}$-BPS class. Indeed,
even though in the case at hand $\mathcal{O}_{non-BPS,Z=0}$ and $\mathcal{O}%
_{\frac{1}{2}-BPS}$ have the same formal coset expression, they actually
correspond to the tips of the two separated branches of the disconnected
manifold $\frac{SU(1,1)\otimes SO(2,2+n)}{SO(2)\otimes SO(2+n)}$, classified
by the sign of the quantity $\widetilde{\mathbf{I}}_{2}\equiv \left|
Z\right| ^{2}-\left| D_{0}Z\right| ^{2}$:
\begin{equation}
\begin{array}{l}
\widetilde{\mathbf{I}}_{2,\frac{1}{2}-BPS,II}=\left| Z\right| _{\frac{1}{2}%
-BPS,II}^{2}>0; \\
\\
\widetilde{\mathbf{I}}_{2,non-BPS,Z=0,II,2}=-\left| D_{0}Z\right|
_{non-BPS,Z=0,II,2}^{2}<0.
\end{array}
\label{CERN3}
\end{equation}
For such classes of critical points the quartic invariant reads \cite{FK2}
\begin{equation}
I_{4,II}=\left( \widetilde{\mathbf{I}}_{2,II}\right) ^{2}>0,
\end{equation}
the two cases corresponding to $\widetilde{\mathbf{I}}_{2,II}\gtrless 0$
(see Eq. (\ref{CERN3})).

This can also be seen in the $n=0$ element $II_{0}$ of the sequence $II$
being treated, i.e. in the $stu$ model, where explicit calculations are
feasible \cite{to-appear}.

Indeed, by setting $n=0$ in the third row of Table 3, one obtains the three
classes of non-degenerate orbits of the $stu$ model $II_{0}$ (the two $Z=0$
orbits coincide in this case):

$\frac{1}{2}$\textbf{-BPS class (}$H_{0}=\left( U(1)\right) ^{2}$):
\begin{equation}
\mathcal{O}_{\frac{1}{2}-BPS,stu}=\frac{G}{H_{0}}=\frac{\left(
SU(1,1)\right) ^{3}}{\left( U(1)\right) ^{2}};
\end{equation}

\textbf{Non-BPS, }$Z=0$\textbf{\ class (}$\widetilde{H}=H_{0}=\left(
U(1)\right) ^{2}$):
\begin{equation}
\mathcal{O}_{non-BPS,Z=0,stu}=\frac{G}{H_{0}}=\frac{\left( SU(1,1)\right)
^{3}}{\left( U(1)\right) ^{2}};
\end{equation}

\textbf{Non-BPS, }$Z\neq 0$\textbf{\ class (}$\widehat{H}=\left(
SO(1,1)\right) ^{2}$):
\begin{equation}
\mathcal{O}_{non-BPS,Z\neq 0,stu}=\frac{G}{\widehat{H}}=\frac{\left(
SU(1,1)\right) ^{3}}{\left( SO(1,1)\right) ^{2}}.
\end{equation}

The orbits $\mathcal{O}_{\frac{1}{2}-BPS,stu}$ and $\mathcal{O}%
_{non-BPS,Z=0,stu}$, despite having the same coset expression, correspond to
different values of the quantity $\left| Z\right| ^{2}-\left| D_{s}Z\right|
^{2}$. Indeed, it can be explicitly computed \cite{to-appear} that:

$\frac{1}{2}$\textbf{-BPS orbit }$\mathcal{O}_{\frac{1}{2}-BPS,stu}$:
\begin{equation}
\left| Z\right| _{\frac{1}{2}-BPS,stu}^{2}-\left| D_{s}Z\right| _{\frac{1}{2}%
-BPS,stu}^{2}=\left| Z\right| _{\frac{1}{2}-BPS,stu}^{2}>0;
\end{equation}

\textbf{Non-BPS, }$Z=0$\textbf{\ orbit\ }$\mathcal{O}_{non-BPS,Z=0,stu}$
(recall Eq. (\ref{stu-stu})):
\begin{equation}
\begin{array}{l}
\left| Z\right| _{non-BPS,Z=0,stu}^{2}-\left| D_{s}Z\right|
_{non-BPS,Z=0,stu}^{2}= \\
\\
=-\left| D_{s}Z\right| _{non-BPS,Z=0,stu}^{2}=-\left| Z\right| _{\frac{1}{2}%
-BPS,stu}^{2}<0.
\end{array}
\end{equation}
Consequently, the orbits $\mathcal{O}_{\frac{1}{2}-BPS,stu}$ and $\mathcal{O}%
_{non-BPS,Z=0,stu}$ correspond to two separated branches of a disconnected
manifold, classified by the local value of the function\linebreak $sgn\left(
\left| Z\right| ^{2}-\left| D_{s}Z\right| ^{2}\right) $. Such a result can
be easily extended to a generic $n\in \mathbb{N}$, i.e. to a generic element
of the sequence $\frac{SU(1,1)}{U(1)}\otimes \frac{SO(2,2+n)}{SO(2)\otimes
SO(2+n)}$.\bigskip\ \textbf{\ }

As for the classes of non-degenerate orbits of cases $VII$ and $VIII$, they
respectively read as follows:\medskip

\textbf{Case }$\mathbf{VII}$ $\left( G=SU(1,1)\right) \medskip $

$\frac{1}{2}$\textbf{-BPS class (}$H_{0}=1$):
\begin{equation}
\mathcal{O}_{\frac{1}{2}-BPS,VII}=\frac{G}{H_{0}}=SU(1,1),~~I_{4}>0;
\end{equation}

\textbf{Non-BPS, }$Z\neq 0$\textbf{\ class (}$\widehat{H}=1$):
\begin{equation}
\mathcal{O}_{non-BPS,Z\neq 0,VII}=\frac{G}{\widehat{H}}=SU(1,1),~~I_{4}<0.
\end{equation}
\medskip

\textbf{Case }$\mathbf{VIII}$ $\left( G=SU(1,1)\otimes SU(1,1)\right) $; it
can actually be obtained by putting $n=-1$ in the third row of Table
3:\medskip

$\frac{1}{2}$\textbf{-BPS class (}$H_{0}=U(1)$):
\begin{equation}
\mathcal{O}_{\frac{1}{2}-BPS,VIII}=\frac{G}{H_{0}}=SU(1,1)\otimes \frac{%
SO(2,1)}{SO(2)}=\frac{SU(1,1)\otimes SU(1,1)}{U(1)};
\end{equation}

\textbf{Non-BPS, }$Z=0$\textbf{\ class (}$\widetilde{H}=H_{0}=U(1)$):
\begin{equation}
\mathcal{O}_{non-BPS,Z=0,VIII}=\frac{G}{H_{0}}=\frac{SU(1,1)\otimes SU(1,1)}{%
U(1)};
\end{equation}

\textbf{Non-BPS, }$Z\neq 0$\textbf{\ class (}$\widehat{H}=SO(1,1)$):
\begin{equation}
\mathcal{O}_{non-BPS,Z\neq 0,VIII}=\frac{G}{\widehat{H}}=\frac{%
SU(1,1)\otimes SU(1,1)}{SO(1,1)}.\bigskip
\end{equation}
\medskip

Let us now consider the mass spectrum of the reducible sequence $II$ at the
regular critical points of $V_{BH}$.\smallskip

From the above analysis, it is clear that for the infinite set of reducible $%
N=2$, $d=4$ symmetric MESGTs based on $\frac{SU(1,1)}{U(1)}\otimes \frac{%
SO(2,2+n)}{SO(2)\otimes SO(2+n)}$, the structure of the Hessian at $\frac{1}{%
2}$-BPS and non-BPS, $Z=0$ critical points will be the same.

As it was pointed out more than once above, in \textit{all} $N=2$, $d=4$
MESGTs (with \textit{regular} special K\"{a}hler geometry of the scalar
manifold) the $2n_{V}\times 2n_{V}$ real form $\mathbf{H}_{\frac{1}{2}%
-BPS,real}^{V_{BH}}$ of the $\frac{1}{2}$-BPS critical Hessian has $2n_{V}$
strictly positive real eigenvalues, and therefore \textit{all} $\frac{1}{2}$%
-BPS critical points of $V_{BH}$ are attractors in a strict sense, i.e. they
are \textit{stable} minima of $V_{BH}$. Consequently, one immediately
obtains that in the case of $\frac{SU(1,1)}{U(1)}\otimes \frac{SO(2,2+n)}{%
SO(2)\otimes SO(2+n)}$ the same holds also for \textit{all} non-BPS, $Z=0$
critical points of $V_{BH}$: they \textit{all} are attractors in a strict
sense, i.e. they are \textit{stable} minima of $V_{BH}$.\smallskip

Concerning the non-BPS, $Z\neq 0$ mass spectrum, one can finally say that,
despite the reducibility of such manifolds, the same analysis performed in
Sect. \ref{N=2-Spectra} for (irreducible) $N=2$, $d=4$ ``magical'' symmetric
MESGTs holds also for this case, and Tripathy and Trivedi's result \cite{TT}
is confirmed.

Thus, it can generally be stated that the mass spectrum along $\mathcal{O}%
_{non-BPS,Z\neq 0}$ of all reducible $N=2$, $d=4$ symmetric $\frac{SU(1,1)}{%
U(1)}\otimes \frac{SO(2,2+n)}{SO(2)\otimes SO(2+n)}$-based MESGTs splits
under $\widehat{h}=m.c.s.\left( \widehat{H}\right) =SO(1+n)$ as follows:
there are $n_{V}-1=n+2$ \textit{massless }and $n_{V}+1=n+4$ \textit{massive }%
non-BPS, $Z\neq 0$ mass ``modes''.

Once again, Tripathy and Trivedi's result can also be confirmed by
performing explicit calculations in the highly symmetric and manageable case
of the $stu$ model\footnote{%
Clearly, since the isolated models $VII$ and $VIII$ are consistent
truncations of the $stu$ model, once the critical mass spectrum for the $stu$
model is known, one can obtain the critical mass spectra for the models $VII$
and $VIII$. Such results will be presented elsewhere \cite{to-appear}.} \cite
{to-appear}.

\end{document}